\documentclass[journal, twoside]{IEEEtran}
\ifCLASSINFOpdf
\else
\fi

\usepackage{times}
\usepackage{epsfig}
\usepackage{graphicx}
\usepackage{amsmath}
\usepackage{amssymb}
\usepackage{multirow}
\usepackage{amsmath,bm,makecell,amsthm}
\usepackage{algorithm}
\usepackage{algorithmic, multirow, url}

\begin{document}

\title{Embedding Bilateral Filter in Least Squares for Efficient Edge-preserving Image Smoothing}

\author{Wei Liu, Pingping Zhang, Xiaogang Chen, Chunhua Shen, Xiaolin Huang \IEEEmembership{Senior Member, IEEE} and Jie Yang
\thanks{Wei Liu, Xiaolin Huang and Jie Yang are with  the Institute of Image Processing and Pattern Recognition, Shanghai Jiao Tong University, China. (e-mail: \{liuwei.1989, xiaolinhuang, jieyang\}@sjtu.edu.cn)}
\thanks{Pingping Zhang is with the Dalian University of Technology, China. (email: jssxzhpp@mail.dlut.edu.cn)}
\thanks{Xiaogang Chen is with the University of Shanghai for Science and Technology, China. (email: xg.chen@live.com)}
\thanks{Chunhua Shen is with the University of Adelaide, Australia. (email: chunhua.shen@adelaide.edu.au)}
\thanks{Jie Yang and Xiaolin Huang are the corresponding authors of this paper.}
\thanks{This research is partly supported by NSFC, China (No: 61876107, 6151101179, 61603248) and 973 Plan, China (No. 2015CB856004), and 1000-Talent Plan (Young Program).}
\thanks{Copyright © 2018 IEEE. Personal use of this material is permitted. However, permission to use this material for any other purposes must be obtained from the IEEE by sending an email to pubs-permissions@ieee.org}
}


\maketitle

\begin{abstract}
Edge-preserving smoothing is a fundamental procedure for many computer vision and graphic applications. This can be achieved with either local methods or global methods. In most cases, global methods can yield superior performance over local ones. However, local methods usually run much faster than global ones. In this paper, we propose a new global method that embeds the bilateral filter in the least squares model for efficient edge-preserving smoothing. The proposed method can show comparable performance with the state-of-the-art global method. Meanwhile, since the proposed method can take advantages of the efficiency of the bilateral filter and least squares model, it runs much faster. In addition, we show the flexibility of our method which can be easily extended by replacing the bilateral filter with its variants. They can be further modified to handle more applications. We validate the effectiveness and efficiency of the proposed method through comprehensive experiments in a range of applications.
\end{abstract}

\begin{IEEEkeywords}
edge-preserving smoothing, bilateral filter, least squares, local methods, global methods, gradient reversals, halos
\end{IEEEkeywords}

\section{Introduction}
\label{SecIntroduction}

\begin{figure*}
  \centering
  \setlength{\tabcolsep}{0.5mm}
  \begin{tabular}{ccccc}
  \includegraphics[width=0.195\linewidth]{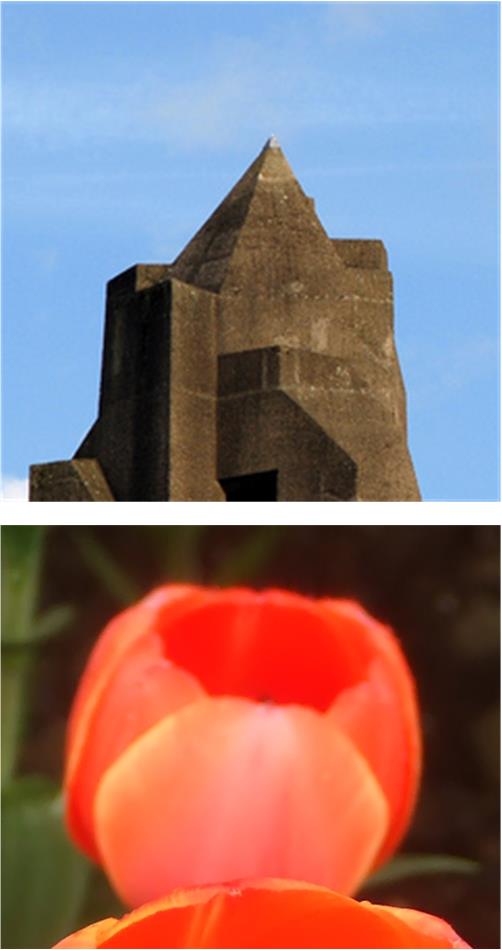} &
  \includegraphics[width=0.195\linewidth]{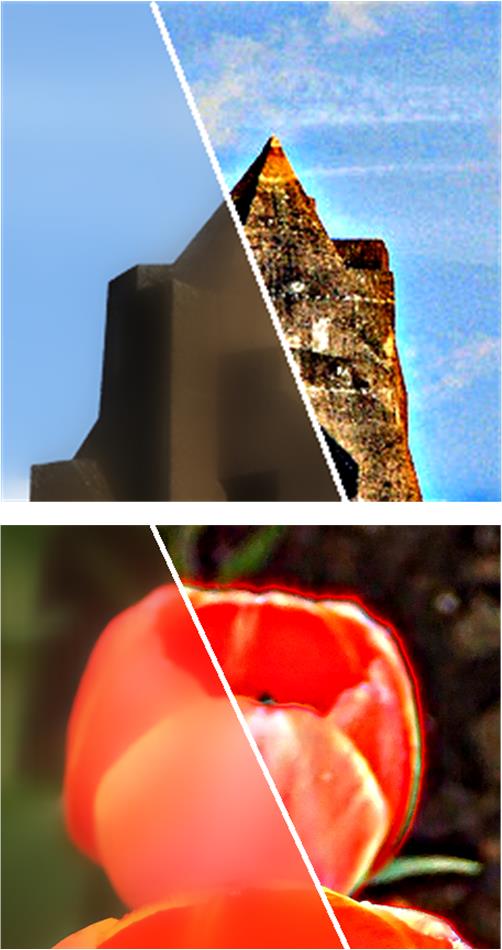} &
  \includegraphics[width=0.195\linewidth]{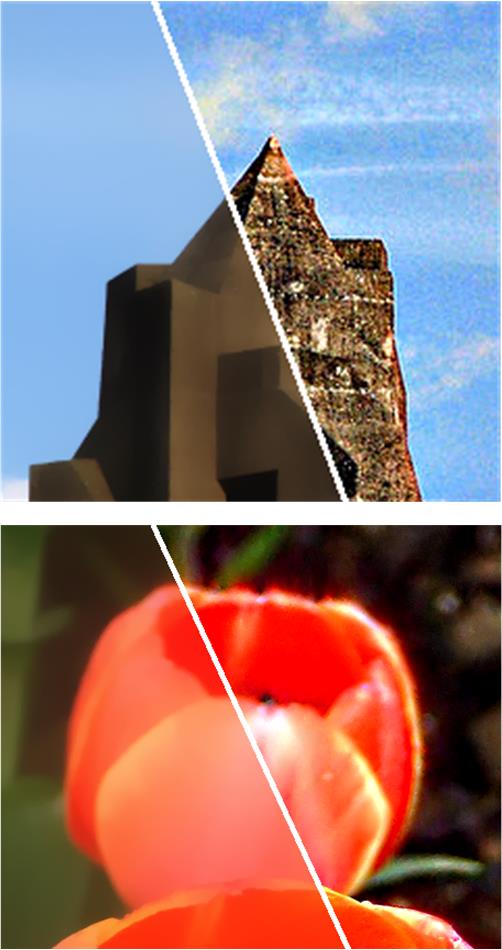} &
  \includegraphics[width=0.195\linewidth]{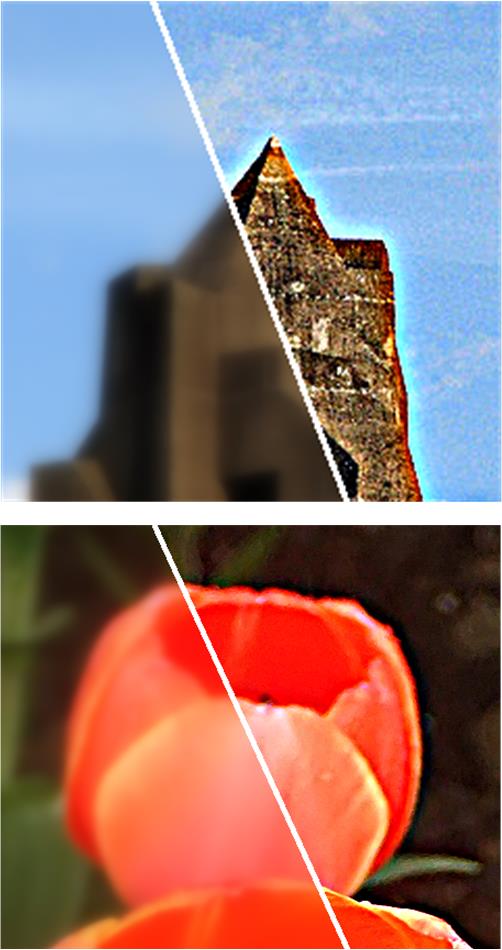} &
  \includegraphics[width=0.195\linewidth]{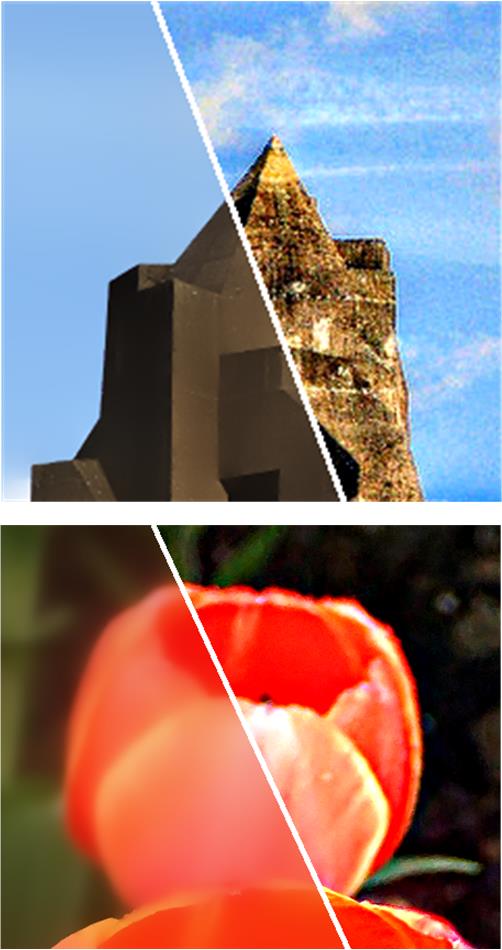}\\
  (a) Input & (b) BLF & (c) WLS & (d) LS & (e) BLF-LS
  \end{tabular}
  \caption{Image detail enhancement comparison of different methods. (a) Input images. Result of (b) BLF \cite{tomasi1998bilateral}, (c) WLS \cite{farbman2008edge}, (d) the LS model in Eq.~(\ref{EqLSObj}) and (e) Our BLF-LS. The left part of each image shows the smoothed image, and the right part illustrates the corresponding detail enhanced image with the detail layer $5\times$ boosted. Parameters of different methods are tuned to yield similar smoothing effects for comparison.}\label{FigHaloAndGradReversal}\vspace{-1em}
\end{figure*}

\IEEEPARstart{E}{dge}-preserving smoothing (EPS) has attracted increasing research interests in the fields of both computer vision and computational photography for decades. The main aim of EPS is to smooth out small details or perturbations in images and preserve the major edges and structures. To this end, many approaches have been proposed in the literature which can be roughly classified into two categories: local methods and global methods. Local methods usually calculate output pixel values as a weighted average of input pixel values inside a local window. Bilateral filter (BLF) \cite{tomasi1998bilateral} is one of the well-known local methods and has been widely used in a number of applications such as HDR tone mapping \cite{durand2002fast}, image detail enhancement \cite{fattal2007multiscale}, mesh smoothing \cite{fleishman2003bilateral}, noise smoothing \cite{tomasi1998bilateral}, artifact removal \cite{zhang2008adaptive}, and etc. As its variant, joint bilateral filter \cite{petschnigg2004digital} has also been used in flash/no flash filtering \cite{petschnigg2004digital, eisemann2004flash} and depth map upsampling \cite{Kopf2007Joint}. There are also methods that built on the BLF for advanced applications such as the rolling guidance filter \cite{zhang2014rolling} and the bilateral texture filtering \cite{cho2014bilateral}. Since the brute-force implementation of the BLF is computationally expensive, a number of approaches have been proposed to either accelerate the BLF \cite{adams2010fast, durand2002fast, paris2006fast, porikli2008constant, yang2009real} or introduce fast alternatives \cite{gastal2011domain, gastal2012adaptive}. There are also other local methods such as weighted median filter \cite{ma2013constant, zhang2014100+}, guided filter (GF) \cite{he2013guided} and its variants \cite{lu2012cross, tan2014multipoint}. Generally, local methods are computationally efficient. Some of them can even achieve real-time or near real-time image smoothing \cite{gastal2011domain, yang2009real}. However, these approaches can cause artifacts such as gradient reversals and halos in their results \cite{farbman2008edge, he2013guided}. Besides, they are less appropriate in preserving image details at arbitrary spatial scales \cite{farbman2008edge}.

Global methods usually formulate the image smoothing as an optimization framework, consisting of a data term and a regularization term that takes global properties of output images into account. The smoothed output image is the solution to the object function. The total variation smoothing \cite{rudin1992nonlinear} is the very early work which regularizes gradients with the $L_1$ norm penalty. Recently, Farbman et~al. \cite{farbman2008edge} proposed to smooth images with a weighted least squares (WLS) framework which imposed a weighted $L_2$ norm penalty on gradients. Their method shows superior performance over BLF \cite{tomasi1998bilateral} and methods based on BLF \cite{chen2007real, fattal2007multiscale} in producing results free of gradient reversals and halos. Gradient $L_0$ norm smoothing was adopted by Xu et~al. \cite{xu2011image} to sharpen salient edges while smooth out weak edges. There are also other global methods that are designed for specific tasks. For example, Xu et~al. \cite{xu2012structure} proposed the relative total variation smoothing which was specifically efficient for image texture removal. Ham et~al. \cite{ham2015robust} proposed the static dynamic filter which was especially effective for guided depth map restoration \cite{liu2017robust}. Generally, global methods can have state-of-the-art performance when compared with local methods. However, the superior performance is achieved with much higher computational cost. Even with the recent acceleration techniques \cite{ afonso2010fast, krishnan2013efficient}, the global methods are still an order of magnitude slower than the local methods.

In this paper, we propose a new global method for efficient edge-preserving smoothing. It can show comparable performance with the state-of-the-art global method but run much faster. Its computational cost is comparable with that of the state-of-the-art local methods. The main contributions of this paper are as follows:
\begin{itemize}
  \item We propose a new global method that embeds the BLF \cite{tomasi1998bilateral} in the least squares model for efficient edge-preserving smoothing. We show that the proposed method can produce results free of gradient reversals and halos which are comparable with the WLS \cite{farbman2008edge}. Meanwhile, it can take advantages of the efficiency of the BLF and least squares model, which makes it more than $10\times$ faster than the WLS.
  \item The proposed method is flexible which can be extended by replacing the BLF with its variants \cite{gastal2011domain, gastal2012adaptive} to yield different smoothing methods. In addition, the extended methods can be modified to handle more complex applications in an efficient manner.
\end{itemize}

The rest of this paper are organized as follows. In Sec.~\ref{SecRelatedWork}, we briefly present three smoothing methods that are highly related to the proposed method. In Sec.~\ref{SecProposedMethod}, we first show our motivation and then present the proposed method together with its possible extensions to handle more applications. Experimental results of our method and comparison with other state-of-the-art methods are shown in Sec.~\ref{SecExp}. We conclude our work in Sec.~\ref{SecConclusion}.

\section{Related Work}
\label{SecRelatedWork}

\begin{figure*}
  \centering
  \setlength{\tabcolsep}{0.5mm}
  \begin{tabular}{cccc}
  \includegraphics[width=0.245\linewidth]{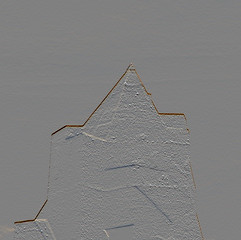} &
  \includegraphics[width=0.245\linewidth]{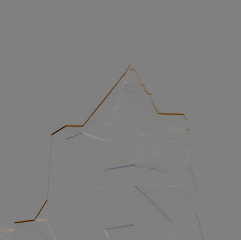} &
  \includegraphics[width=0.245\linewidth]{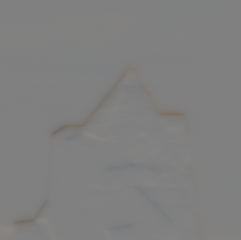} &
  \includegraphics[width=0.245\linewidth]{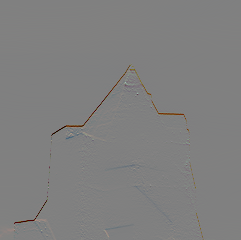} \\
  (a) Input & (b) WLS & (c) LS & (d) BLF-LS
  \end{tabular}
  \caption{Image gradients comparison. Gradients in (a)$\sim$(c) correspond to the $y$-axis direction gradients of the input and smoothed images in the first row of Fig.~\ref{FigHaloAndGradReversal}. Gradients of (a) the input image, (b) the image smoothed by WLS \cite{farbman2008edge}, (c) the image smoothed by the LS model in Eq.~(\ref{EqLSObj}). (d) The result of smoothing the gradients in (a) with the BLF \cite{tomasi1998bilateral}.}\label{FigGradComp}\vspace{-1em}
\end{figure*}

In this section, we briefly present three smoothing methods that are highly related to our method: the BLF \cite{tomasi1998bilateral}, the WLS framework \cite{farbman2008edge} and the least squares (LS) framework which is an un-weighted version of the WLS framework.

\subsection{The Bilateral Filter}
\label{SecBLF}

Given an input image $g$, each pixel in the output image $u$ of BLF \cite{tomasi1998bilateral} is a weighted average of its neighborhoods in $g$:
\begin{equation}\label{EqBLF}
\begin{split}
  & u_s = \frac{1}{Z_s}\sum\limits_{t\in N(s)}G_{\sigma_s}(\|s-t\|)G_{\sigma_r}(\|g_s-g_t\|)g_t,\\
  & \ \ Z_s = \sum\limits_{t\in N(s)}G_{\sigma_s}(\|s-t\|)G_{\sigma_r}(\|g_s-g_t\|),
\end{split}
\end{equation}
where $s$ and $t$ denote pixel positions. The spatial kernel $G_{\sigma_s}$ and the range kernel $G_{\sigma_r}$ are typically Gaussians, where ${\sigma_s}$ determines the spatial support and ${\sigma_r}$ controls the sensitivity to edges.

The advantage of BLF is that it has fast approximations \cite{adams2010fast, durand2002fast, paris2006fast, porikli2008constant, yang2009real} and alternatives \cite{gastal2011domain, gastal2012adaptive} which can achieve real-time or near real-time image processing. The disadvantage is that it can produce results with gradient reversals and halos which are also shared by its variants \cite{gastal2011domain, gastal2012adaptive}. The reason of gradient reversals is that edges are sharpened in the smoothed image and these edges are boosted in a reverse direction in the enhanced image. This usually happens when small $\sigma_r$ is adopted. Halos may occur when large $\sigma_s$ and $\sigma_r$ are adopted. Fig.~\ref{FigHaloAndGradReversal}(b) shows examples of these two kinds of artifacts with image detail enhancement.

\subsection{The WLS Model}
\label{SecWLS}

For an input image $g$, the output image $u$ of the WLS \cite{farbman2008edge} smoothing is the optimum of the following objective function:
\begin{equation}\label{EqWLSObj}
    \min\limits_u\sum\limits_{s}\left((u_s - g_s)^2 +\lambda\left(\omega_{x,s} (\nabla u_{x,s})^2 + \omega_{y,s} (\nabla u_{y,s})^2 \right)\right),
\end{equation}
where $s$ denotes the pixel position, $\nabla u_{x,s}$ and $\nabla u_{y,s}$ represent the gradients of $u$ at $s$ along $x$-axis and $y$-axis, respectively. The weights $\omega_{x,s}$ and $\omega_{y,s}$ are defined as:
\begin{equation}\label{EqWLSWeight}
    \omega_{x,s} = \left(|\nabla \mathcal{\ell}_{x,s}|^\alpha + \varepsilon\right)^{-1},\ \ \omega_{y,s} = \left(|\nabla \mathcal{\ell}_{y,s}|^\alpha + \varepsilon\right)^{-1},
\end{equation}
where $\mathcal{\ell}$ is the log-luminance channel of the input image $g$. The exponent $\alpha$ determines the sensitivity to the gradients of $\mathcal{\ell}$, $\varepsilon$ is a small constant that prevents division by zero in areas where $g$ is constant.

The objective function in Eq.~(\ref{EqWLSObj}) is quadratic and has closed-form solution which can be obtained by solving the following large sparse linear system:
\begin{equation}\label{EqWLSSolution}
    (I + \lambda L_g)u = g \ \rightarrow \ u = (I + \lambda L_g)^{-1}g,
\end{equation}
where $L_g=D_x^TW_xD_x + D_y^TW_yD_y$. $W_x$ and $W_y$ are diagonal matrixes containing $\omega_{x,s}$ and $\omega_{y,s}$. $D_x$ and $D_y$ are forward discrete difference operators along $x$-axis and $y$-axis direction. $D_x^T$ and $D_y^T$ are backward difference operators. Hence, $L_g$ is a five-point non-homogenous Laplacian matrix.

The advantage of WLS smoothing is that it can produce results free of gradient reversals and halos. Fig.~\ref{FigHaloAndGradReversal}(c) shows examples of image detail enhancement produced by the WLS. The disadvantage is that solving Eq.~(\ref{EqWLSSolution}) is time-consuming. Even with recent acceleration techniques \cite{afonso2010fast, krishnan2013efficient}, it is still an order of magnitude slower than the state-of-the-art local methods \cite{gastal2011domain, gastal2012adaptive, he2013guided}. Note that the recently proposed bilateral solver \cite{barron2016fast} is reported to enable a fast solution to the WLS problem like Eq.~(\ref{EqWLSObj}). However, it is designed to only handle Gaussian guidance weights in Eq.~(\ref{EqWLSObj}) which are in exponential form. It thus cannot solve the problem in Eq.~(\ref{EqWLSObj}) of which the guidance weights are in fractional form as defined in Eq.~(\ref{EqWLSWeight}).

\subsection{The LS model}

In fact, the high computational cost of Eq.~(\ref{EqWLSSolution}) results from the non-homogeneous Laplacian matrix $L_g$. This makes solving Eq.~(\ref{EqWLSSolution}) have to be performed in the image domain which requires an inverse of a very large matrix. However, if $L_g$ is a homogeneous Laplacian matrix, Eq.~(\ref{EqWLSSolution}) can be solved in the Fourier domain, which is much more efficient. By setting $\omega_{x,s}=\omega_{y,s}=1$ for all the pixel position $s$, we can get a homogeneous Laplacian matrix $L_g$. This leads Eq.~(\ref{EqWLSObj}) into an un-weighted one:
\begin{equation}\label{EqLSObj}
    \min\limits_u\sum\limits_{s}\left((u_s - g_s)^2 +\lambda\left((\nabla u_{x,s})^2 + (\nabla u_{y,s})^2 \right)\right).
\end{equation}

Since Eq.~(\ref{EqLSObj}) is un-weighted, we denote it as \emph{least squares} (LS) model in this paper. Solving the LS model in Eq.~(\ref{EqLSObj}) can be achieved as follows:
\begin{equation}\label{EqLSSolution}
\small
    u=\mathcal{F}^{-1}\left(\frac{\mathcal{F}(g)}{\mathcal{F}(1) + \lambda(\overline{\mathcal{F}(\partial_x)}\cdot\mathcal{F}(\partial_x) + \overline{\mathcal{F}(\partial_y)}\cdot\mathcal{F}(\partial_y))}\right),
\end{equation}
where $\mathcal{F}(\cdot)$ and $\mathcal{F}^{-1}(\cdot)$ are the fast Fourier transform (FFT) and inverse fast Fourier transform (IFFT) operators. $\overline{\mathcal{F}(\cdot)}$ denotes the complex conjugate of $\mathcal{F}(\cdot)$. $\mathcal{F}(1)$ is the FFT of the delta function. The plus, multiplication and division are all point-wise operations.

The advantage of Eq.~(\ref{EqLSObj}) is that it can be solved efficiently. This is because Eq.~(\ref{EqLSSolution}) transforms the matrix inverse in image domain into a point-wise division in the Fourier domain. And since both FFT and IFFT can be implemented very efficiently, solving Eq.~(\ref{EqLSObj}) with Eq.~(\ref{EqLSSolution}) is much more efficient than solving Eq.~(\ref{EqWLSSolution}) in the image domain. For example, for a $1024\times1024$ color image, solving Eq.~(\ref{EqLSObj}) with Eq.~(\ref{EqLSSolution}) in the Fourier domain only needs $\sim0.12$ seconds while solving Eq.~(\ref{EqWLSSolution}) in the image domain needs $\sim6.8$ seconds. The disadvantage is that Eq.~(\ref{EqLSObj}) is not an edge-preserving smoothing operator which can cause halos in the results. Since the LS model penalizes all the gradients, the gradients in the smoothed image cannot be larger than the corresponding ones in the input image. It thus does not cause gradient reversals. We show the results produced by the LS model in Fig.~\ref{FigHaloAndGradReversal}(d) where clear halos exist.

\section{The Proposed Method}
\label{SecProposedMethod}

\subsection{The BLF-LS Model}

All the three approaches mentioned in Sec.~\ref{SecRelatedWork} directly perform the smoothing on the intensities of each image. Each of them has its advantage and disadvantage where a tradeoff between the smoothing quality and processing speed exists. In contrast, we propose to first smooth the image gradients with the BLF and then embed the smoothed gradients into the LS framework for efficient edge-preserving smoothing. This can enable our method to maintain both the smoothing quality and processing efficiency.

Our method is motivated by the following two facts. First, we illustrate the gradients (along $y$-axis direction) of the image smoothed by the WLS in Fig.~\ref{FigGradComp}(b), which correspond to the smoothed image in the first row of Fig.~\ref{FigHaloAndGradReversal}(c). Compared with the gradients of the input image in Fig.~\ref{FigGradComp}(a), the small gradients in Fig.~\ref{FigGradComp}(b) have been removed while the large gradients remain almost un-changed and sharp. This motivates us that we can directly smooth the gradients of the input image with an edge-preserving smoothing operator, for example, the BLF \cite{tomasi1998bilateral}, to yield similar gradients. Fig.~\ref{FigGradComp}(d) shows the result of smoothing the gradients in Fig.~\ref{FigGradComp}(a) with the BLF. Clearly, large gradients are properly preserved with small gradients being removed, which is similar to the one in Fig.~\ref{FigGradComp}(b). Second, the LS model in Eq.~(\ref{EqLSObj}) is in fact a framework that penalizes all the gradients close to zero. It thus cannot preserve the large gradients, as shown in Fig.~\ref{FigGradComp}(c). To solve this problem, we propose to modify the LS model in Eq.~(\ref{EqLSObj}) to make the gradients of the smoothed image close to that in Fig.~\ref{FigGradComp}(d) rather than zero. This procedure can be achieved with the following optimization framework:
\begin{equation}\label{EqMyLSObj}
\small
  \min\limits_u\sum\limits_{s}\left((u_s - g_s)^2 +\lambda\sum\limits_{\ast\in\{x,y\}}\left(\nabla u_{\ast,s} - \left(f_{BLF}(\nabla g_\ast)\right)_s\right)^2\right),
\end{equation}
where $f_{BLF}(\nabla g_x)$ and $f_{BLF}(\nabla g_y)$ denote smoothing the $x$-axis and $y$-axis direction gradients of the input image $g$ with the BLF, respectively. With a sufficient large value of $\lambda$, the gradients of the output $u$, i.e., $\nabla u_\ast$, will be close to $f_{BLF}(\nabla g_\ast) (\ast\in\{x,y\})$. In this way, the smoothed image $u$ in Eq.~(\ref{EqMyLSObj}) will be similar to the one produced by the WLS. Fig.~\ref{FigHaloAndGradReversal}(e) shows the results obtained with our method. Specially, the image in the first row of Fig.~\ref{FigHaloAndGradReversal}(e) is obtained with the smoothed gradients in Fig.~\ref{FigGradComp}(d). Our method can produce results free of gradient reversals and halos.

Similar to Eq.~(\ref{EqLSSolution}), solving Eq.~(\ref{EqMyLSObj}) can also be implemented with a few FFTs and IFFTs. The output is obtained through:
\begin{equation}\label{EqMyLSSolution}
\small
u=\mathcal{F}^{-1}\left(\frac{\mathcal{F}(g) + \lambda\sum\limits_{\ast\in\{x,y\}}\overline{\mathcal{F}(\partial_\ast)}\cdot\mathcal{F}(f_{BLF}(\nabla g_\ast))}{\mathcal{F}(1) + \lambda\sum\limits_{\ast\in\{x,y\}}\overline{\mathcal{F}(\partial_\ast)}\cdot\mathcal{F}(\partial_\ast)}\right).
\end{equation}

Compared with Eq.~(\ref{EqLSSolution}), only two additional FFTs are needed in Eq.~(\ref{EqMyLSSolution}). Thus, Eq.~(\ref{EqMyLSSolution}) can also maintain the efficiency of Eq.~(\ref{EqLSSolution}). For example, it only takes $\sim0.16$ seconds to process a $1024\times1024$ color image.

Since the BLF smoothing can also be achieved with its fast approximations \cite{adams2010fast, durand2002fast, paris2006fast, porikli2008constant, yang2009real}, our method can be implemented in an efficient manner. In our experiments, we adopt the fast approximation proposed by Paris et~al. \cite{paris2006fast} for BLF smoothing. In this way, our method is able to achieve high smoothing quality and high processing efficiency as well.

In summary, our method can be implemented with the following two steps: (1) smoothing the $x$-axis and $y$-axis gradients of the input image with the BLF and (2) solving Eq.~(\ref{EqMyLSObj}) with Eq.~(\ref{EqMyLSSolution}) to obtain the smoothed image. Since our method embeds the BLF into the LS model, we denote our method in Eq.~(\ref{EqMyLSObj}) as \emph{BLF-LS}.

\subsection{Mathematical Explanation}
\label{SecMathExp}

To have a better understanding of our model in Eq.~(\ref{EqMyLSObj}), we present its mathematical explanation in this subsection. It is well-known that in the BLF \cite{tomasi1998bilateral} smoothing, output is assumed to be piecewise constant inside a local support, which can be formulated as:
\begin{equation}\label{EqPieceConstantModel}
    u_s=c_k, \ s\in N(k),
\end{equation}
where $c_k$ is the expected constant value of output pixel values inside the local support $N(k)$.

We assume that the output image is spatially piecewise linear, which is formulated as:
\begin{equation}\label{EqPieceLinearModel}
    u_s=a_k\cdot s + b_k, \ s\in N(k),
\end{equation}
where $a_k$ and $b_k$ are the linear coefficients that are assumed to remain constant inside $N(k)$. $N(k)$ is defined the same as that in Eq.~(\ref{EqPieceConstantModel}).

By taking the derivative of $u_s$ with respect to $s$, we have:
\begin{equation}\label{EqPieceLinearModelDerivative}
    \frac{\partial u_s}{\partial s}=a_k, \ s \in N(k).
\end{equation}

Note that $\frac{\partial u_s}{\partial s}$ is in fact the gradient of $u$ at $s$, i.e. $\frac{\partial u_s}{\partial s}=\nabla u_s$. Based on the above facts, we can have the following conclusion: \emph{when the image is spatially piecewise linear as formulated in Eq.~(\ref{EqPieceLinearModel}), its gradients are piecewise constant as formulated in Eq.~(\ref{EqPieceLinearModelDerivative}).}

The comparison between Eq.~(\ref{EqPieceLinearModelDerivative}) and Eq.~(\ref{EqPieceConstantModel}) can motivate us with the following facts: When the image is piecewise constant as formulated in Eq.~(\ref{EqPieceConstantModel}), we can adopt the BLF to smooth the intensities of the input image to get the intensities of the output image. When the image is spatially piecewise linear as formulated in Eq.~(\ref{EqPieceLinearModel}), its gradients are piecewise constant as formulated in Eq.~(\ref{EqPieceLinearModelDerivative}). Similarly, we can also adopt the BLF to smooth the gradients of the input image to get the gradients of the output image. By denoting the input image as $g$, the output image as $u$, their gradients as $\nabla g$ and $\nabla u$, respectively, we can formulate the above statement as:
\begin{equation}\label{EqPieceConstantDerivative}
    \nabla u = f_{BLF}(\nabla g),
\end{equation}
where both $\nabla u$ and $\nabla g$ represent gradients along either $x$-axis or $y$-axis.

Having stated the key conclusion, we also have a problem at hand: we cannot get the final smoothed image $u$ only from its gradients $\nabla u$ which are equal to $f_{BLF}(\nabla g)$. Note that the input image $g$ is also available at the same time. Hence, the smoothed image $u$ can be obtained through the following optimization problem:
\begin{eqnarray}\label{EqReconstructOri}
\begin{array}{c}
\underset{u}{\min}\|u-g\|^2\\
s.t.~~\nabla u = f_{BLF}(\nabla g).
\end{array}
\end{eqnarray}

However, the optimization problem in Eq.~(\ref{EqReconstructOri}) is essentially intractable because of the constraint. To make the problem tractable, we relax the constraint $\nabla u = f_{BLF}(\nabla g)$ with $\|\nabla u - f_{BLF}(\nabla g)\|^2\leq\epsilon$. Then for a sufficient small $\epsilon$, we have $\nabla u \approx f_{BLF}(\nabla g)$. The smoothed image $u$ now can be obtained via the following optimization problem:
\begin{eqnarray}\label{EqReconstructOriRelax}
\begin{array}{c}
\underset{u}{\min}\|u-g\|^2\\
s.t.~~\|\nabla u - f_{BLF}(\nabla g)\|^2\leq\epsilon.
\end{array}
\end{eqnarray}
Then for a proper value of $\lambda\propto\frac{1}{\epsilon}$, Eq.~(\ref{EqReconstructOriRelax}) can be exactly transformed into our model in Eq.~(\ref{EqMyLSObj}). In this way, we show our model in Eq.~(\ref{EqMyLSObj}) is in fact an approximation of the smoothing operator that assumes the intensities of the output image are spatially piecewise linear as formulated in Eq.~(\ref{EqPieceLinearModel}). Note that the trilateral filter in \cite{choudhury2005trilateral} also adopted a piecewise linear assumption, however, our method differs from theirs in both the motivation and the mathematical formulations of the approach.

\begin{figure*}
  \centering
  \setlength{\tabcolsep}{0.5mm}
  \begin{tabular}{cccc}
  \includegraphics[width=0.275\linewidth]{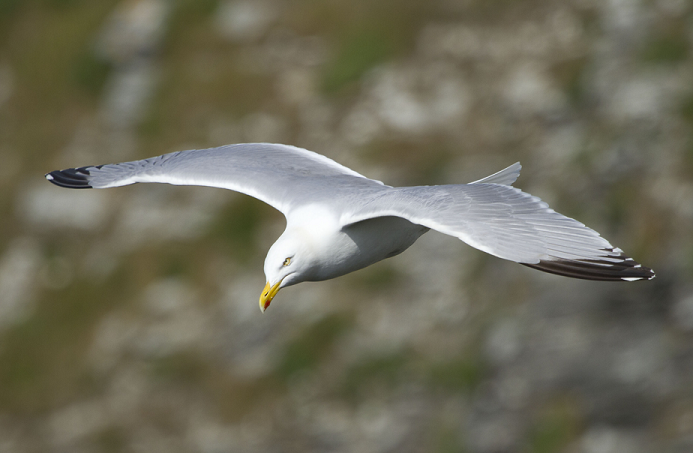}&
  \includegraphics[width=0.275\linewidth]{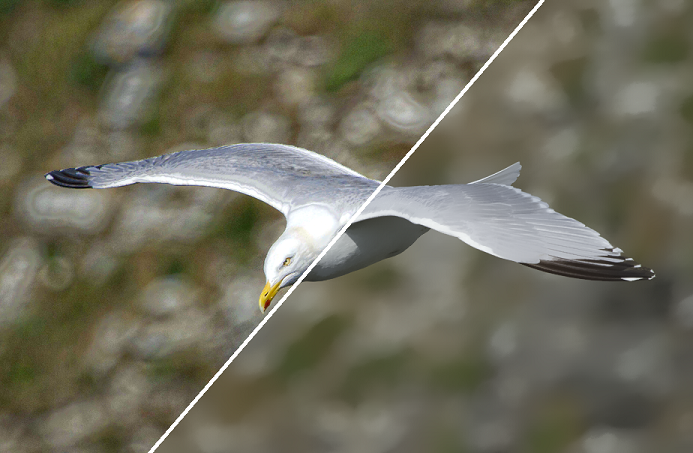}&
  \includegraphics[width=0.275\linewidth]{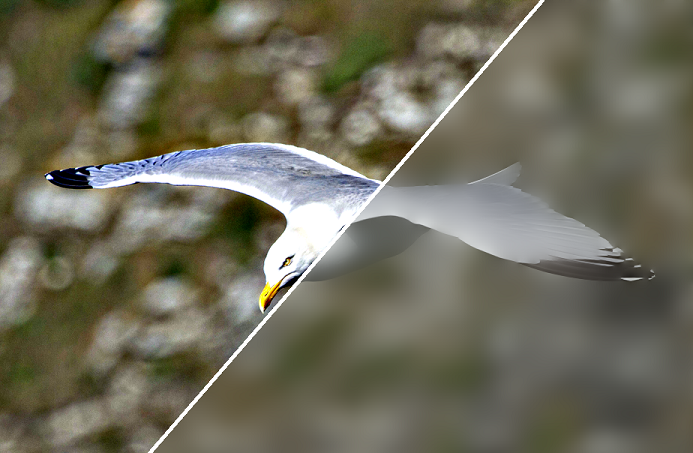}&
  \includegraphics[width=0.169\linewidth]{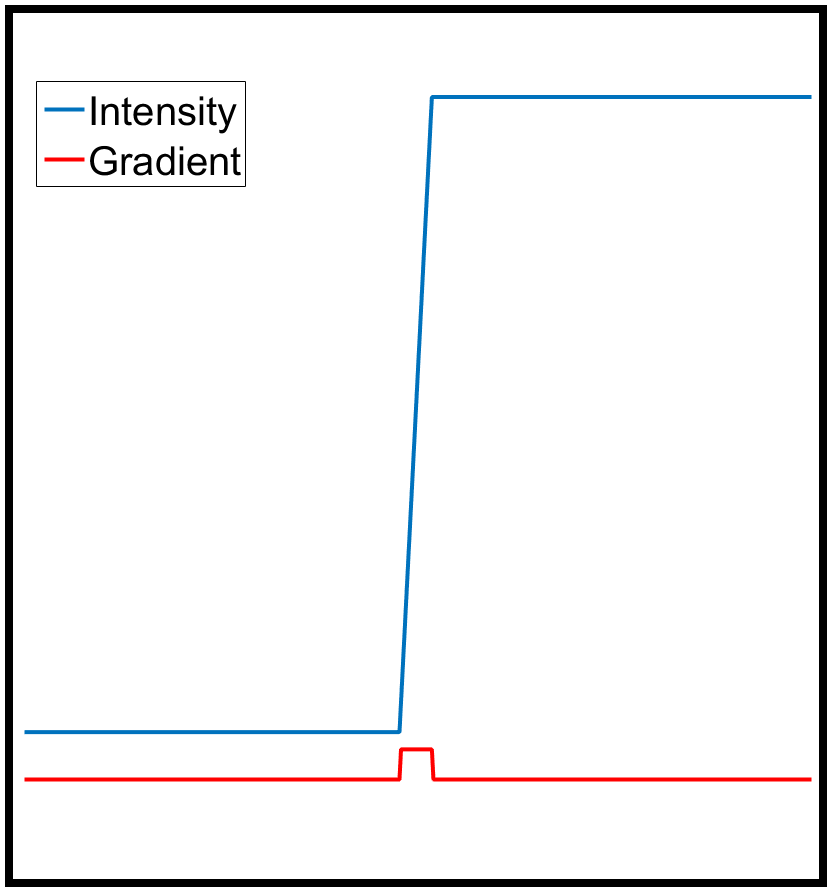}\\
  (a) Input & (b) BLF & (c) BLF-LS & (d)
  \end{tabular}
  \caption{Comparison of the BLF and our BLF-LS. (a) Input image. Smoothed image and the corresponding detail enhanced image with the detail layer $5\times$ boosted by (b) BLF with $\sigma_s=16,\sigma_r=0.06$, (c) our BLF-LS with $\sigma_s=16,\sigma_r=0.06$. For the same parameters, our BLF-LS can have stronger smoothing on the input image than the BLF. (d) A 1D illustration of a 1D signal in the intensity domain and its corresponding gradient domain signal. Two distinct points with larger difference in the intensity domain have much smaller difference in the gradient domain.}\label{FigBLFvsMy}
\end{figure*}

\begin{figure*}
  \centering
  \includegraphics[width=1\linewidth]{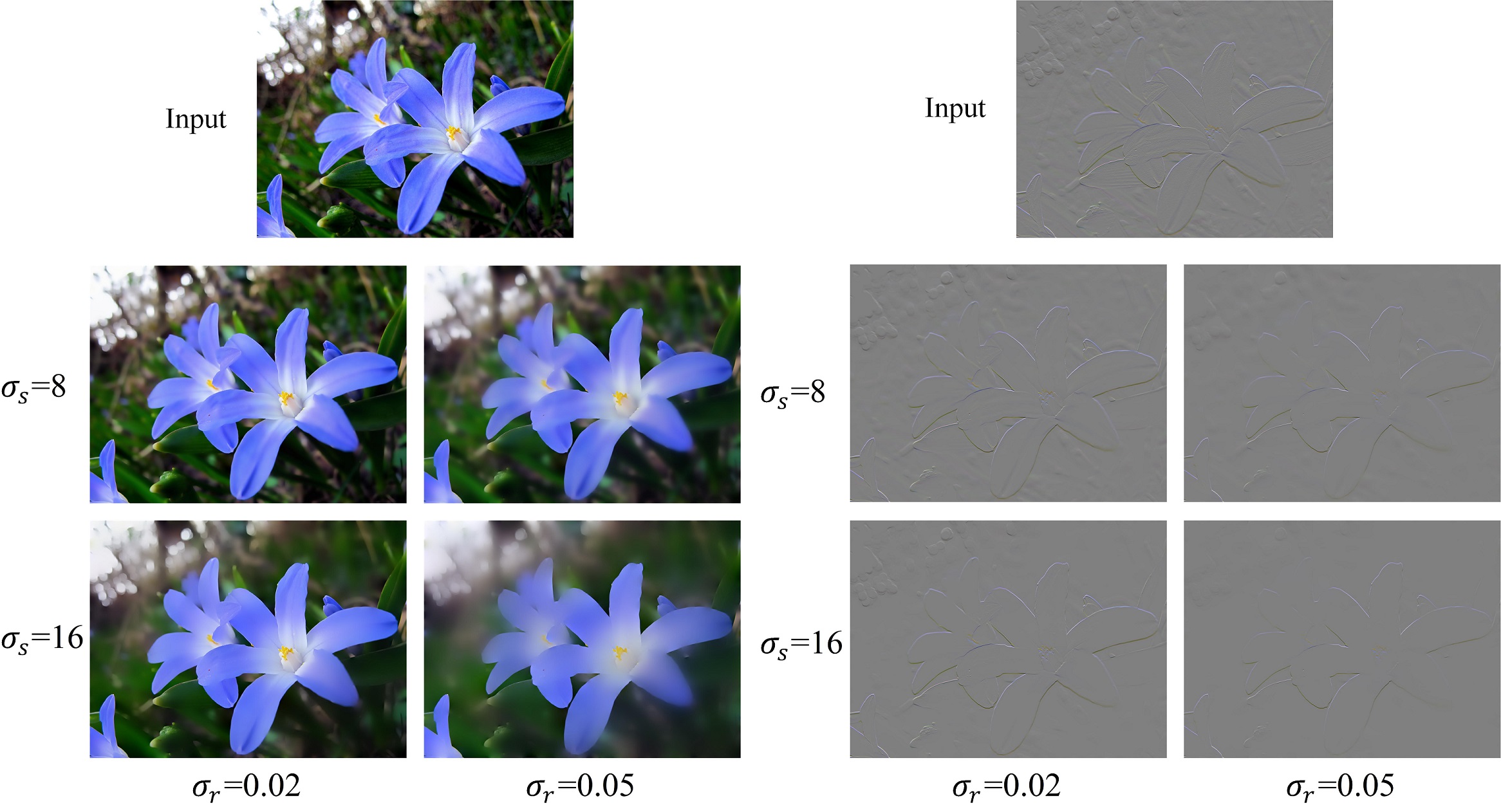}\\
  \vspace{-1em}\caption{The smoothing effect of our BLF-LS can be controlled in a similar way as that in BLF \cite{tomasi1998bilateral}. Larger $\sigma_s$ and $\sigma_r$ can yield stronger smoothing on the input image. The right part shows the $y$-axis direction gradients corresponding to the images on the left. We fix $\lambda=1024$ for all the images.}\label{FigParaDiscussSigma}\vspace{-1em}
\end{figure*}

\begin{figure*}
  \centering
  \includegraphics[width=1\linewidth]{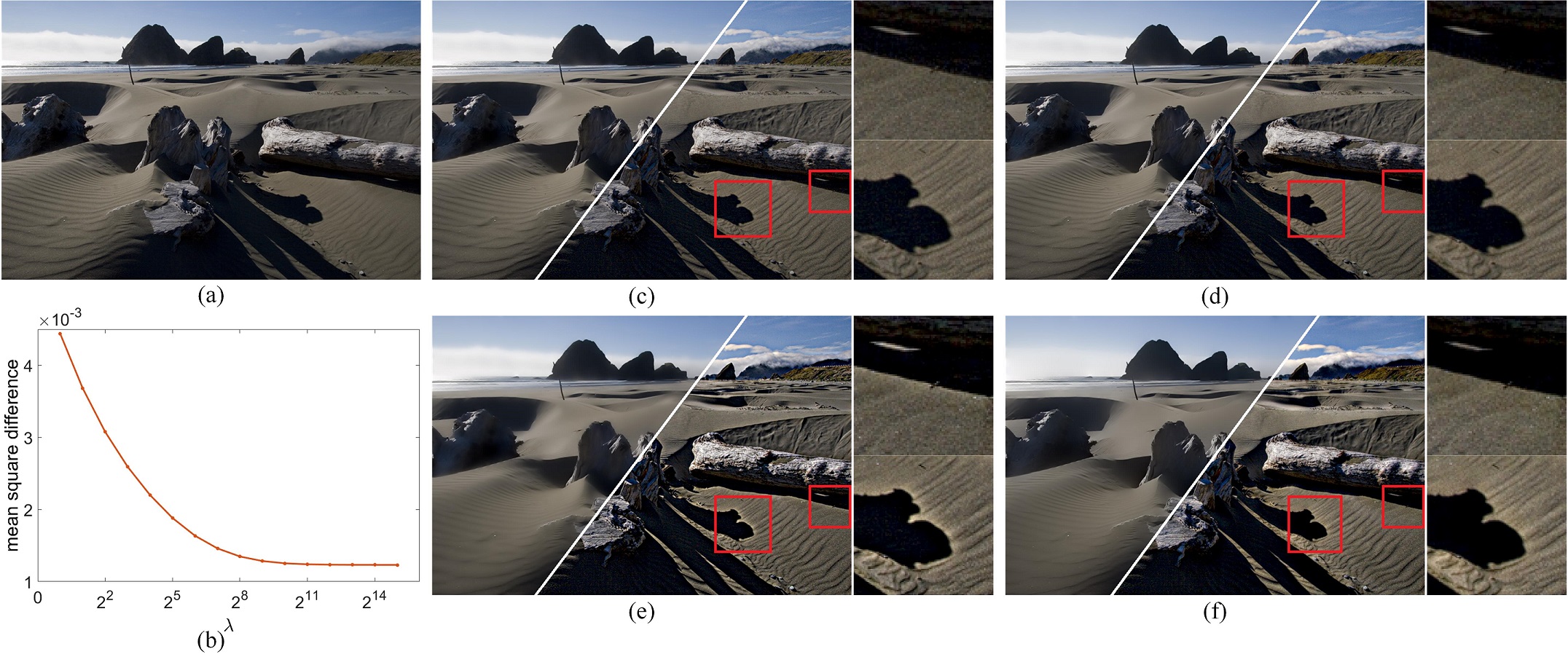}
  \vspace{-2em}\caption{(a) Input image. (b) Mean square difference between the result of smoothing the input gradients with the BLF and the gradients of the final smoothed image by our BLF-LS.  Smoothed image and the corresponding detail enhanced image with the detail layer $5\times$ boosted by our BLF-LS using (c) $\sigma_s=6, \sigma_r=0.015, \lambda =32$, (d) $\sigma_s=6, \sigma_r=0.015, \lambda =1024$, (e) $\sigma_s=12, \sigma_r=0.03, \lambda =32$, (f) $\sigma_s=12, \sigma_r=0.03, \lambda =1024$. For large $\sigma_s$ and $\sigma_r$, a small $\lambda$ can cause halos in the results which can be eliminated by using a larger value of $\lambda$.}\label{FigParaDiscussLambda}\vspace{-1em}
\end{figure*}

\begin{figure*}
  \centering
  \setlength{\tabcolsep}{0.5mm}
  \begin{tabular}{ccc}
  \includegraphics[width=0.33\linewidth]{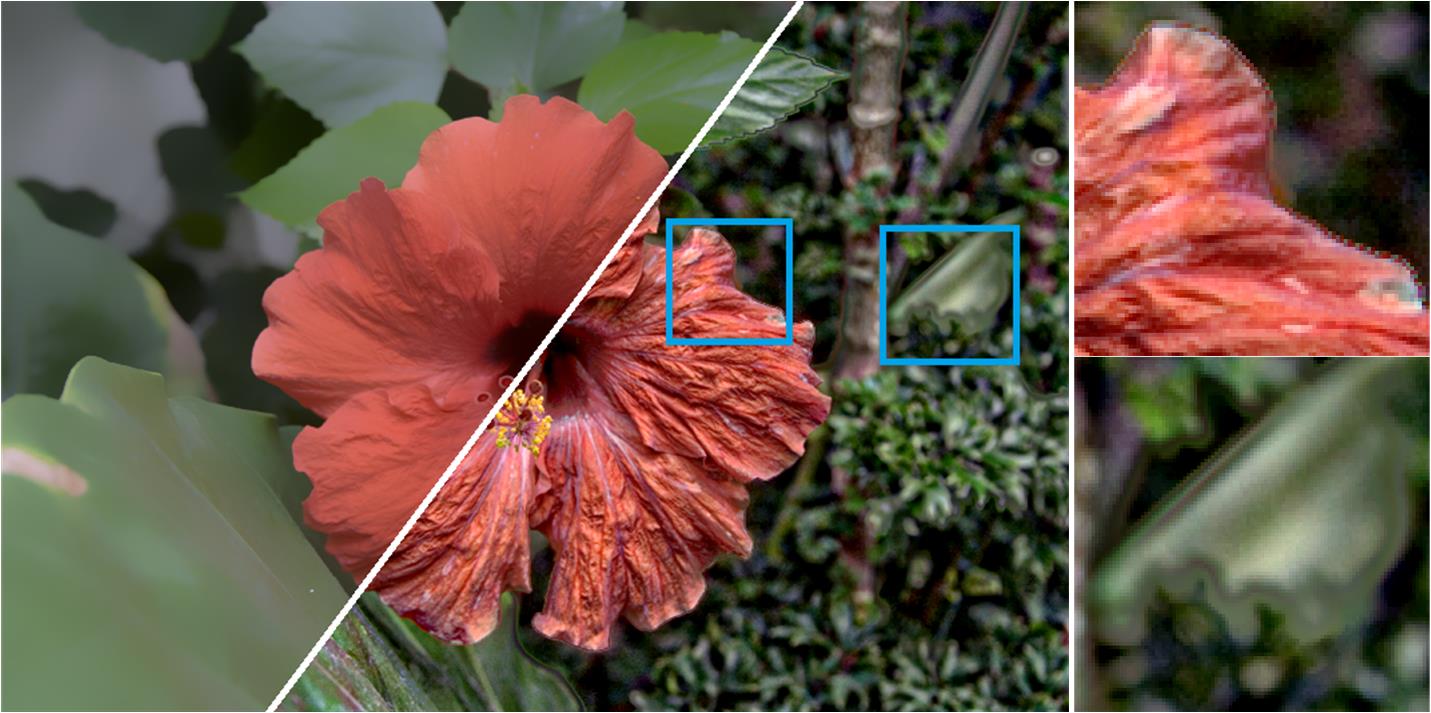}&
  \includegraphics[width=0.33\linewidth]{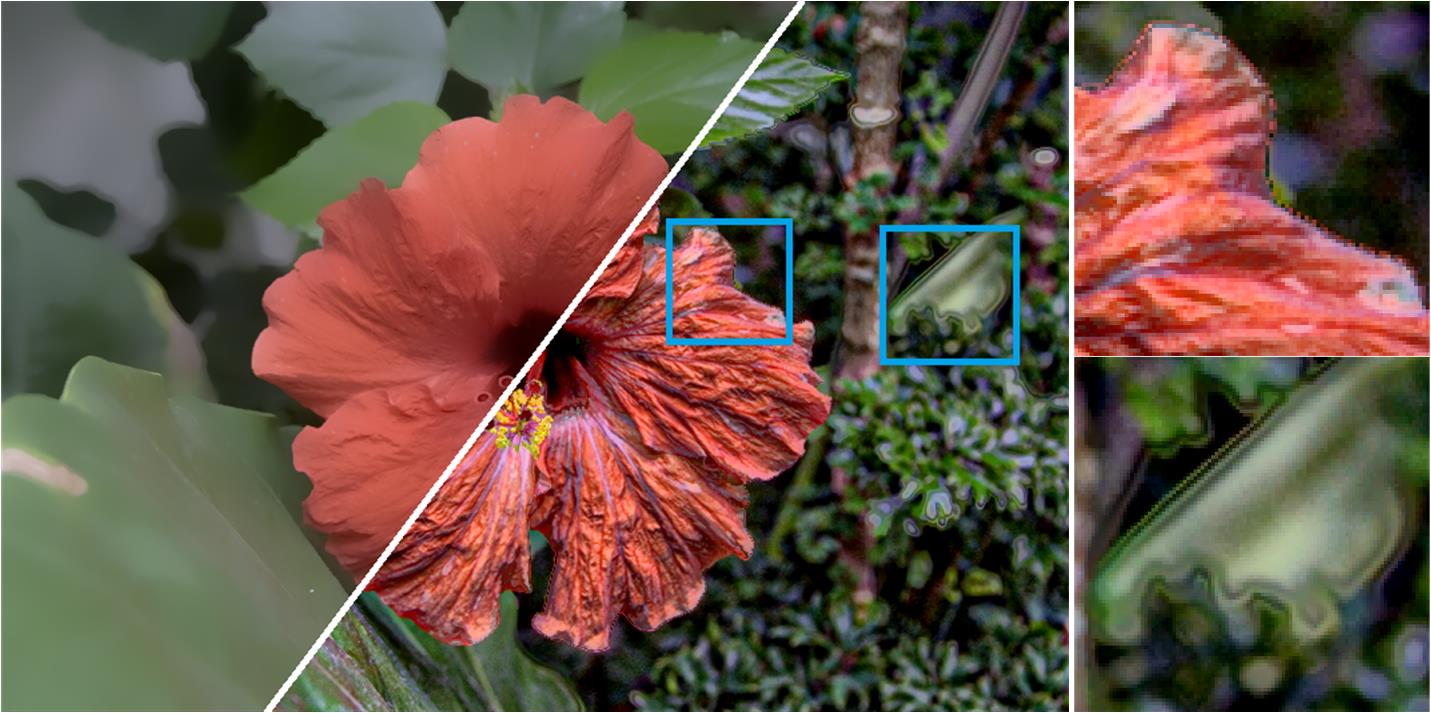}&
  \includegraphics[width=0.33\linewidth]{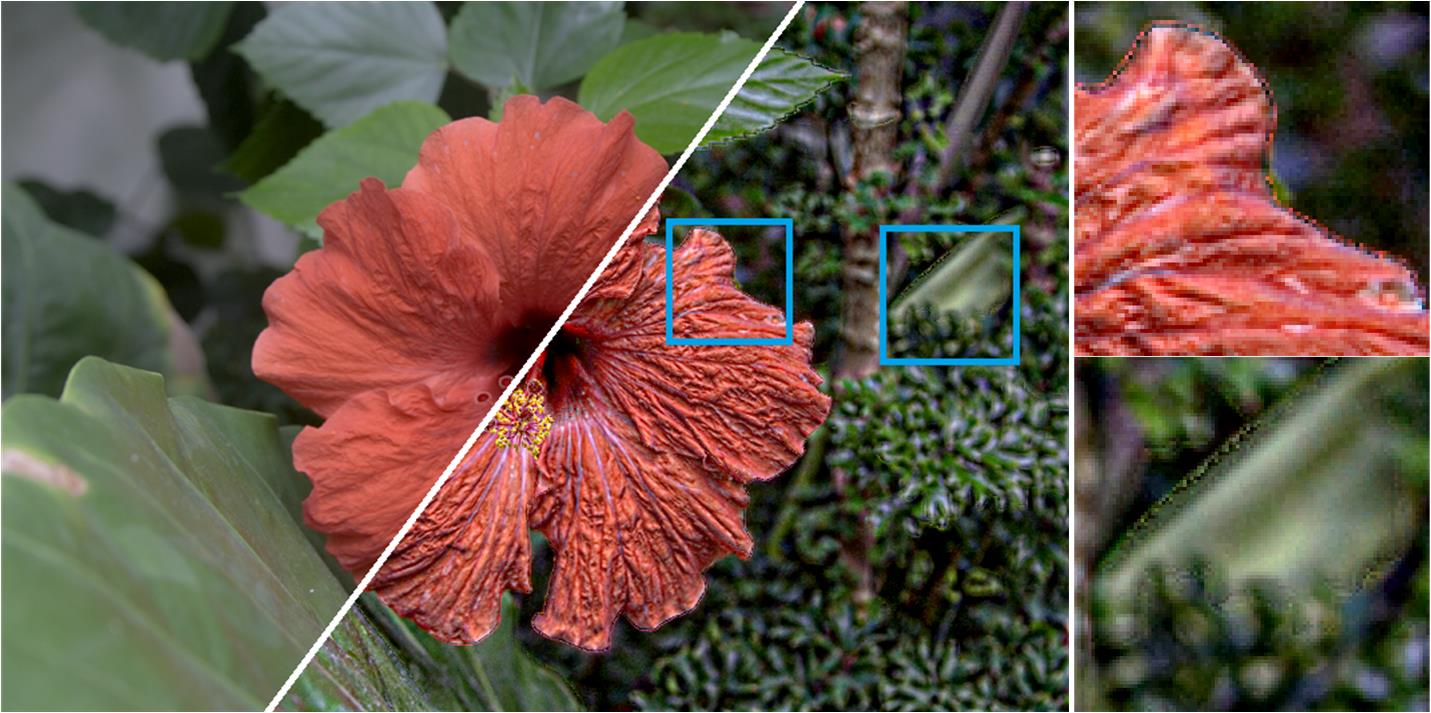}\\
  (a) AMF & (b) BLF & (c) NC\\

  \includegraphics[width=0.33\linewidth]{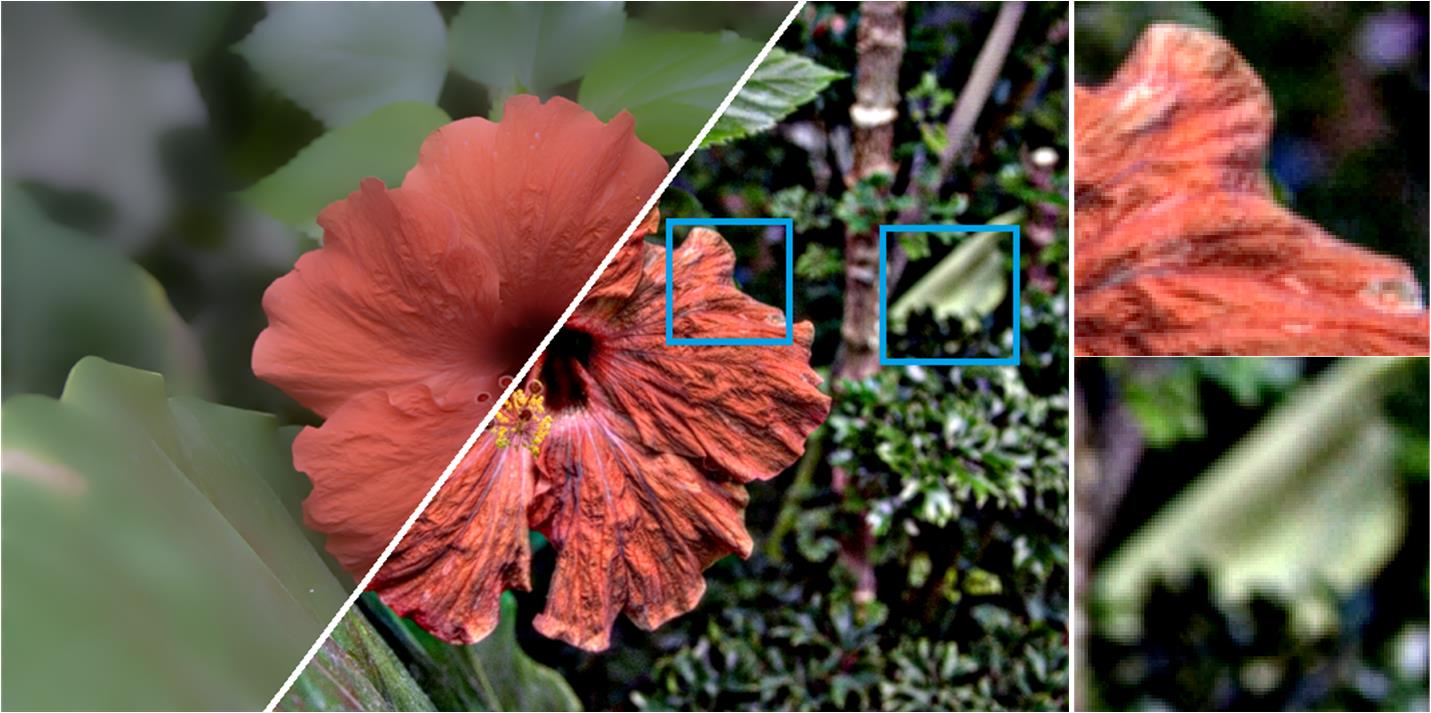}&
  \includegraphics[width=0.33\linewidth]{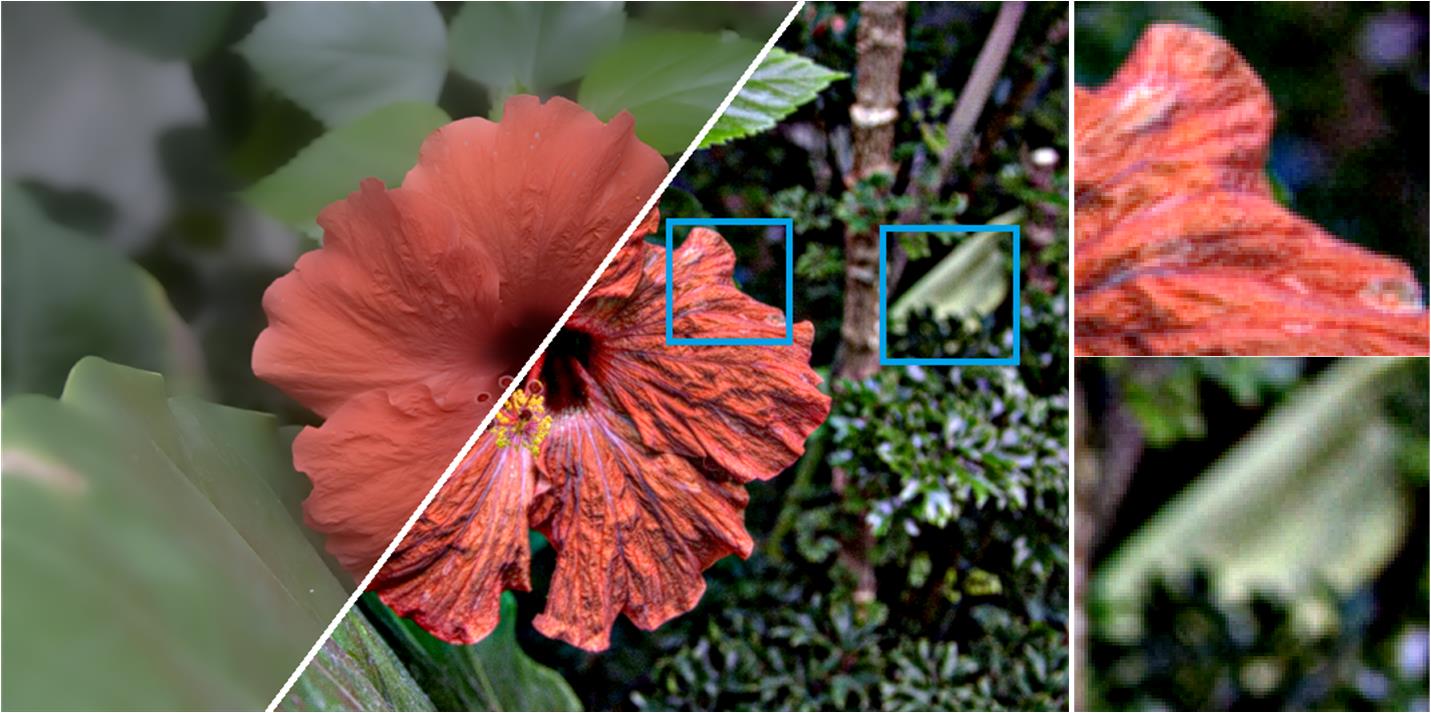}&
  \includegraphics[width=0.33\linewidth]{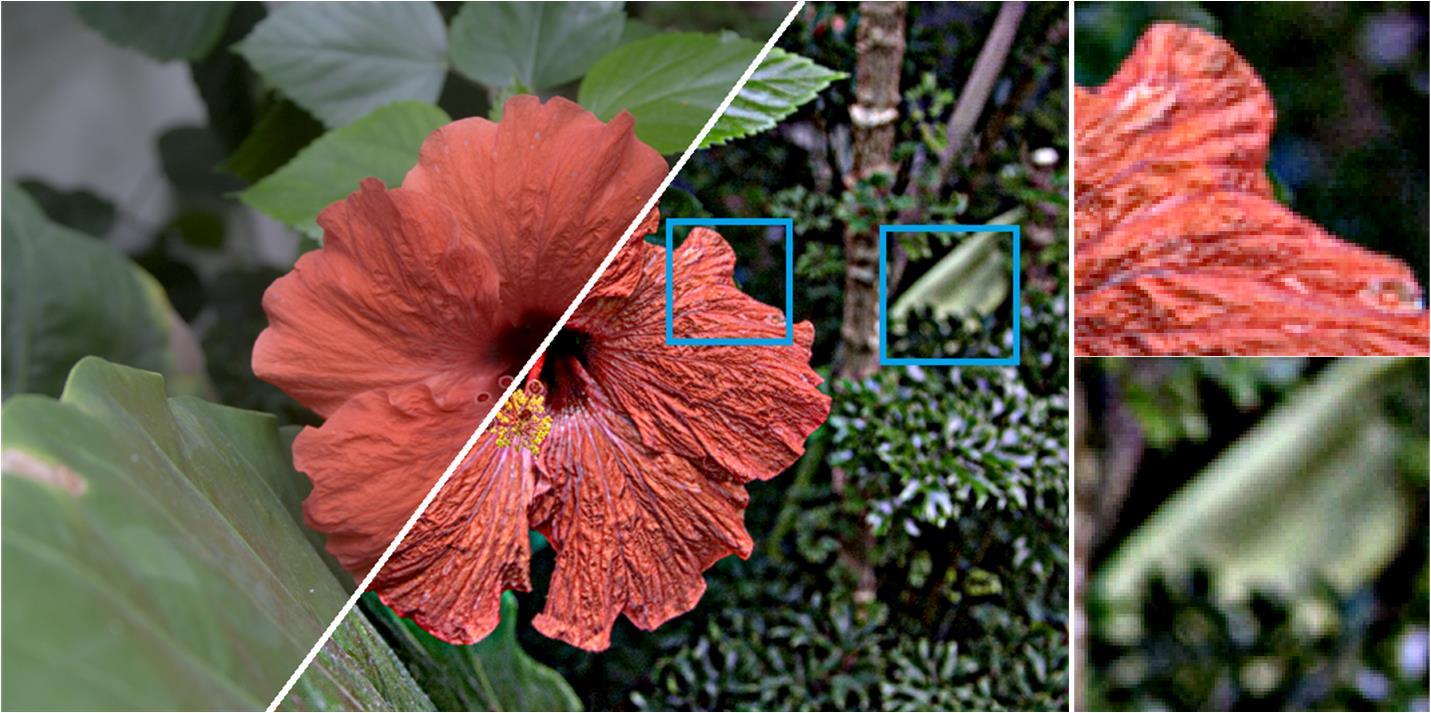}\\
  (a) AMF-LS & (b) BLF-LS & (c) NC-LS\\
  \end{tabular}
  \vspace{-0.5em}\caption{Smoothed images and the corresponding detail enhanced image with the detail layer $5\times$ boosted by (a) BLF \cite{tomasi1998bilateral} with $\sigma_s=12,\sigma_r=0.1$, (b) AMF \cite{gastal2012adaptive} with $\sigma_s=12,\sigma_r=0.1$ and (c) NC filter \cite{gastal2011domain} with $\sigma_s=12,\sigma_r=0.2$, (d) our BLF-LS with $\sigma_s=12,\sigma_r=0.02$, (e) our AMF-LS with $\sigma_s=12,\sigma_r=0.02$ and (f) our NC-LS with $\sigma_s=12,\sigma_r=0.07$. Our method can properly eliminate the gradient reversals in the images in the first row.}\label{FigExtensionDetailEnhancement}\vspace{-1em}
\end{figure*}

\begin{figure*}
  \centering
  \setlength{\tabcolsep}{0.5mm}
  \begin{tabular}{ccc}
  \includegraphics[width=0.33\linewidth]{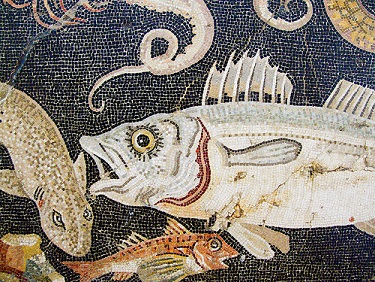}&
  \includegraphics[width=0.33\linewidth]{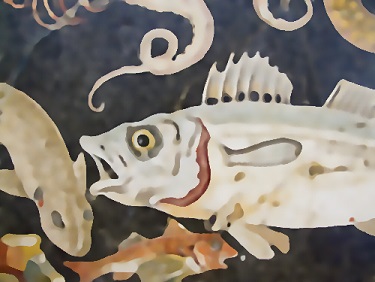}&
  \includegraphics[width=0.33\linewidth]{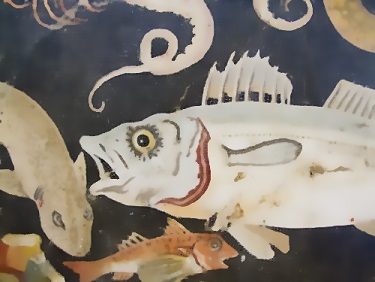}\\
  (a) Input & (b) RGF & (c) NC-LS
  \end{tabular}
  \vspace{-0.5em}\caption{(a) Input image. Its textures are removed by (b) rolling guidance filter \cite{zhang2014rolling} with $\sigma_s=5,\sigma_r=0.05,n^{iter}=6$, (c) our rolling guidance NC-LS with $\sigma_s=8,\sigma_r=0.02, n=3$, the initial guidance image is obtained by smoothing the input with a Gaussian filter of standard deviation $\sigma=2.5$. Our method can better smooth the textures while preserve image structures.}\label{FigExtensionTextureSmooth}
\end{figure*}

\subsection{Parameter Discussion}

As our method is highly related to the BLF \cite{tomasi1998bilateral}, we first show a simple comparison of the smoothing ability between our method and the BLF. Fig.~\ref{FigBLFvsMy} shows two smoothed images and the corresponding detail enhanced images obtained by the BLF and our method with the same $\sigma_s$ and $\sigma_r$. As observed from Fig.~\ref{FigBLFvsMy}(b) and Fig.~\ref{FigBLFvsMy}(c), our method can have stronger smoothing on the input image than the BLF. We can explain this phenomenon with Fig.~\ref{FigBLFvsMy}(d) and show that this results from the change in the range kernel of the BLF. More specifically, for two distinct pixels $a$ and $b$ in the intensity domain, the corresponding pixels in the gradient domain are denoted as $a'$ and $b'$. It is clear that the difference between $a$ and $b$ is much larger than the difference between $a'$ and $b'$. Note that the squared difference is used to calculate the weight between two pixels (see Eq.~(\ref{EqBLF}) for more details). Thus, for the same $\sigma_r$, the weight between $a'$ and $b'$ will be larger than the one between $a$ and $b$. This results in a stronger smoothing between $a'$ and $b'$ than that between $a$ and $b$. The above analysis also shows that we need to set a smaller value of $\sigma_r$ in our method than its usually required value in the BLF.

We then elaborate how to control the smoothing strength of our method. There are three parameters in our method: $\lambda$, $\sigma_s$ and $\sigma_r$. For a fixed $\lambda$, the smoothing strength of our method can be controlled by $\sigma_s$ and $\sigma_r$ in a similar way as that in BLF. Fig.~\ref{FigParaDiscussSigma} shows smoothed examples of our method with different values of $\sigma_s$ and $\sigma_r$. The visual comparison indicates that larger $\sigma_s$ and $\sigma_r$ result in stronger smoothing on the input image. The reason is that larger $\sigma_s$ and $\sigma_r$ can have stronger smoothing on the gradients which accordingly results in stronger smoothing on the input image. Fig.~\ref{FigParaDiscussSigma} also shows the $y$-axis gradients corresponding to the smoothed images. Clearly, small gradients are gradually smoothed out as $\sigma_s$ and $\sigma_r$ get larger while large gradients remain sharp. This allows our method to avoid halos when the smoothing strength is enlarged.

For the value of $\lambda$, we generally have $\lambda\propto\frac{1}{\epsilon}$ as stated in Sec.~\ref{SecMathExp}. Fig.~\ref{FigParaDiscussLambda}(b) shows the mean square difference between $\nabla u$ (gradients of the smoothed image) and $f_{BLF}(\nabla g)$ (the result of smoothing the gradients of the input image with the BLF). As the value of $\lambda$ increases, the difference between $\nabla u$ and $f_{BLF}(\nabla g)$ becomes smaller, which is consistent with the fact $\lambda\propto\frac{1}{\epsilon}$. According to our experimental results, for small $\sigma_s$ and $\sigma_r$ (i.e., weak smoothing on the image), our model with a smaller $\lambda=32$ can produce similar results as it with a larger $\lambda=1024$. We show two examples in Fig.~\ref{FigParaDiscussLambda}(c) and Fig.~\ref{FigParaDiscussLambda}(d). However, when $\sigma_s$ and $\sigma_r$ become large (i.e., strong smoothing on the image), our model with a smaller $\lambda=32$ can produce results with halos along salient edges as illustrated in Fig.~\ref{FigParaDiscussLambda}(e). When a larger $\lambda=1024$ is adopted, the halo artifacts are properly eliminated as shown in Fig.~\ref{FigParaDiscussLambda}(f). Thus, if not specified, we fix $\lambda=1024$ in all the experiments in this paper. This is different from the one in the WLS model where a quite small value of $\lambda$ is usually adopted. 

\vspace{-0.5em}
\subsection{Model Extensions}

Our method described in the above subsections can be extended in several ways. First, as there are many variants of the BLF such as the adaptive manifold filter (AMF) \cite{gastal2012adaptive} and the domain transform filter \cite{gastal2011domain}, we can replace the BLF in our method with these alternatives. Especially, we find the normalized convolution (NC) of the domain transform filter \cite{gastal2011domain} can well suit our method and produce promising results. We use \emph{AMF-LS} and \emph{NC-LS} to denote our method embedding the AMF and the NC filter, respectively. Fig.~\ref{FigExtensionDetailEnhancement} shows examples of detail enhancement produced by our AMF-LS and NC-LS. When compared with the results produced by the AMF and the NC filter, our AMF-LS and NC-LS can properly eliminate the gradient reversals. We will further show our method in handling the halos in the experimental section.

In addition, as can be observed in Fig.~\ref{FigExtensionDetailEnhancement}, our AMF-LS can produce results quite similar to our BLF-LS, while the results of our NC-LS are different from those of AMF-LS and BLF-LS. This is because the weighting scheme of the AMF is quite similar to that of the BLF but the weighting scheme of the NC filter is different from that of the other two. This means embedding different methods in the proposed framework can result in methods with different smoothing properties.

Besides the above extension, we can also use our method to perform joint image filtering other than the single image filtering. One straightforward extension can be achieved in a similar way as the joint bilateral filter \cite{petschnigg2004digital}. In this case, the guidance weights of the BLF, AMF and NC filter in our method are based on the gradients of another guidance image. They can be applied to tasks such as flash/no flash image filtering \cite{petschnigg2004digital}, which will be shown in our experimental part.

In addition, inspired by the rolling guidance filter (RGF) \cite{zhang2014rolling}, our NC-LS can also be used to handle more complex applications such as image texture removal and clip-art compression artifacts removal. We can use our NC-LS to iteratively jointly smooth the input image with the guidance of the updated image at each iteration. The initial guidance image is set as the output of smoothing the input image with a Gaussian filter of standard deviation $\sigma$. We show our NC-LS can achieve better performance than the RGF. Fig.~\ref{FigExtensionTextureSmooth}(c) illustrates one texture removal example of our method. When compared with the result of the RGF in Fig.~\ref{FigExtensionTextureSmooth}(b), our method can better smooth out textures while properly preserve image structures. In a similar way, we denote the above modification as rolling guidance NC-LS. We use $n$ to denote the iteration number of our rolling guidance NC-LS. In general, we find our rolling guidance NC-LS can achieve promising results within three iterations, i.e., $n\leq3$.

\section{Applications and Experimental Results}
\label{SecExp}

\begin{figure*}
  \centering
  \setlength{\tabcolsep}{0.25mm}
  \begin{tabular}{cccc}
  \includegraphics[width=0.245\linewidth]{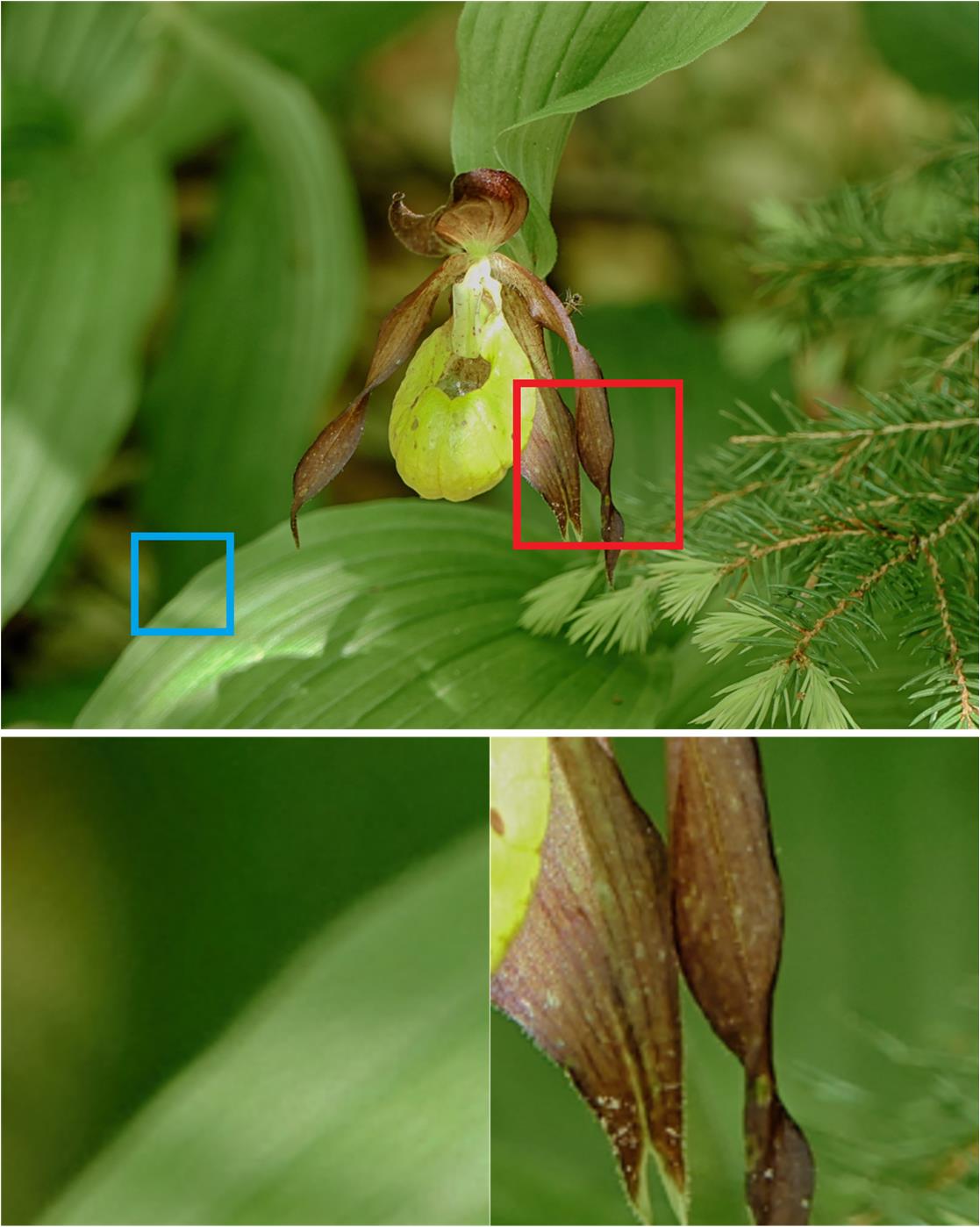}&
  \includegraphics[width=0.245\linewidth]{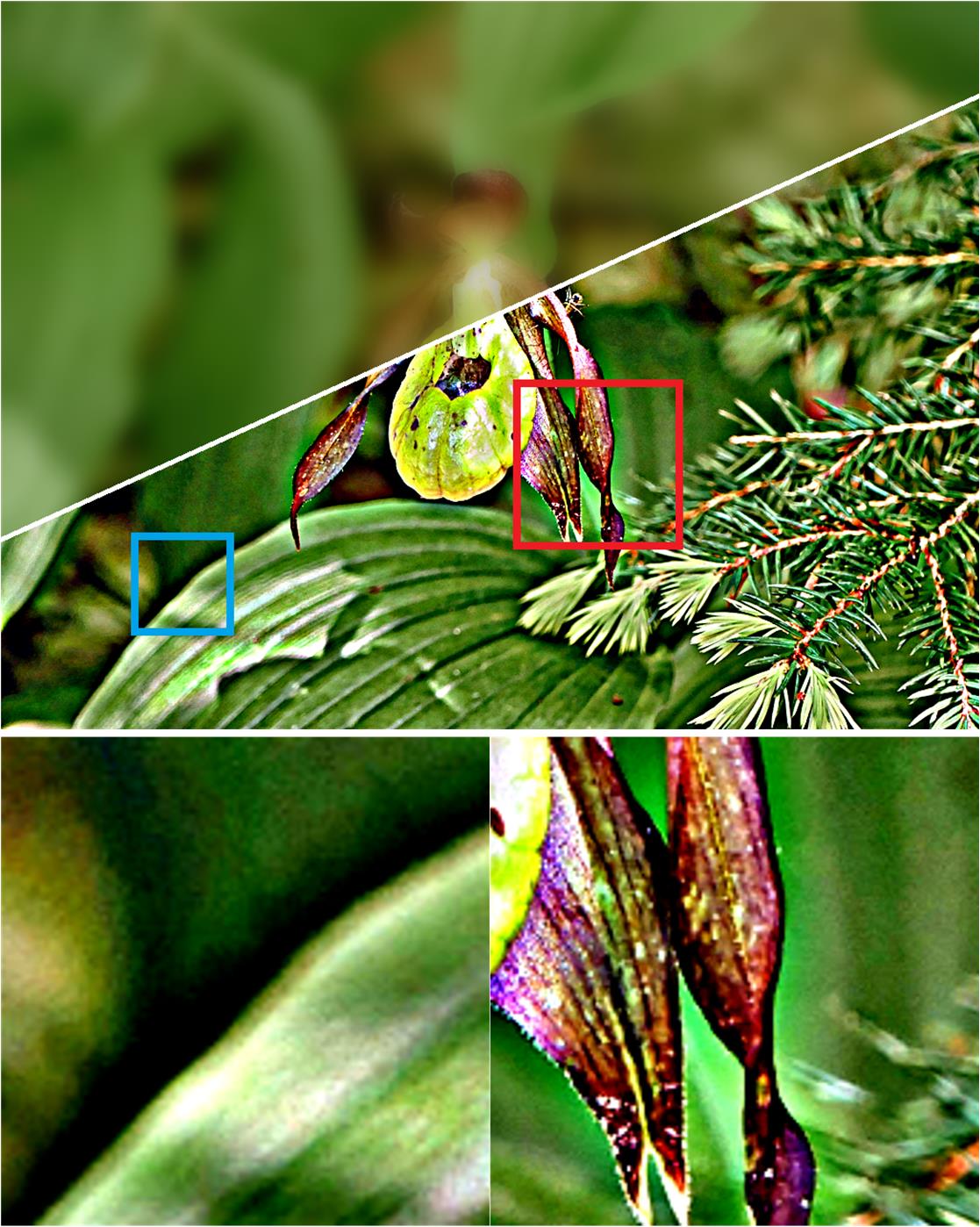}&
  \includegraphics[width=0.245\linewidth]{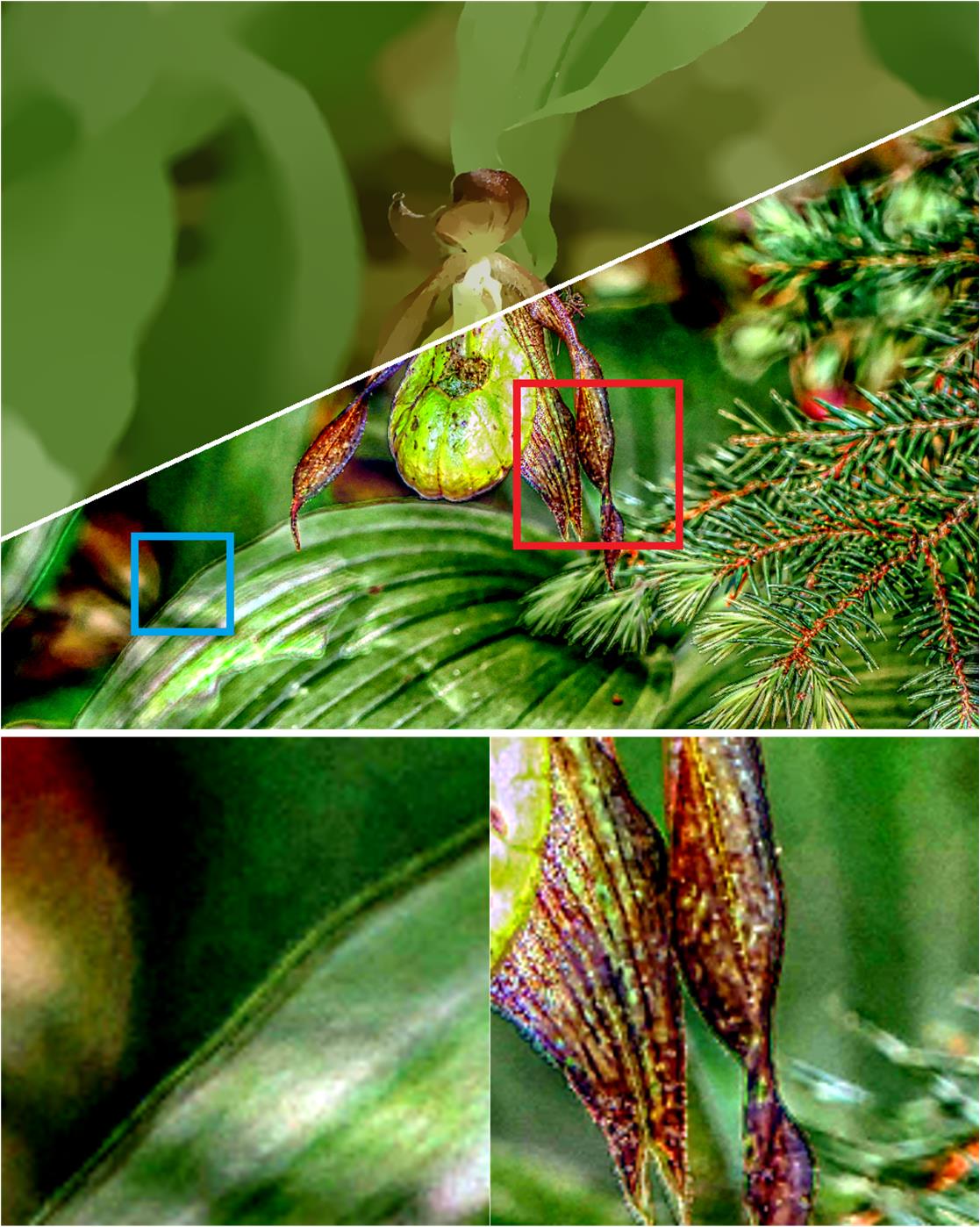}&
  \includegraphics[width=0.245\linewidth]{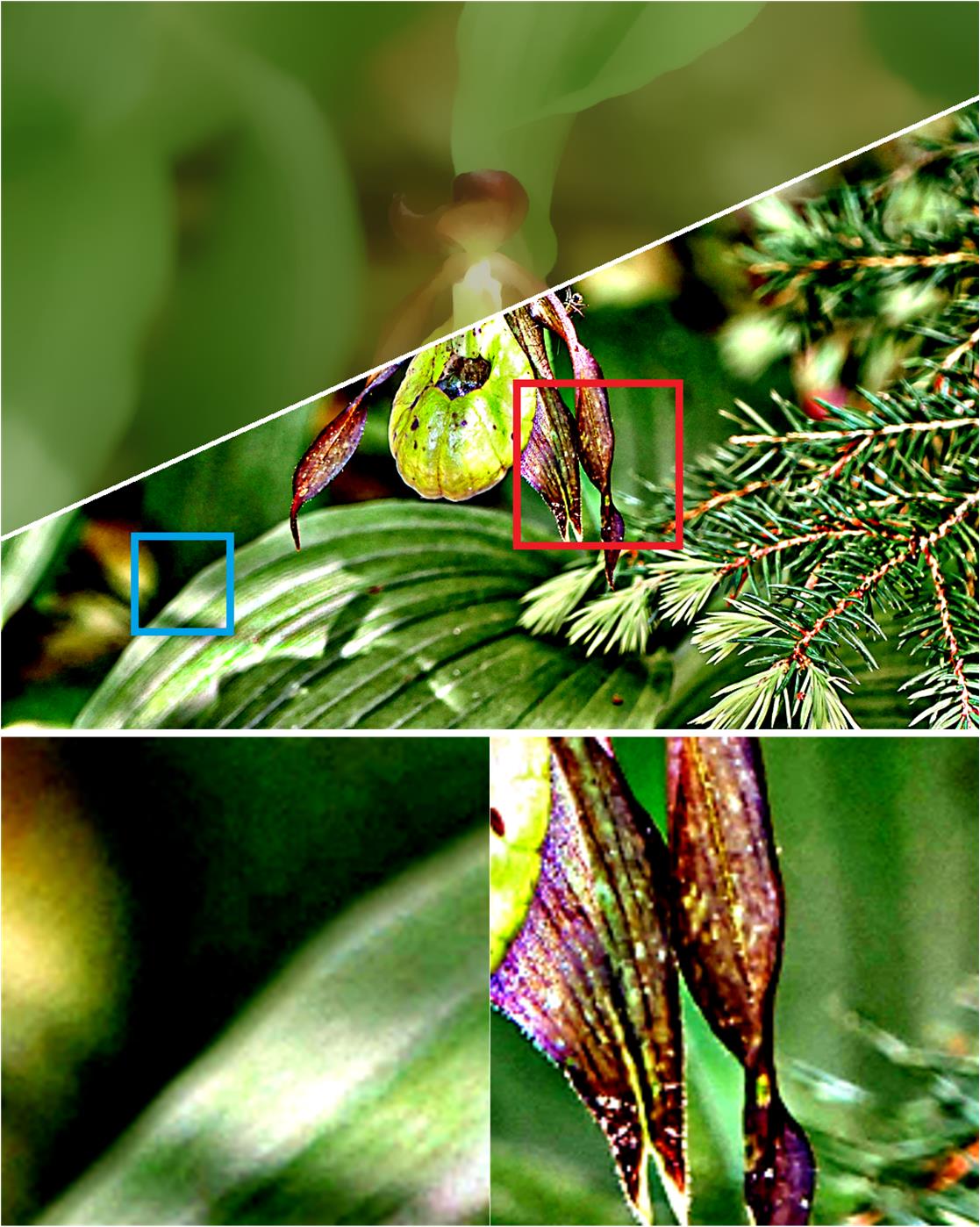}\\
  (a) Input & (b) GF & (c) $L_0$ norm & (d) WLS \\

  \includegraphics[width=0.245\linewidth]{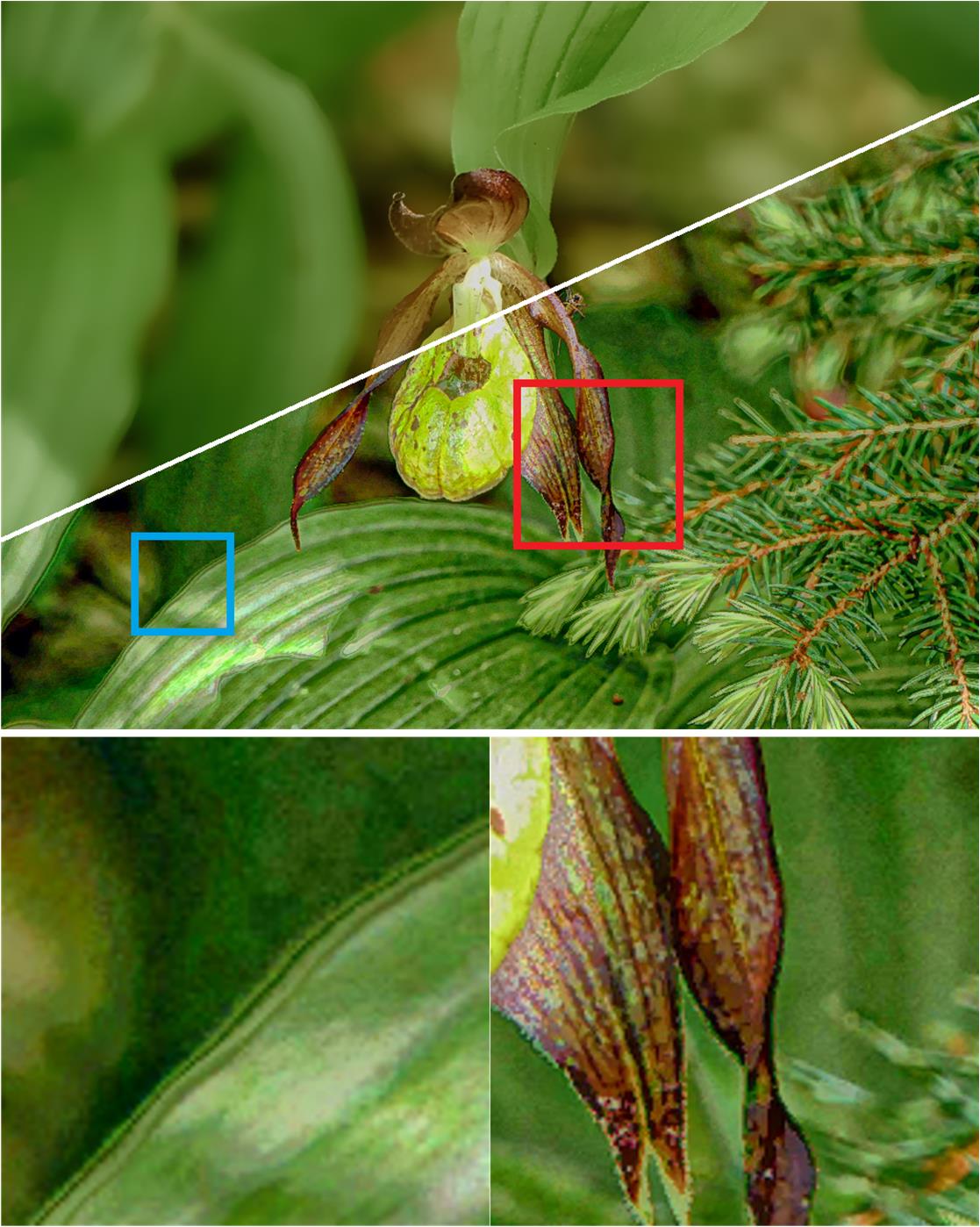}&
  \includegraphics[width=0.245\linewidth]{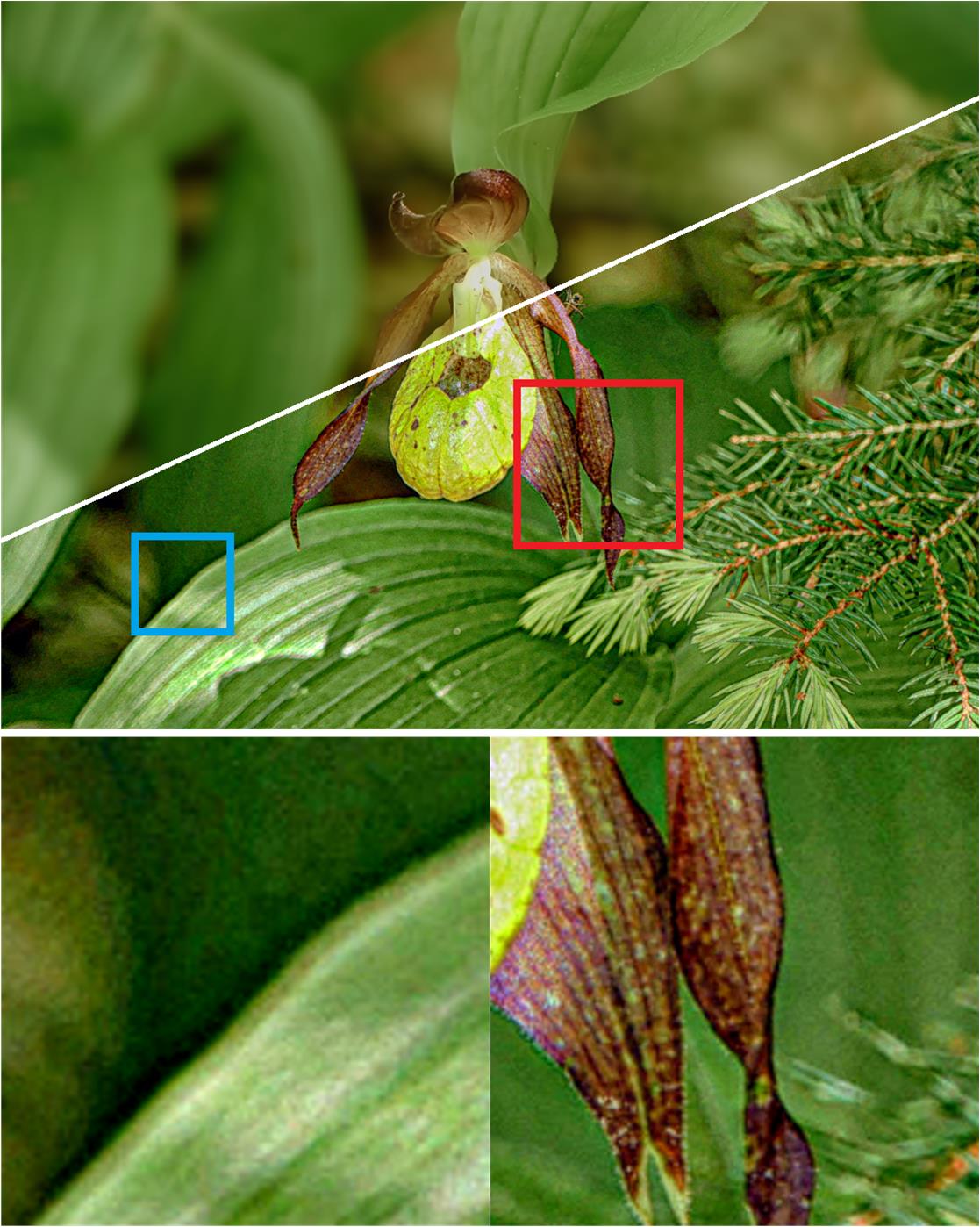}&
  \includegraphics[width=0.245\linewidth]{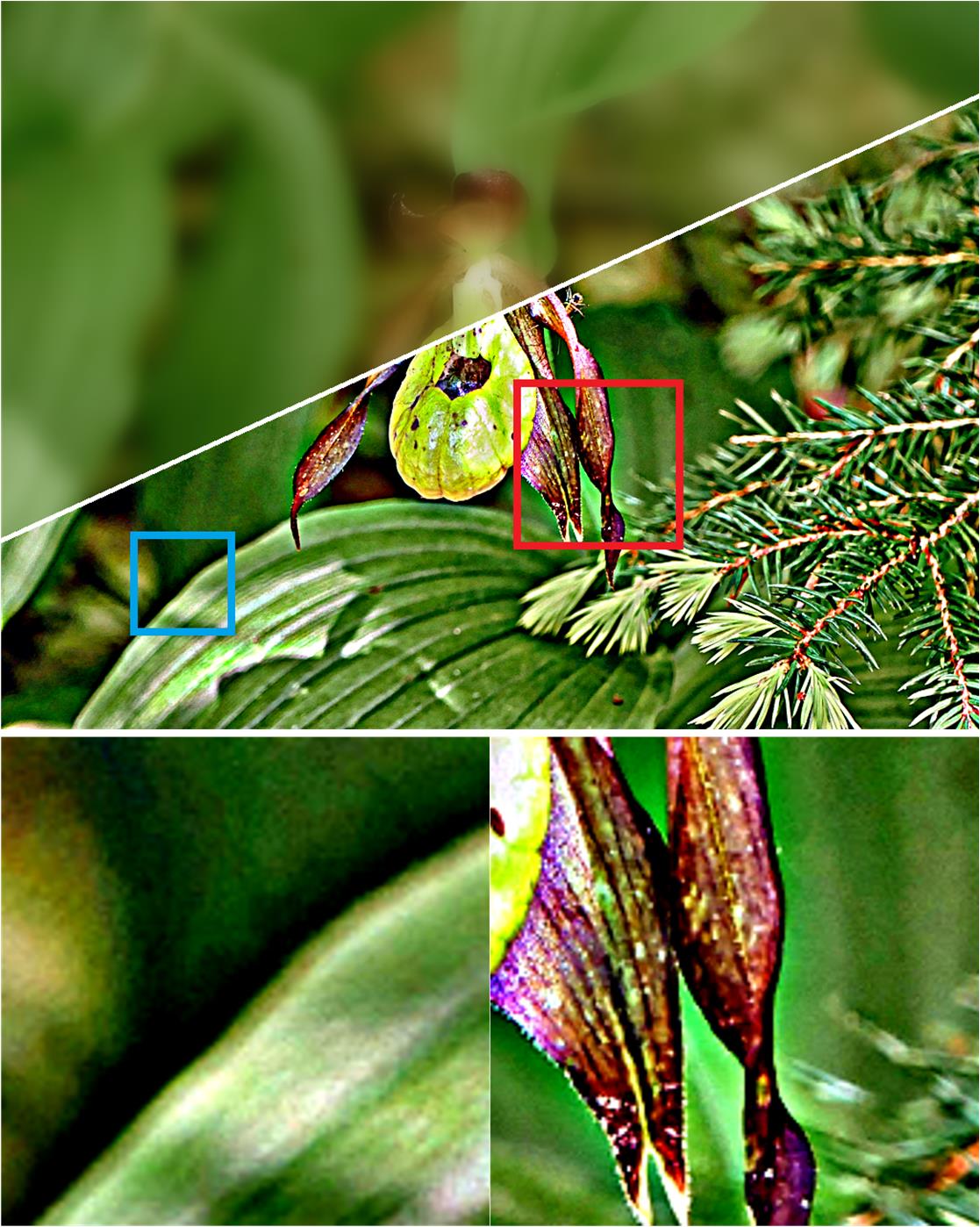}&
  \includegraphics[width=0.245\linewidth]{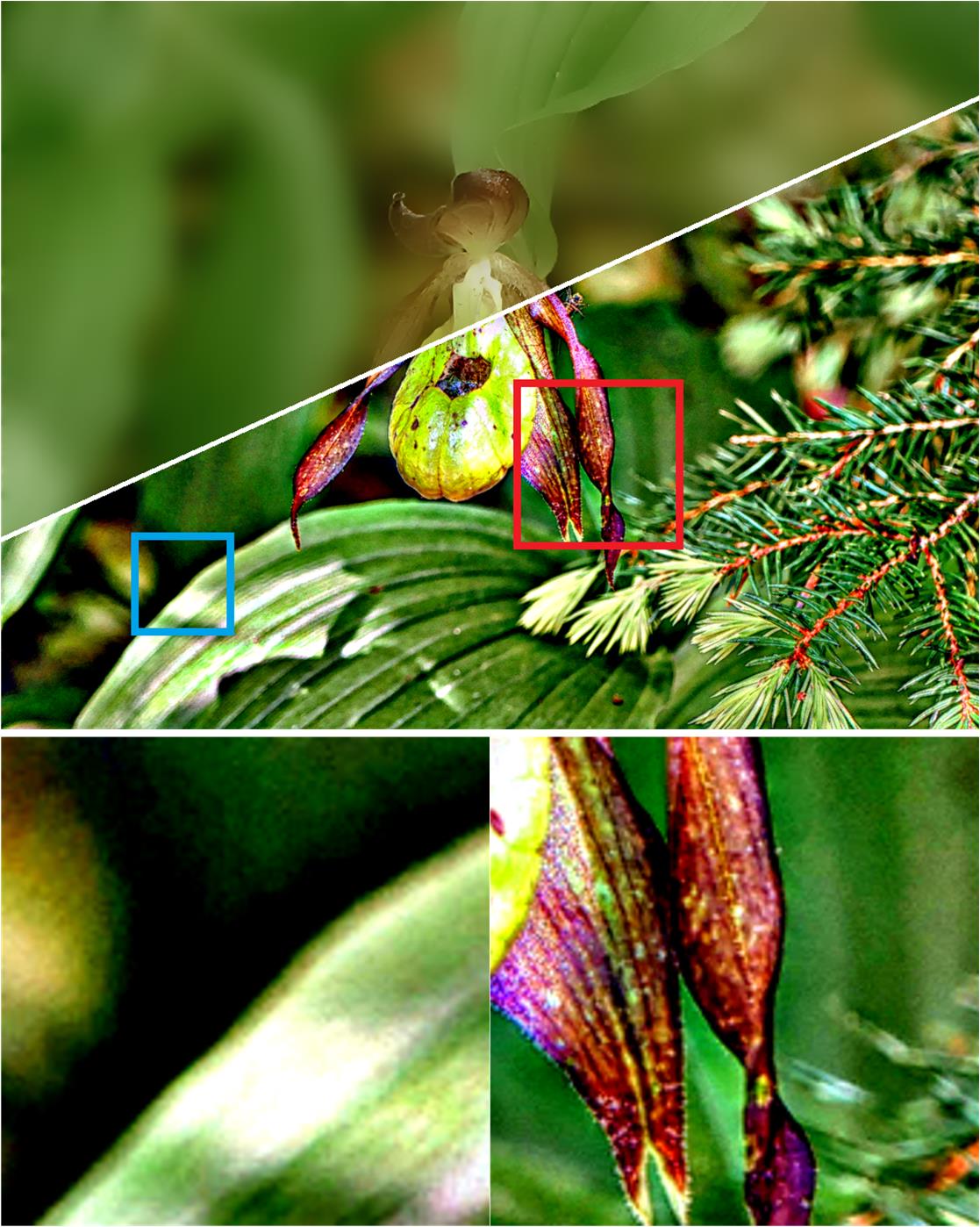}\\
  (e) BLF & (f) BLF-LS & (g) BLF & (h) BLF-LS \\

  \includegraphics[width=0.245\linewidth]{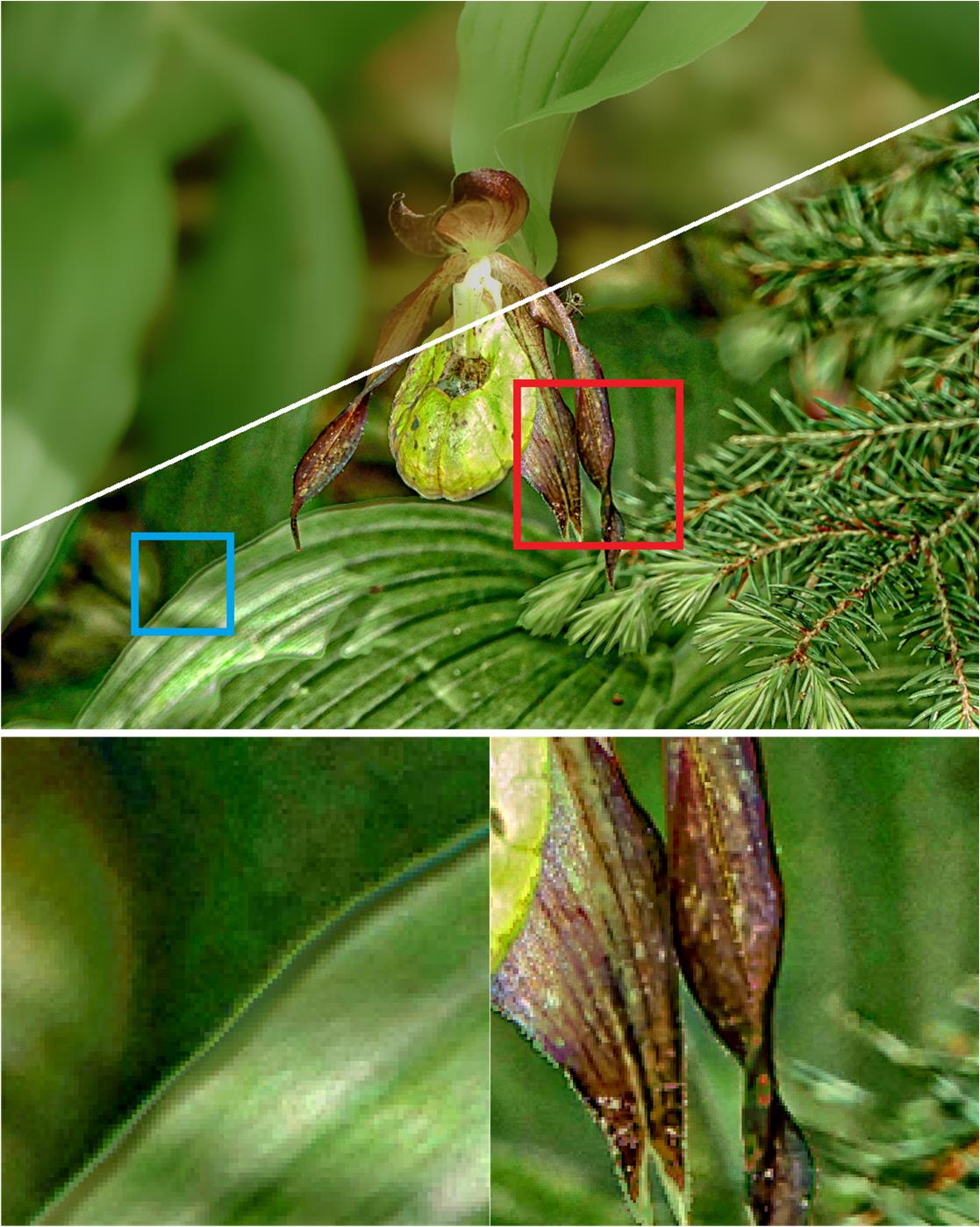}&
  \includegraphics[width=0.245\linewidth]{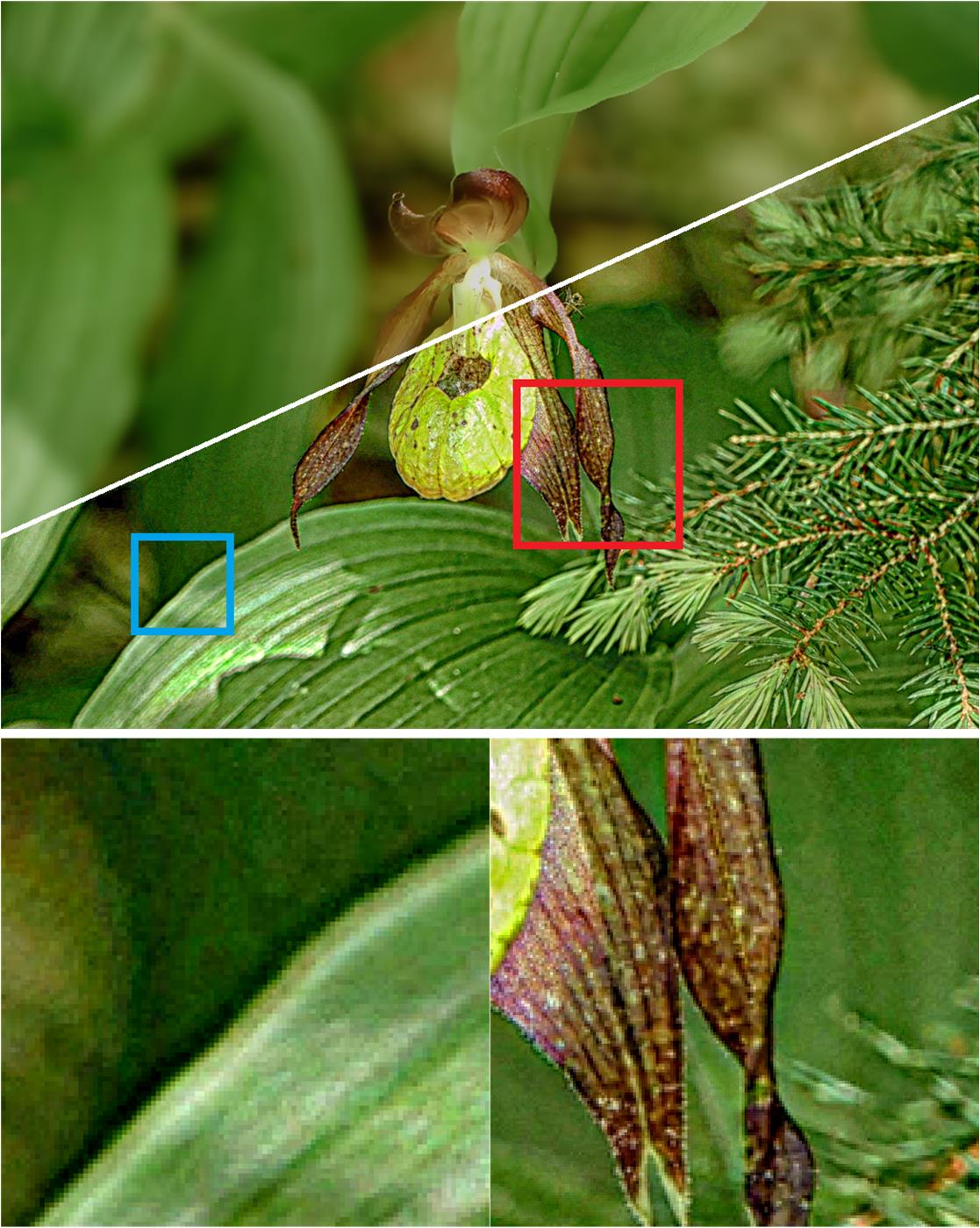}&
  \includegraphics[width=0.245\linewidth]{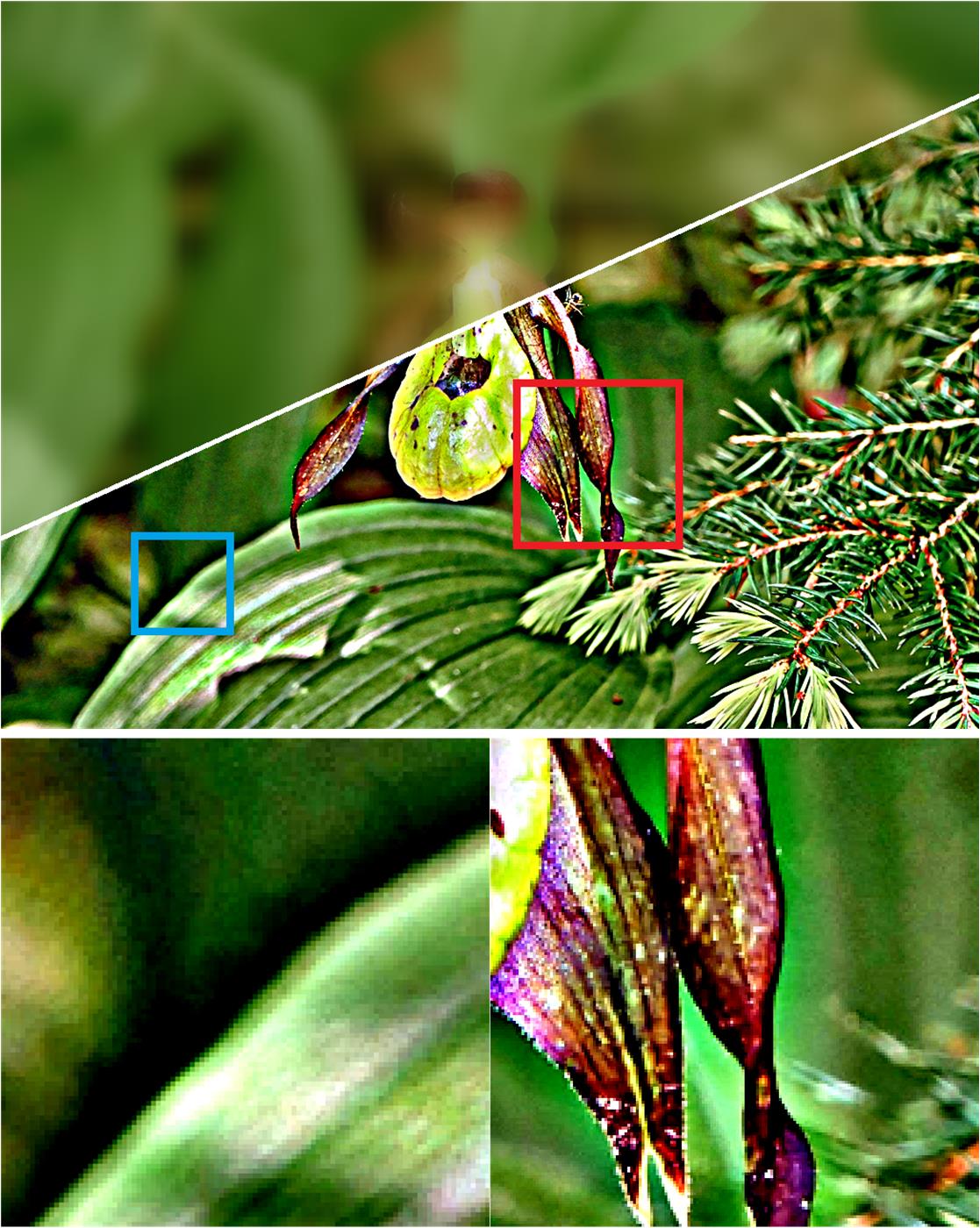}&
  \includegraphics[width=0.245\linewidth]{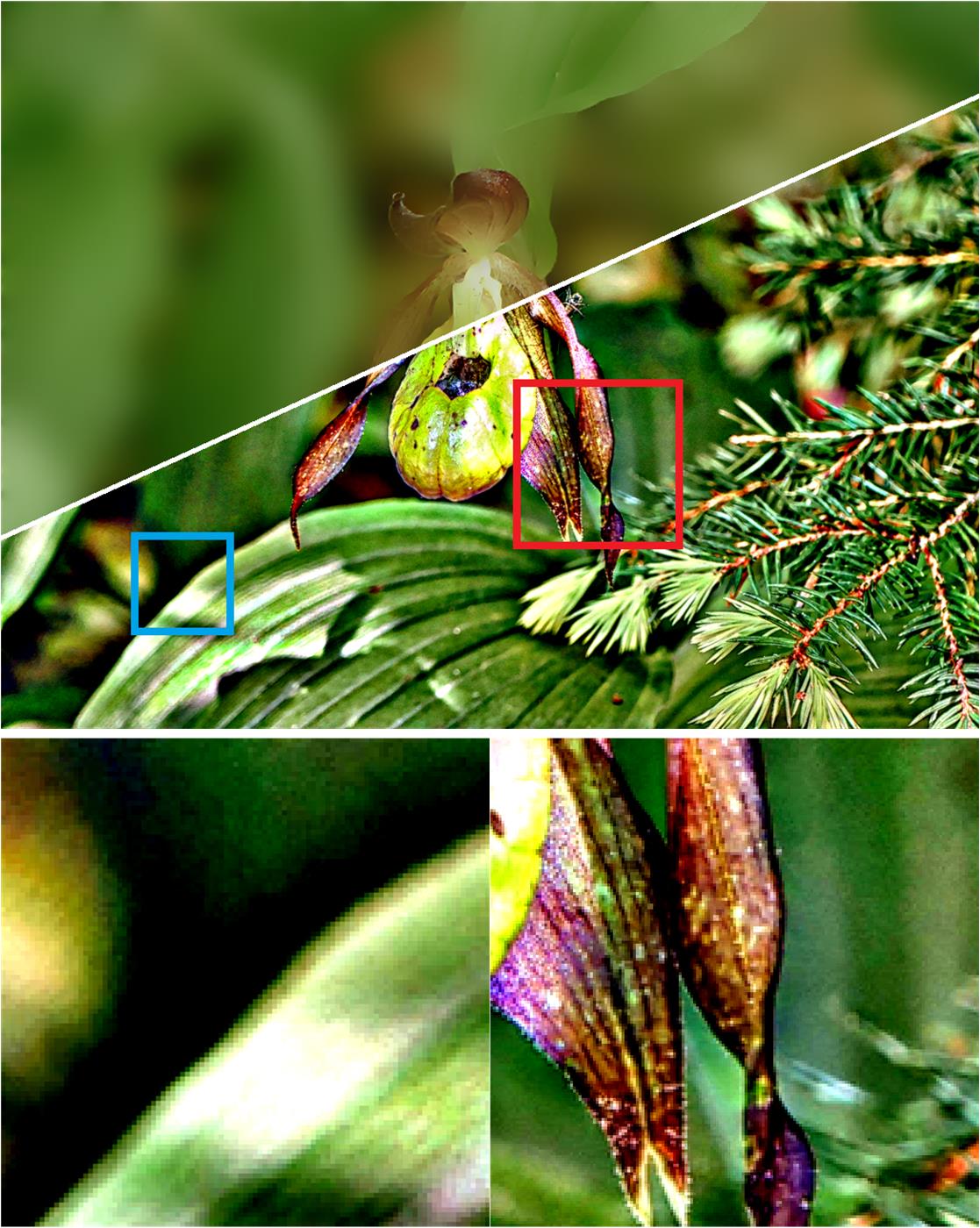}\\
  (e) AMF & (f) AMF-LS & (g) AMF & (h) AMF-LS \\
  \end{tabular}
  \caption{Image detail enhancement comparison of different methods. (a) Input image. Smoothed image and the corresponding detail enhanced image with the detail layer $5\times$ boosted by (b) GF \cite{he2013guided} ($r=12,\varepsilon=0.2^2$), (c) gradient $L_0$ norm smoothing \cite{xu2011image} ($\lambda=0.02$), (d) WLS \cite{farbman2008edge} ($\lambda=0.8,\alpha=1.2$), (e) BLF \cite{tomasi1998bilateral} ($\sigma_s=12,\sigma_r=0.08$), (f) our BLF-LS ($\sigma_s=6,\sigma_r=0.02$), (g) BLF \cite{tomasi1998bilateral} ($\sigma_s=12,\sigma_r=0.3$), (h) our BLF-LS ($\sigma_s=12,\sigma_r=0.04$), (i) AMF \cite{gastal2012adaptive} ($\sigma_s=12,\sigma_r=0.08$), (j) our AMF-LS ($\sigma_s=6,\sigma_r=0.02$), (k) AMF \cite{gastal2012adaptive} ($\sigma_s=12,\sigma_r=0.3$) and (l) our AMF-LS ($\sigma_s=12,\sigma_r=0.04$).}\label{FigDetailEnhancement}\vspace{-1em}
\end{figure*}

\begin{figure*}
  \centering
  \setlength{\tabcolsep}{0.25mm}
  \begin{tabular}{ccc}
  \includegraphics[width=0.33\linewidth]{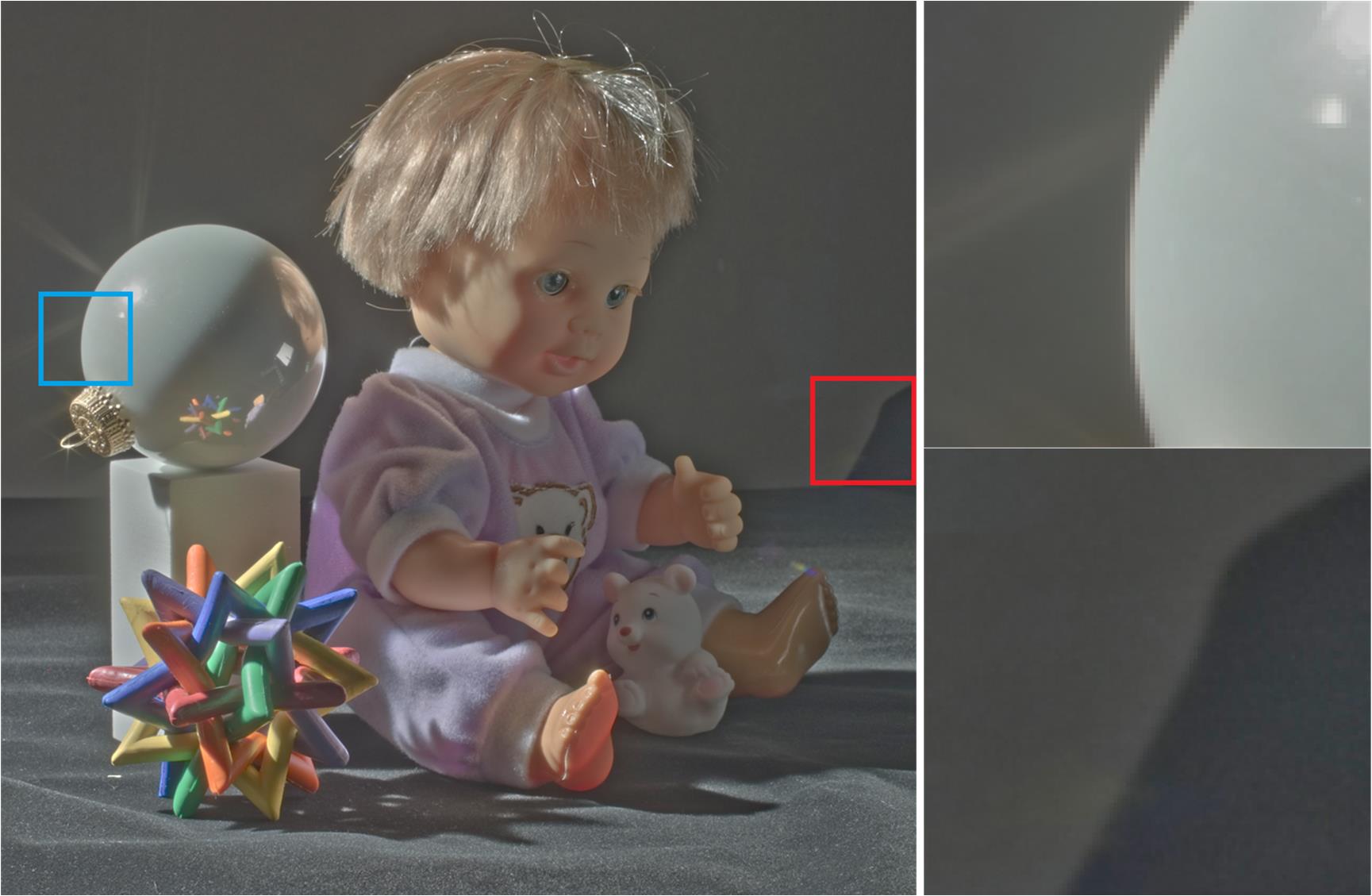} &
  \includegraphics[width=0.33\linewidth]{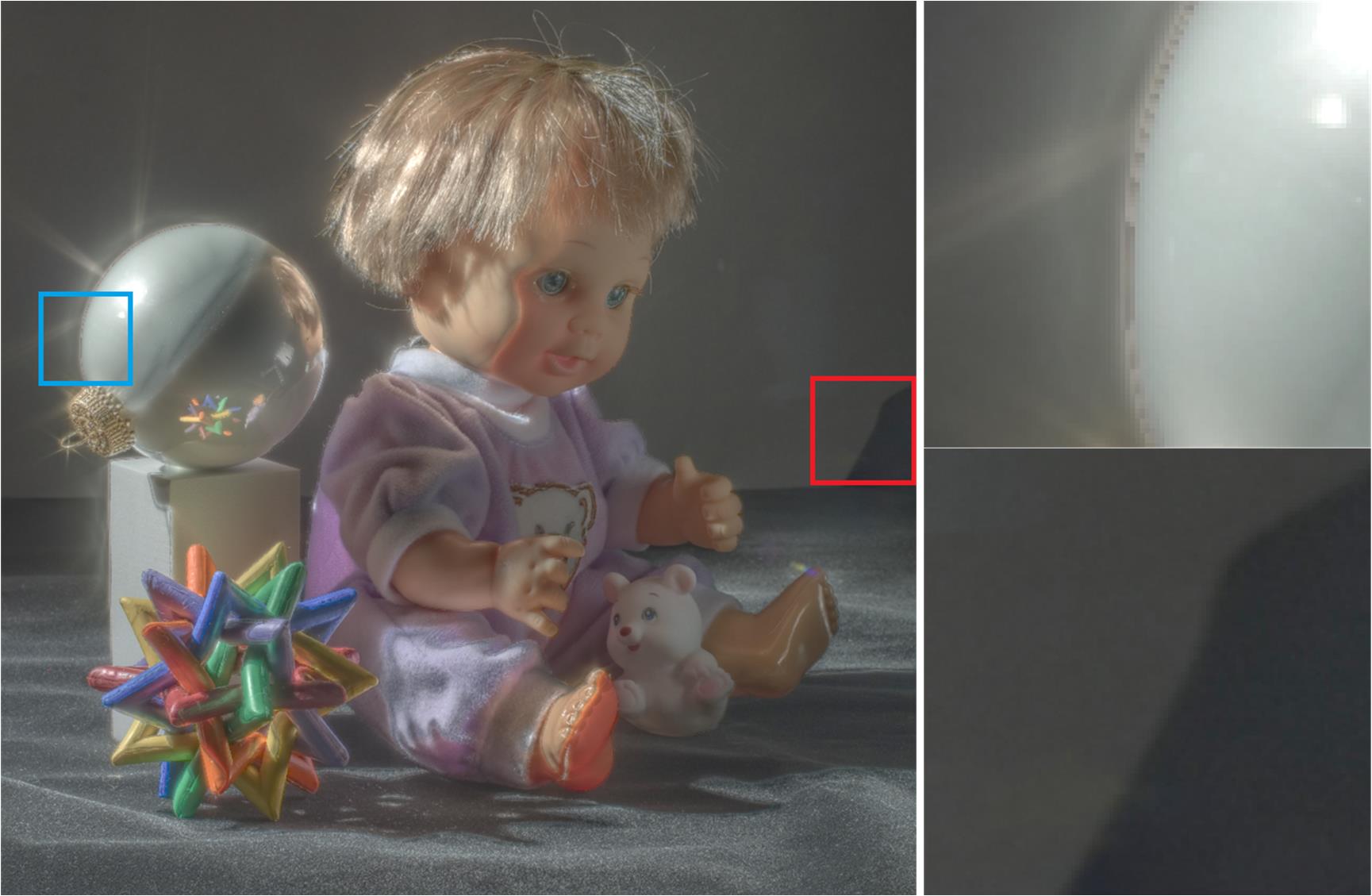} &
  \includegraphics[width=0.33\linewidth]{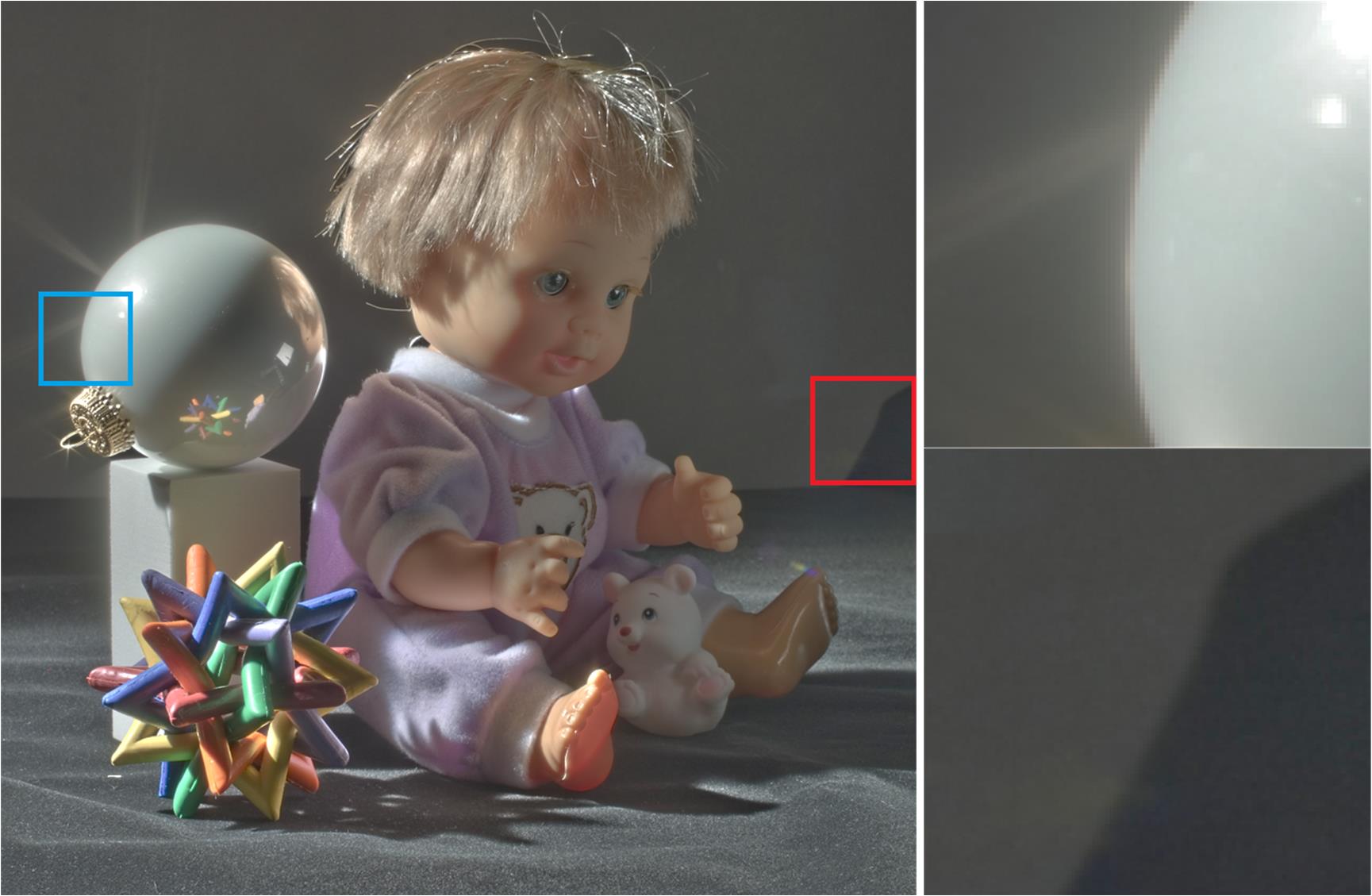} \\
  (a) GF & (b) $L_0$ norm & (c) WLS\\

  \includegraphics[width=0.33\linewidth]{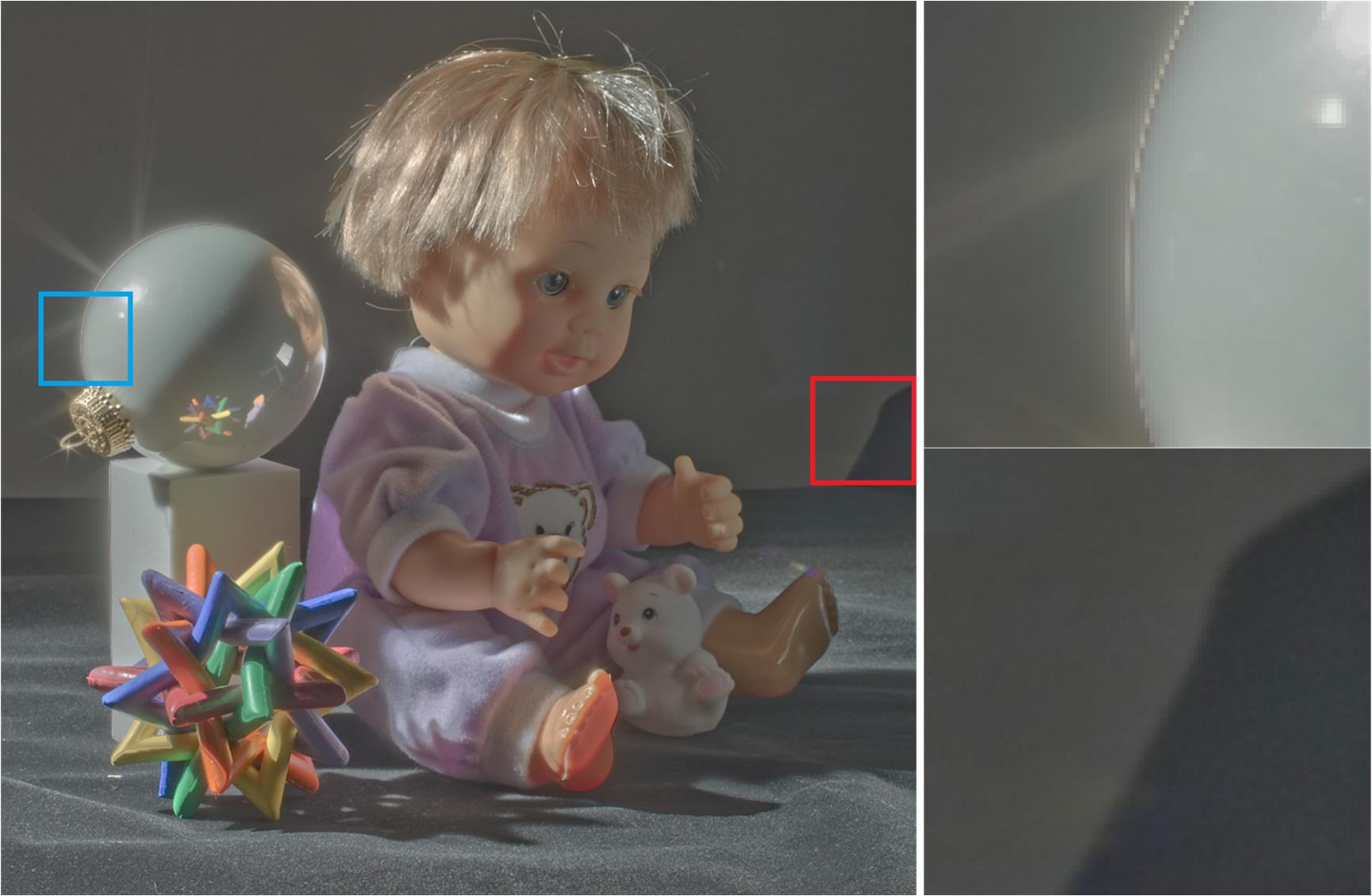} &
  \includegraphics[width=0.33\linewidth]{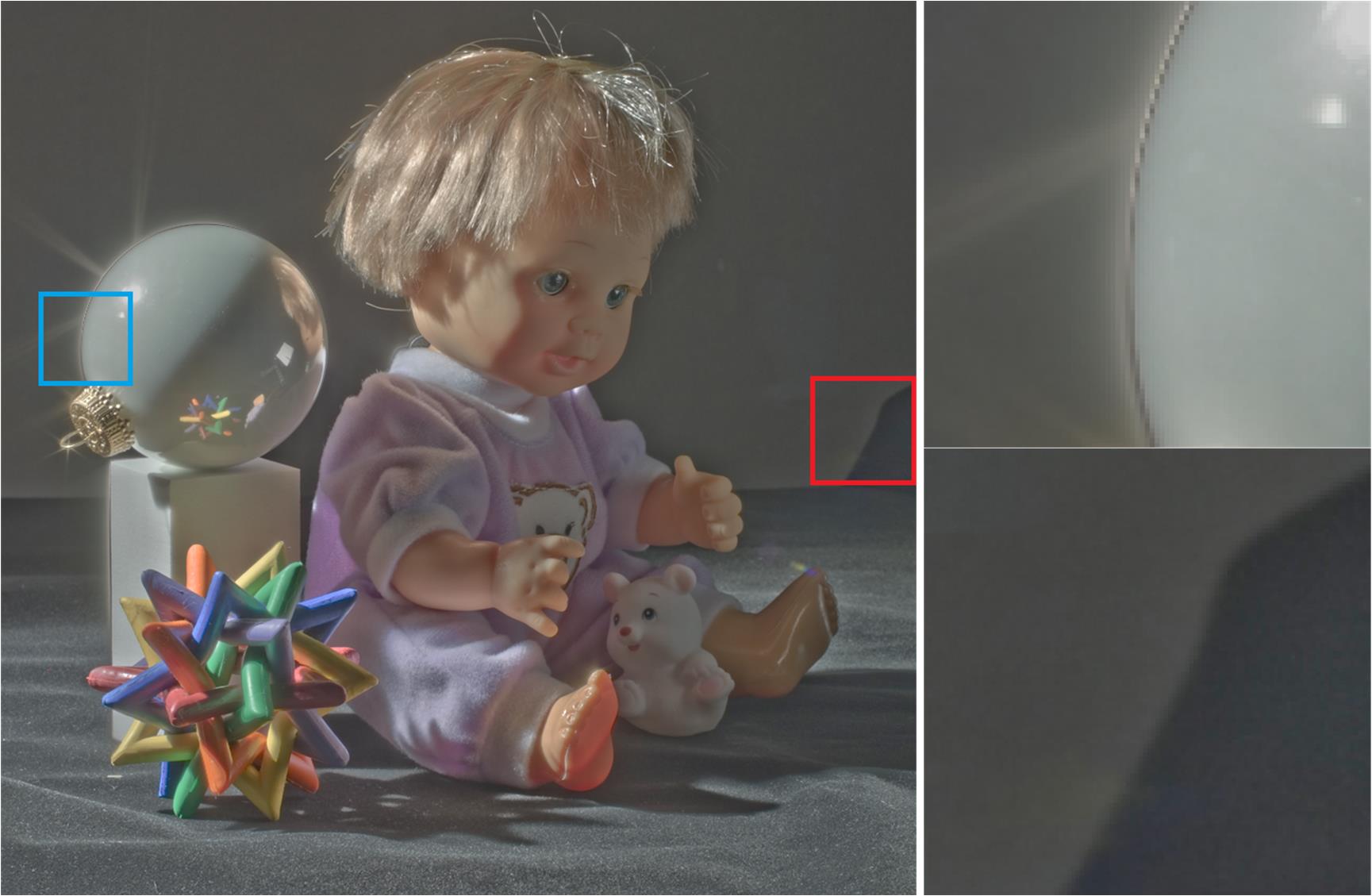} &
  \includegraphics[width=0.33\linewidth]{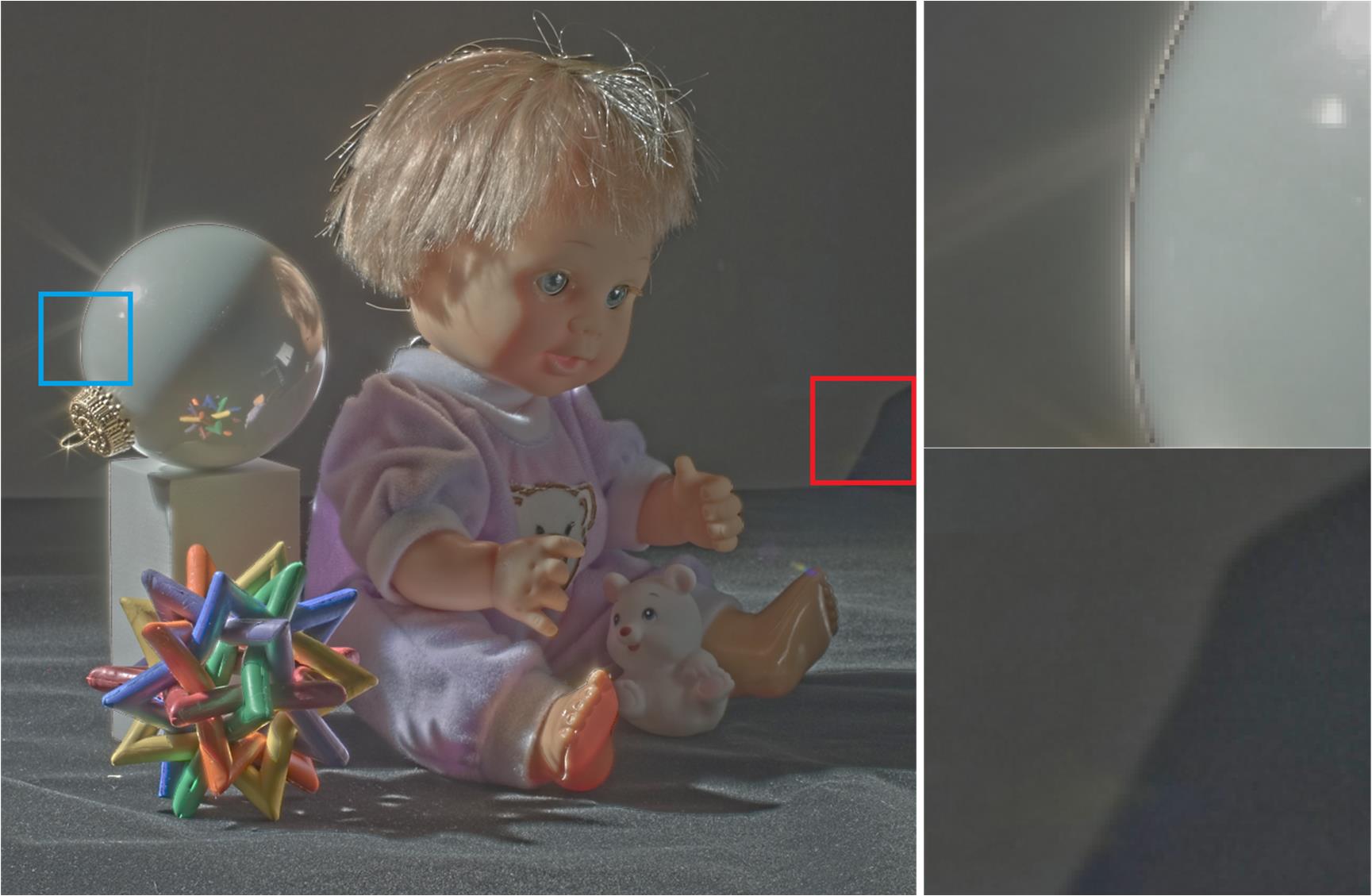} \\
  (a) AMF & (b) BLF & (c) NC\\

  \includegraphics[width=0.33\linewidth]{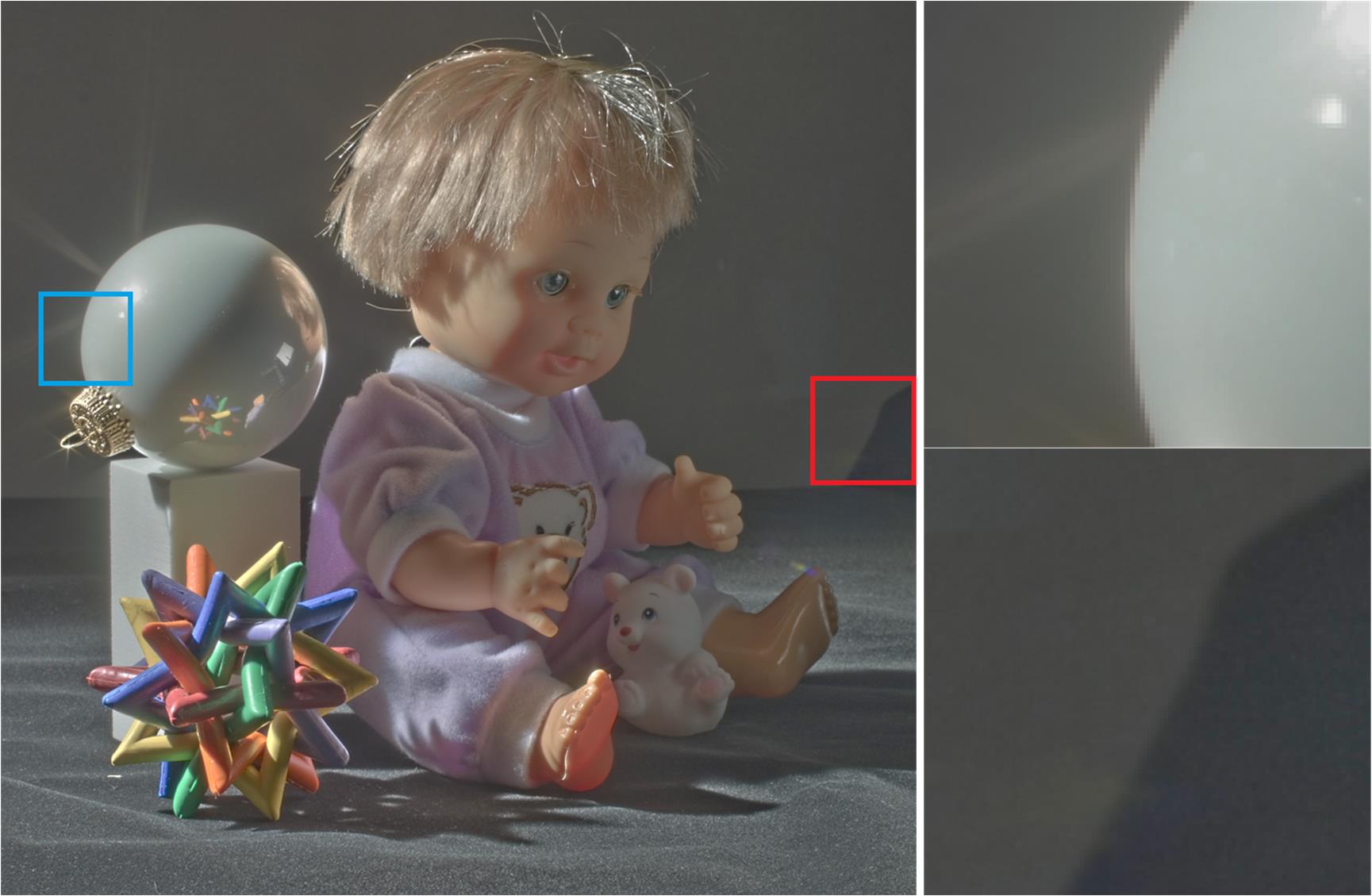} &
  \includegraphics[width=0.33\linewidth]{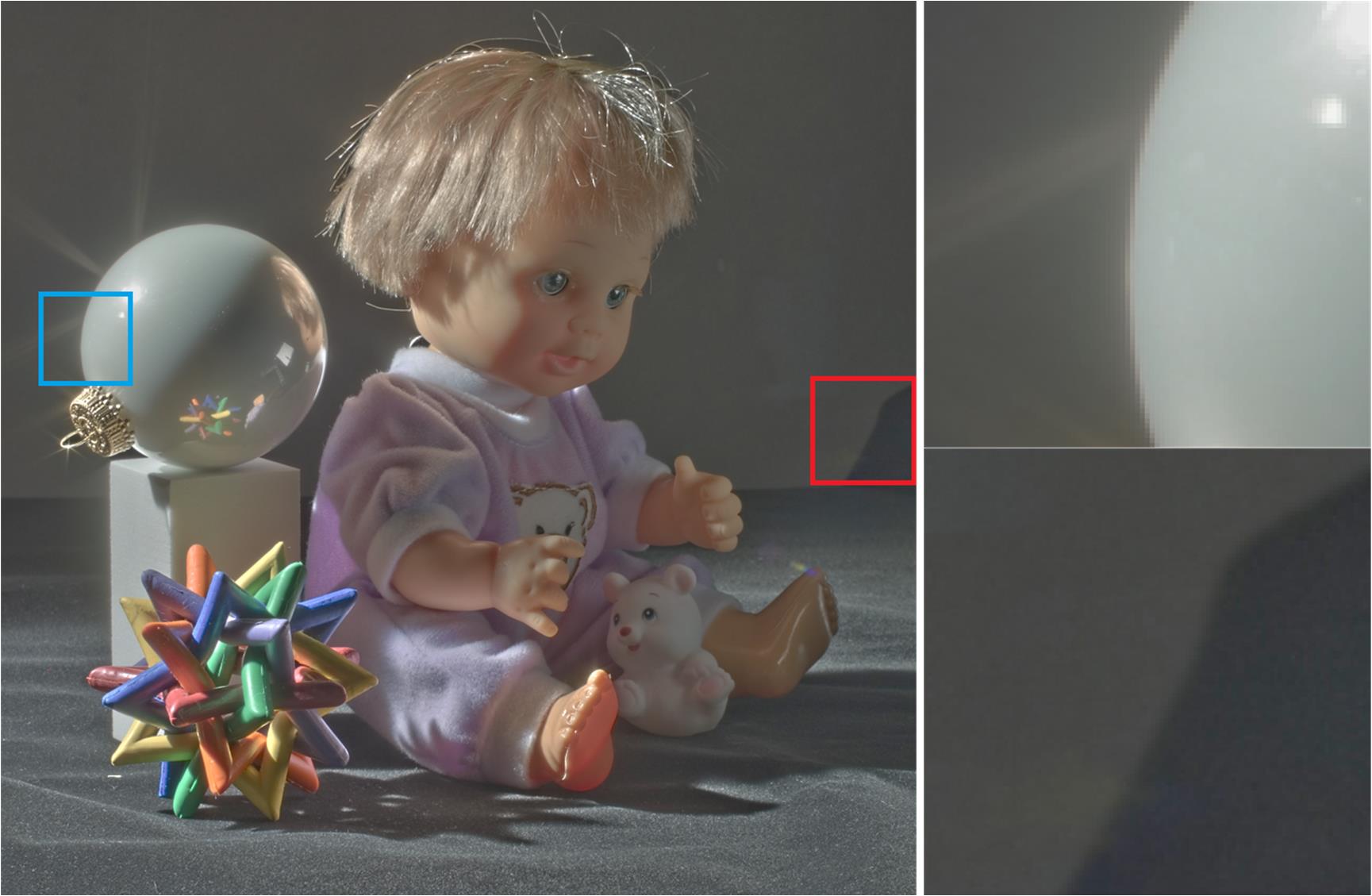} &
  \includegraphics[width=0.33\linewidth]{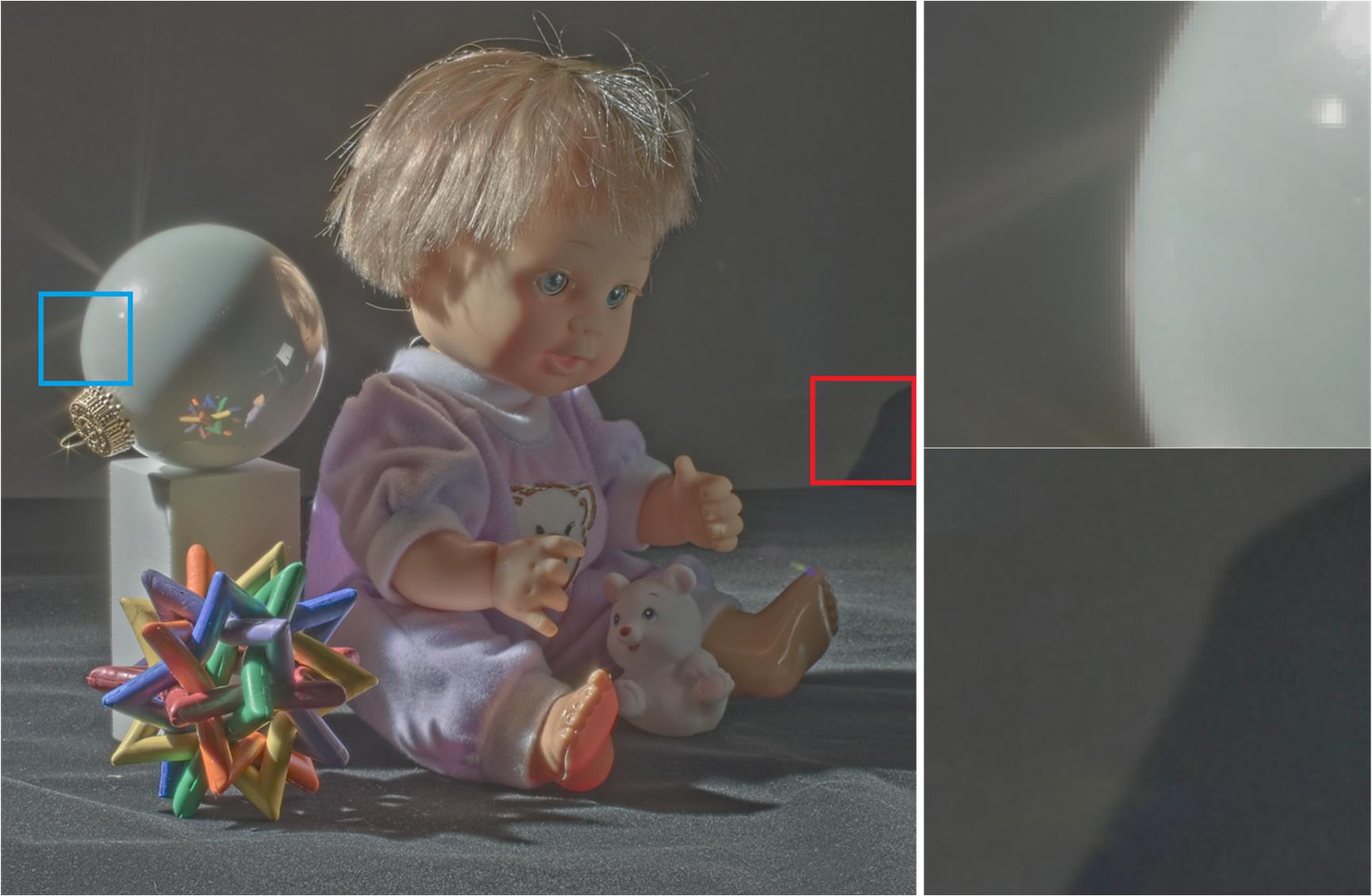} \\
  (a) AMF-LS & (b) BLF-LS & (c) NC-LS\\
  \end{tabular}
  \caption{HDR tone maping comparison of different methods. Result of (a) GF \cite{he2013guided} ($r=16,\varepsilon=0.12^2$), (b) gradient $L_0$ norm smoothing ($\lambda=0.07$), (c) WLS \cite{farbman2008edge} ($\lambda=1,\alpha=1.2$), (d) AMF \cite{gastal2012adaptive} ($\sigma_s=16,\sigma_r=0.12$), (e) BLF \cite{tomasi1998bilateral} ($\sigma_s=16,\sigma_r=0.12$), (f) NC filter \cite{gastal2011domain} ($\sigma_s=16,\sigma_r=0.2$), (g) our AMF-LS ($\sigma_s=8,\sigma_r=0.03$), (h) our BLF-LS ($\sigma_s=8,\sigma_r=0.03$) and (i) our NC-LS ($\sigma_s=12,\sigma_r=0.05$).}\label{FigHDRToneMapping}\vspace{-1em}
\end{figure*}

\begin{figure*}
  \centering
  \setlength{\tabcolsep}{0.25mm}
  \begin{tabular}{ccc}
  \includegraphics[width=0.33\linewidth]{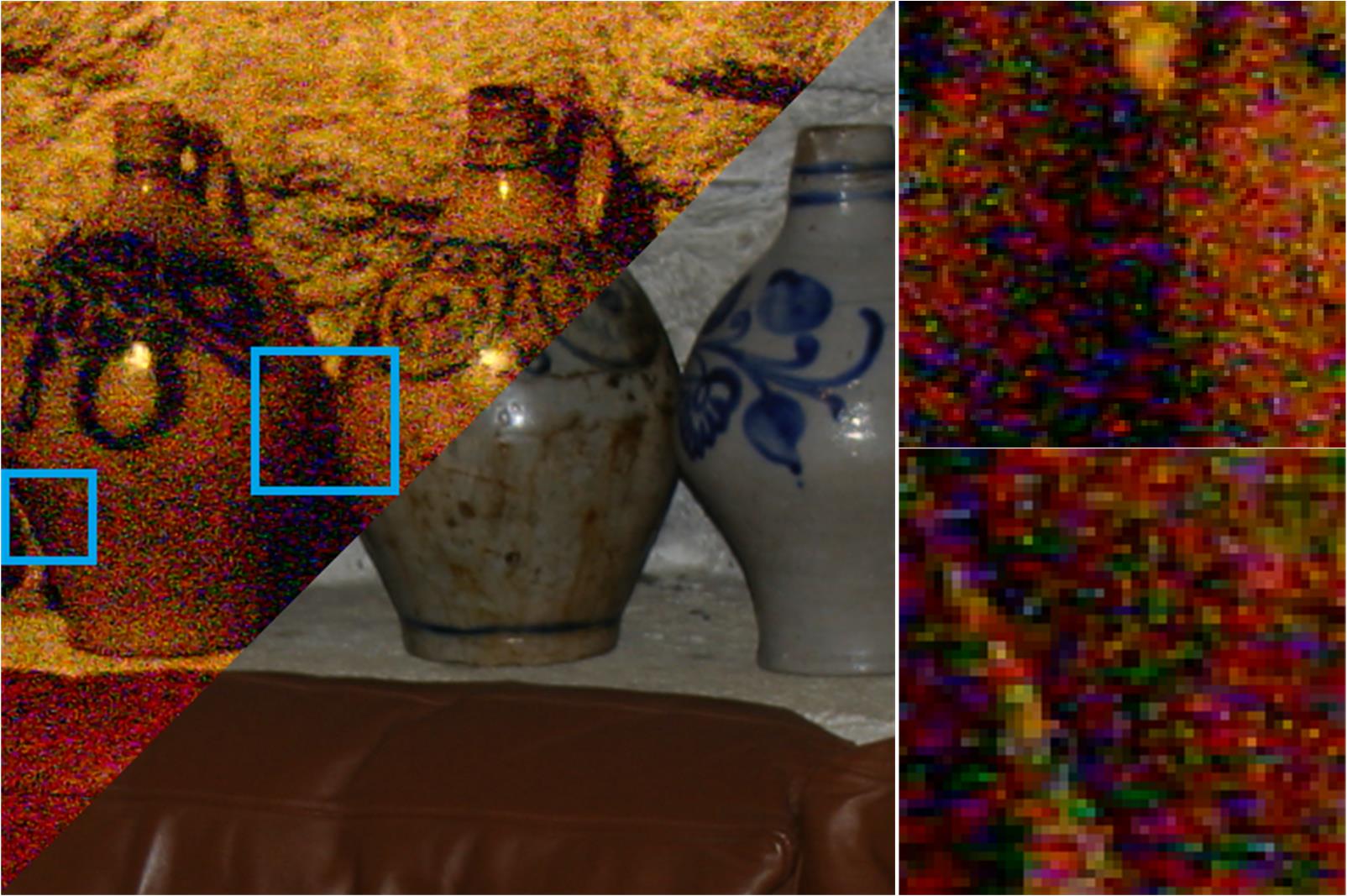} &
  \includegraphics[width=0.33\linewidth]{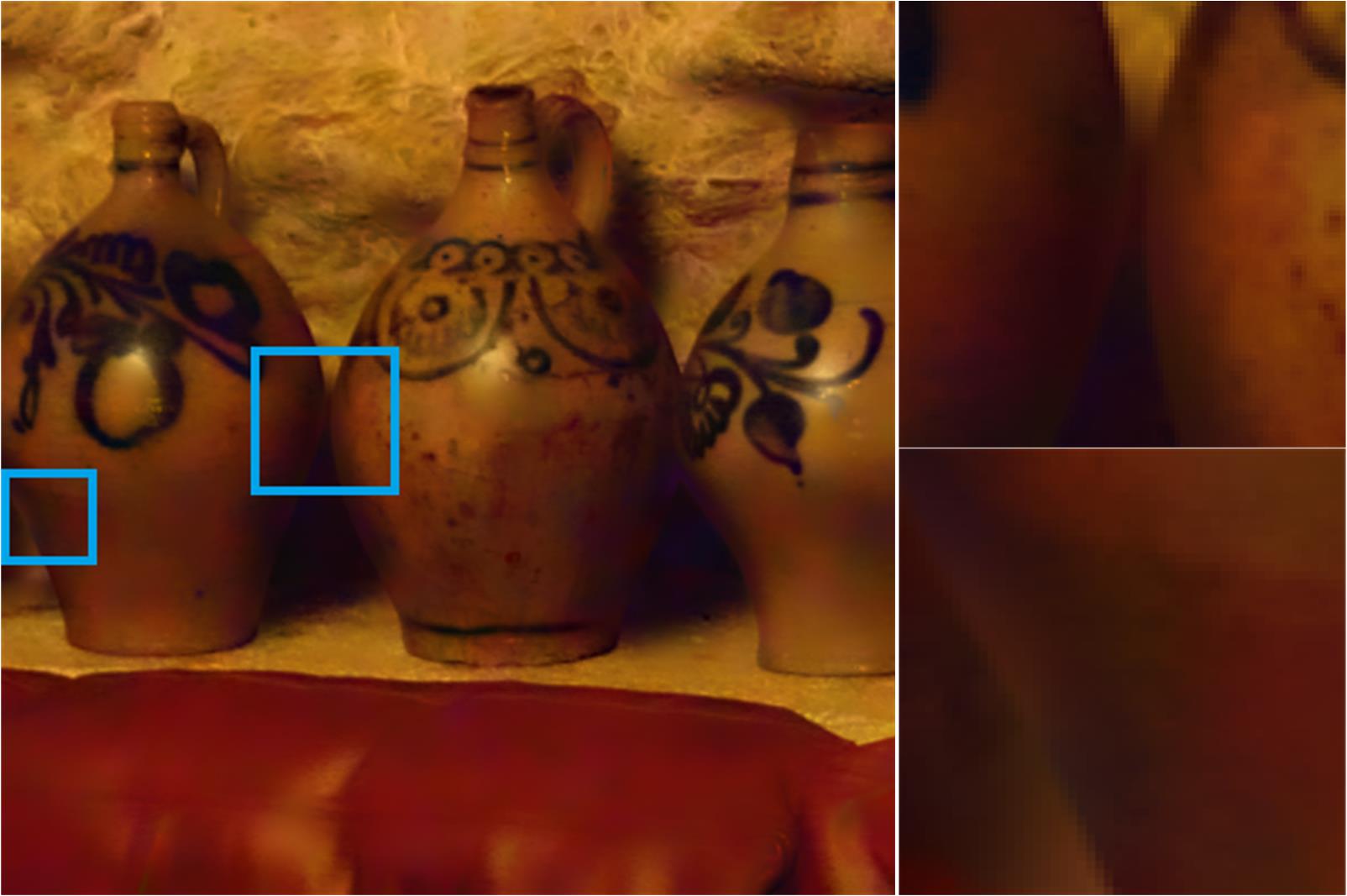} &
  \includegraphics[width=0.33\linewidth]{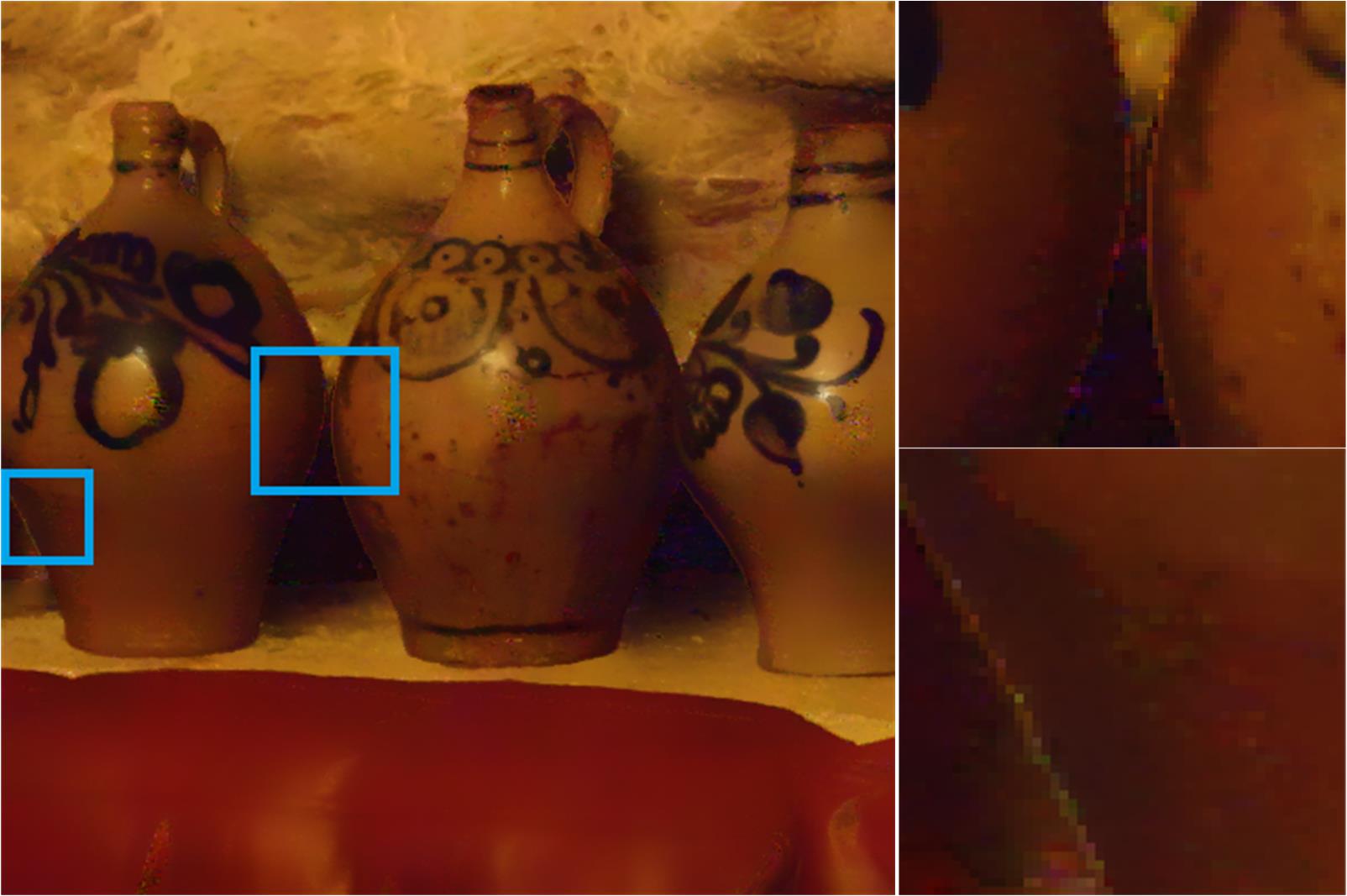} \\
  (a) Input & (b) GF & (c) AMF\\

  \includegraphics[width=0.33\linewidth]{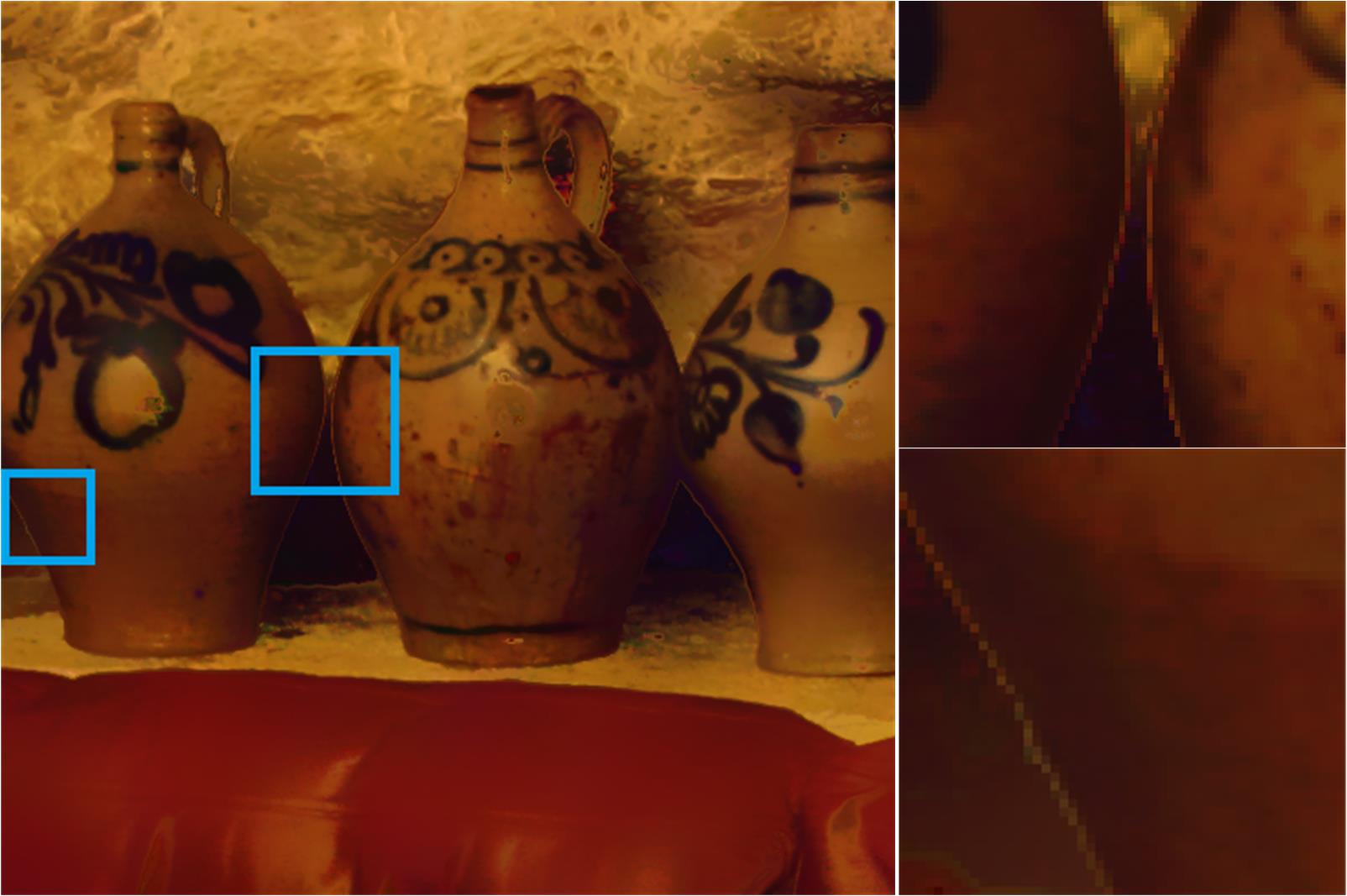} &
  \includegraphics[width=0.33\linewidth]{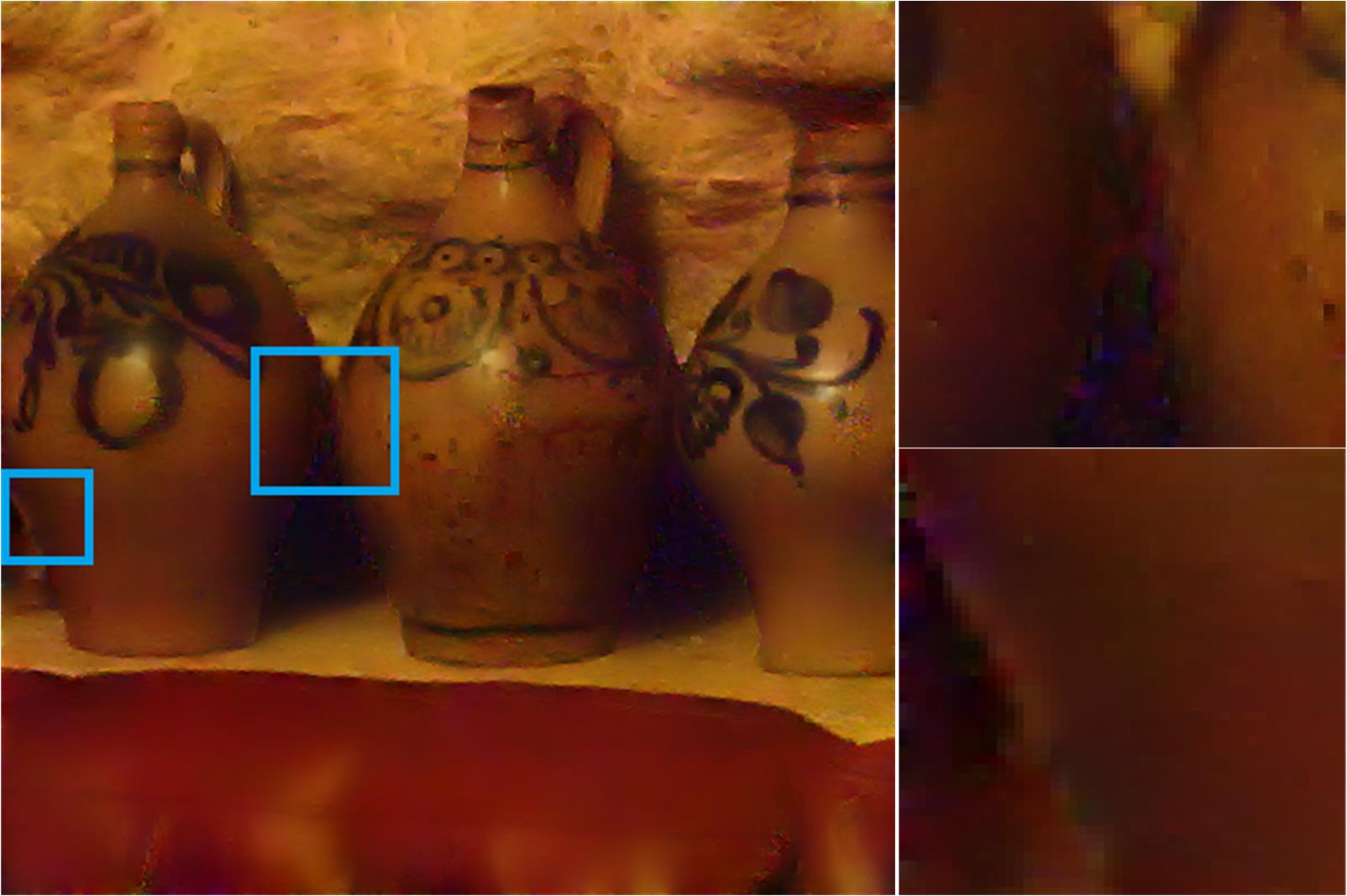} &
  \includegraphics[width=0.33\linewidth]{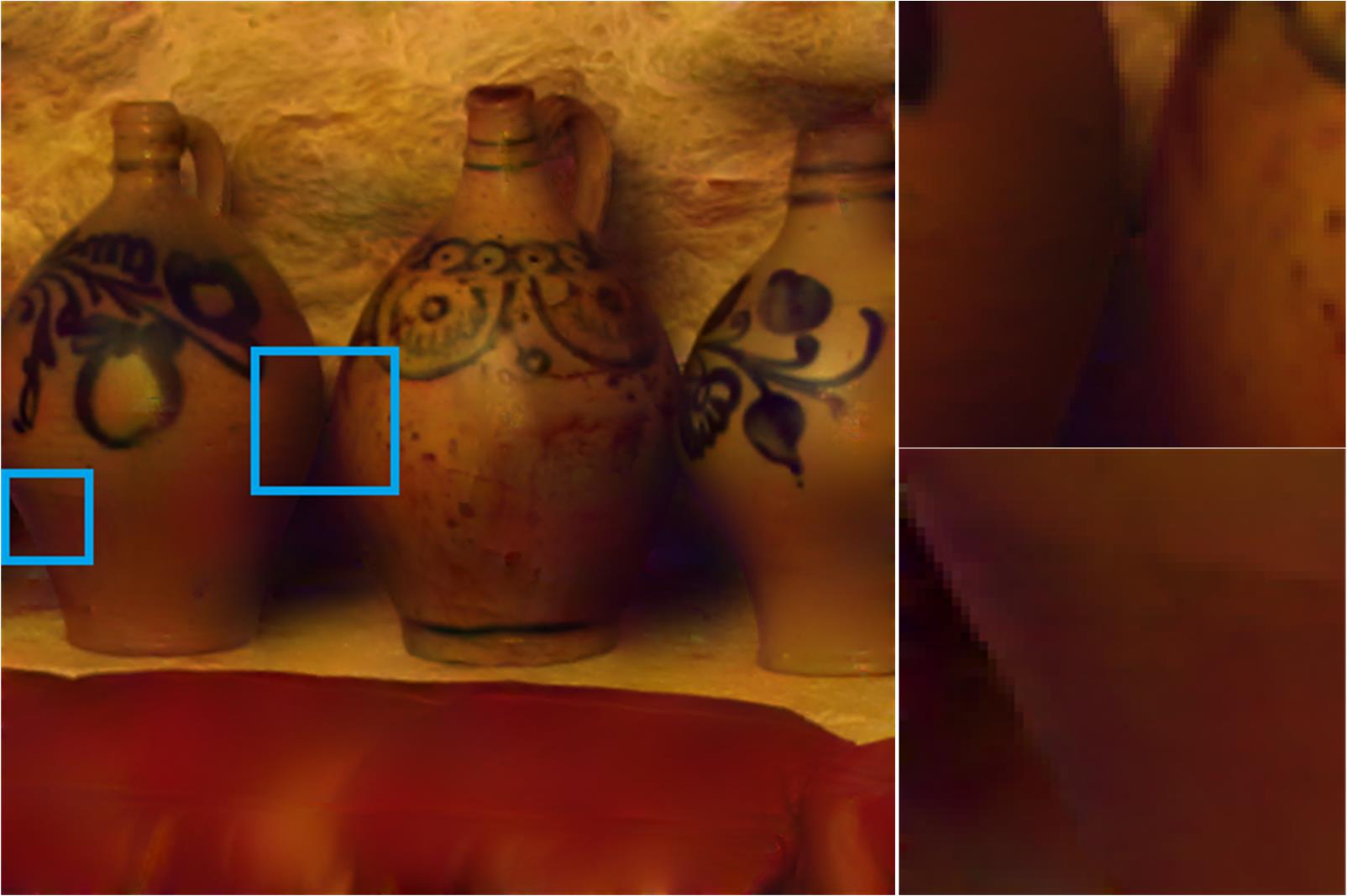} \\
  (a) BLF & (b) AMF-LS & (c) BLF-LS\\
  \end{tabular}
  \caption{Flash/no flash image filtering results of different methods. (a) Input flash/no flash image pair. Smoothing result of (b) GF \cite{he2013guided} ($r=12,\varepsilon=0.02^2$), (c) AMF \cite{gastal2012adaptive} ($\sigma_s=16,\sigma_r=0.06$), (d) joint BLF \cite{petschnigg2004digital} ($\sigma_s=16,\sigma_r=0.02$), (e) our AMF-LS ($\sigma_s=12,\sigma_r=0.006$) and (f) our BLF-LS ($\sigma_s=12,\sigma_r=0.003$).}\label{FigFlashNoFlash}\vspace{-1em}
\end{figure*}

\begin{figure*}
  \centering
  \setlength{\tabcolsep}{0.25mm}
  \begin{tabular}{ccccc}
  \includegraphics[width=0.1975\linewidth]{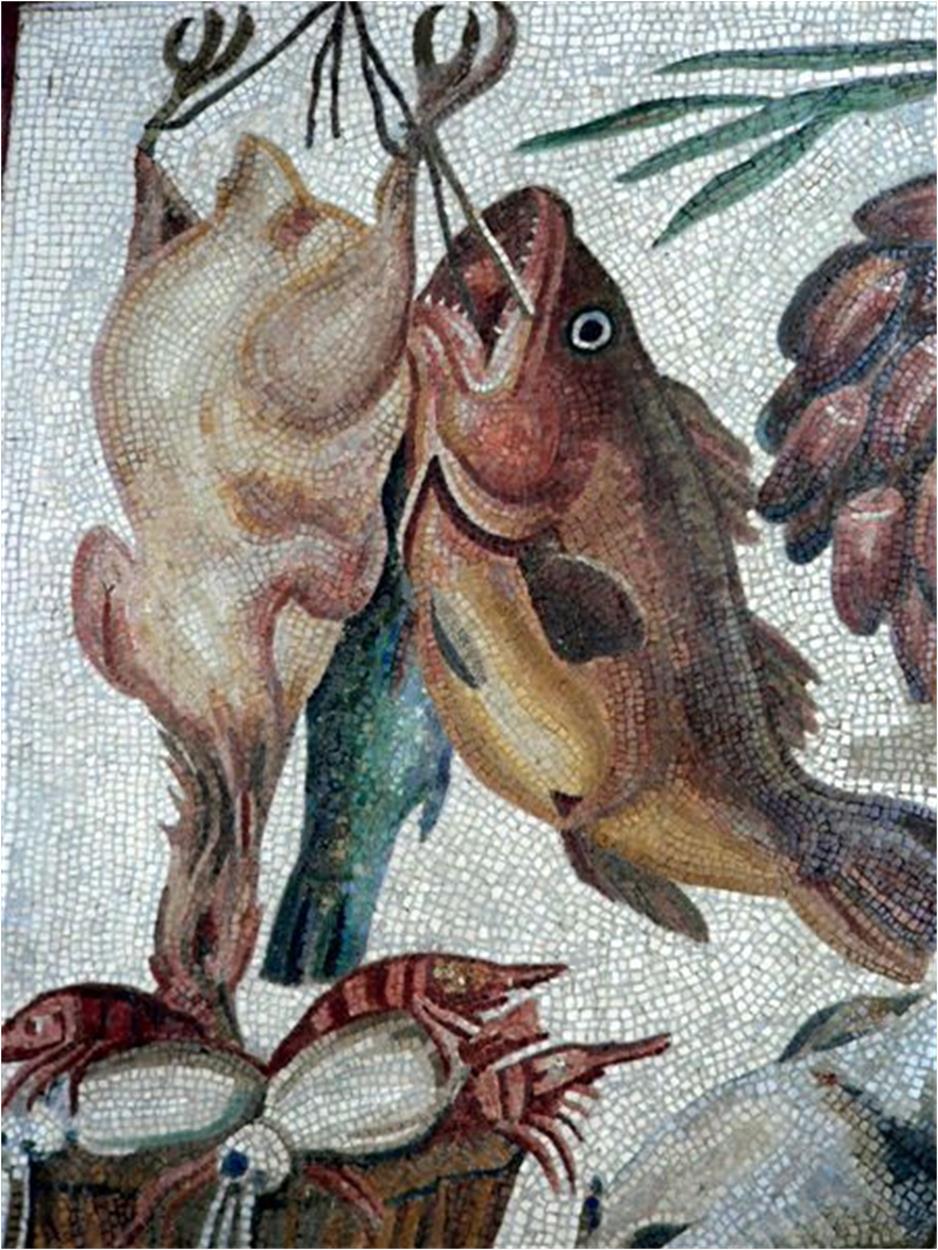} &
  \includegraphics[width=0.1975\linewidth]{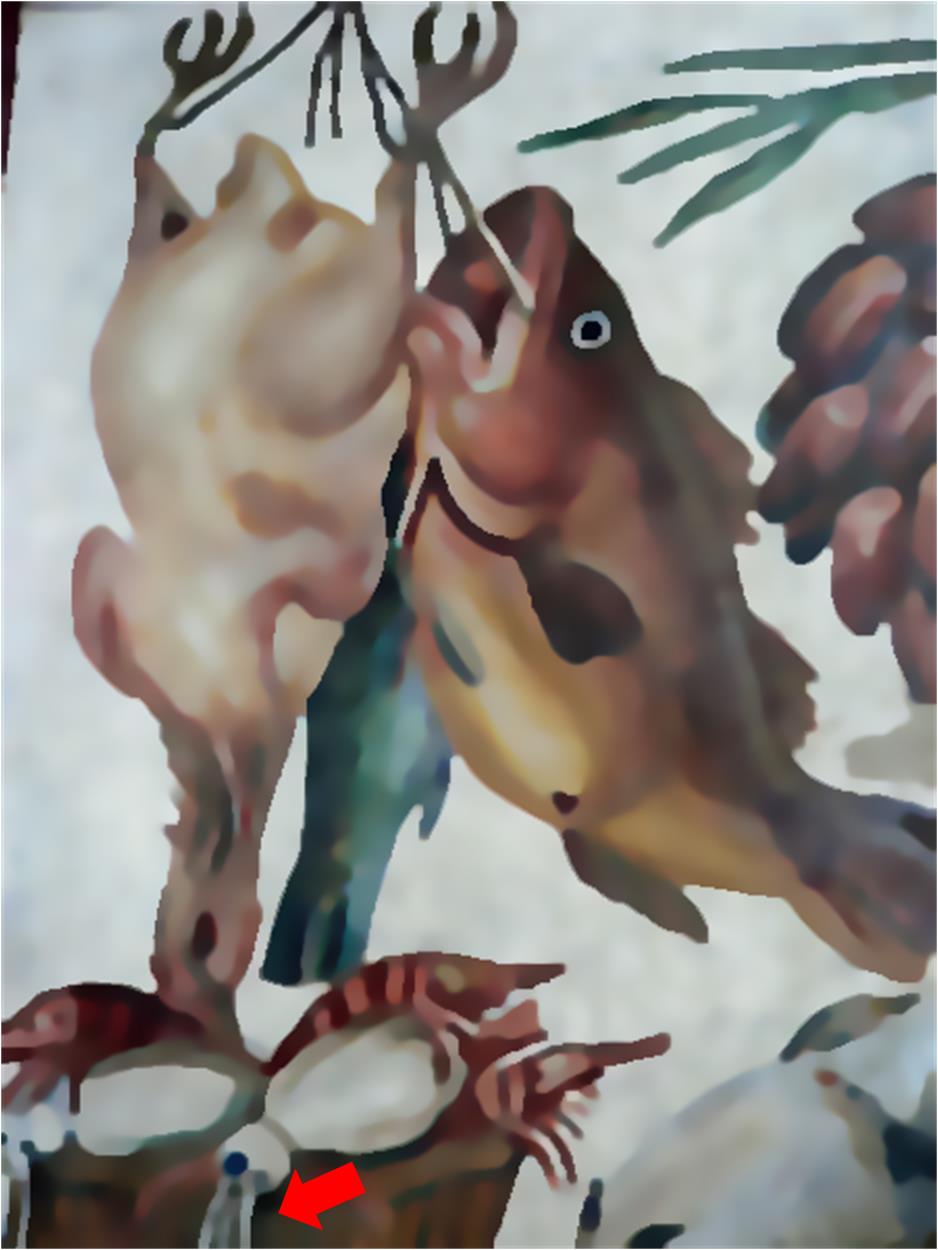} &
  \includegraphics[width=0.1975\linewidth]{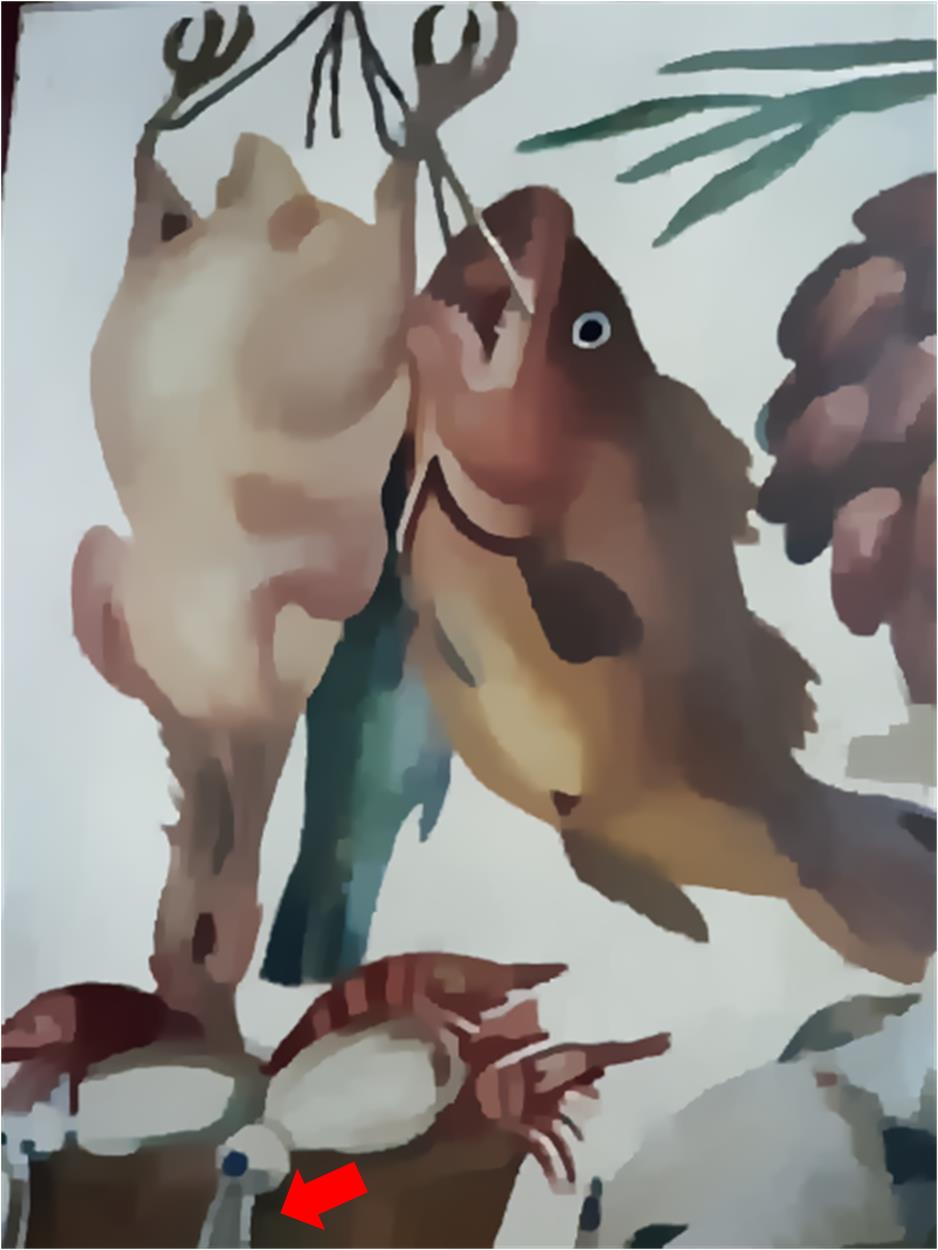} &
  \includegraphics[width=0.1975\linewidth]{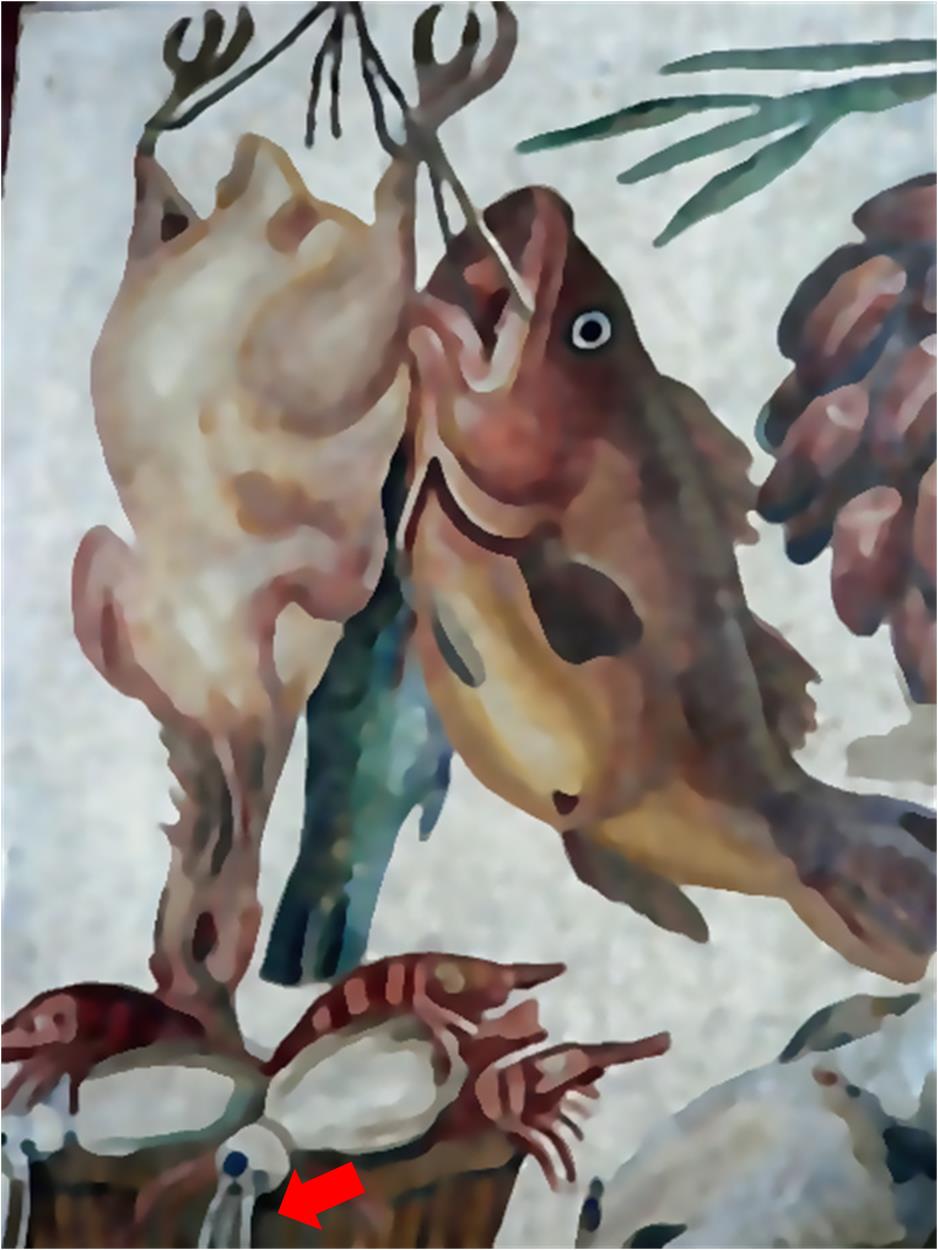} &
  \includegraphics[width=0.1975\linewidth]{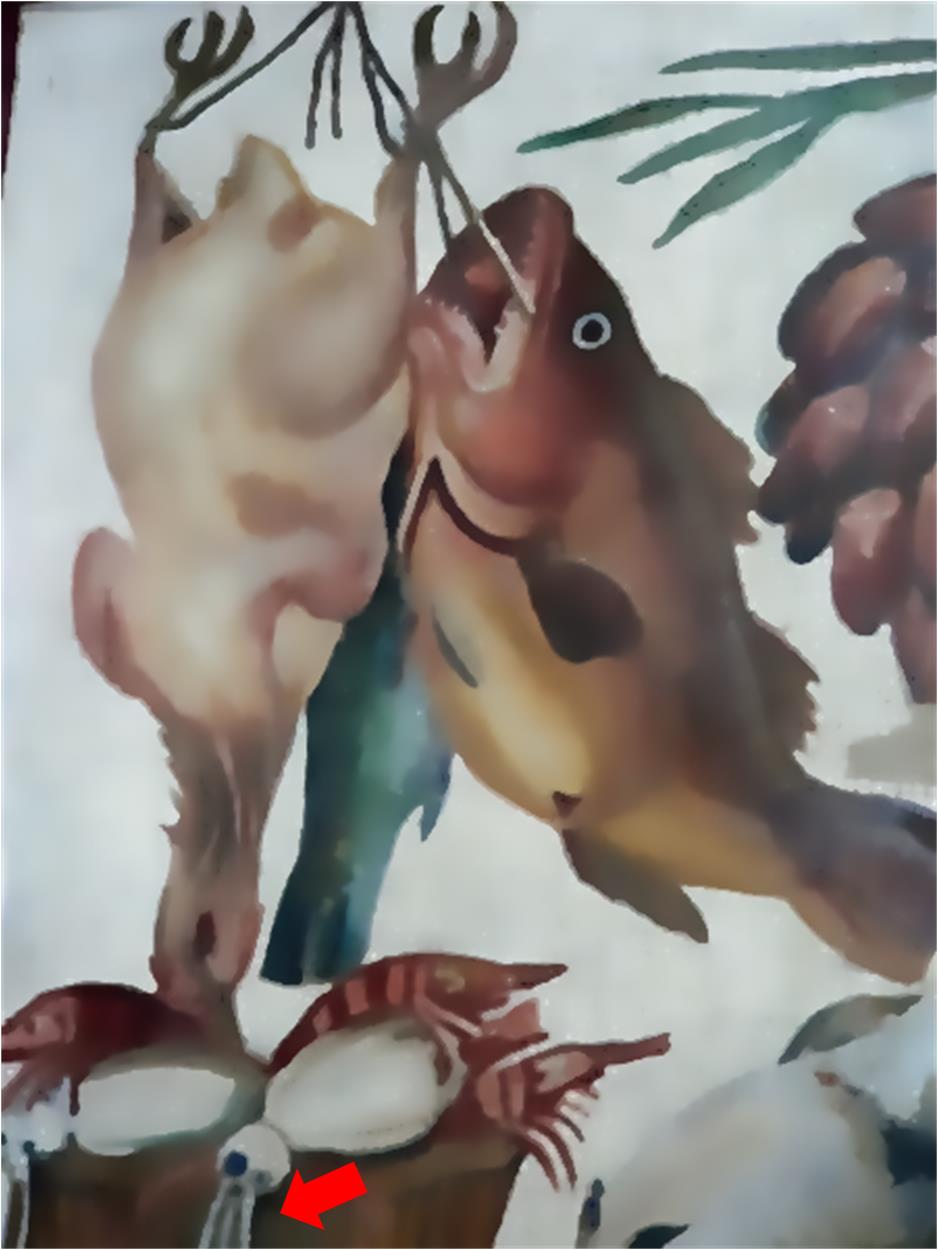} \\
  (a) Input & (b) BTF & (c) RTV & (d) RGF & (e) NC-LS
  \end{tabular}
  \caption{Image texture removal results of different methods. (a) Input image. Texture removal result of (b) bilateral texture filtering \cite{cho2014bilateral} ($k=5, n_{itr=5}$), (c) relative total variation \cite{xu2012structure} ($\lambda=0.025,\sigma=3$), (d) rolling guidance filter \cite{zhang2014rolling} ($\sigma_s=5,\sigma_r=0.07,n^{iter}=5$) and (e) our rolling guidance NC-LS ($\sigma_s=8,\sigma_r=0.02, n=3$), the initial guidance image is obtained by smoothing the input with a Gaussian filter of standard deviation $\sigma=2.5$.}\label{FigTextureSmooth}\vspace{-1em}
\end{figure*}

\begin{figure*}
  \centering
  \setlength{\tabcolsep}{0.25mm}
  \begin{tabular}{ccc}
  \includegraphics[width=0.33\linewidth]{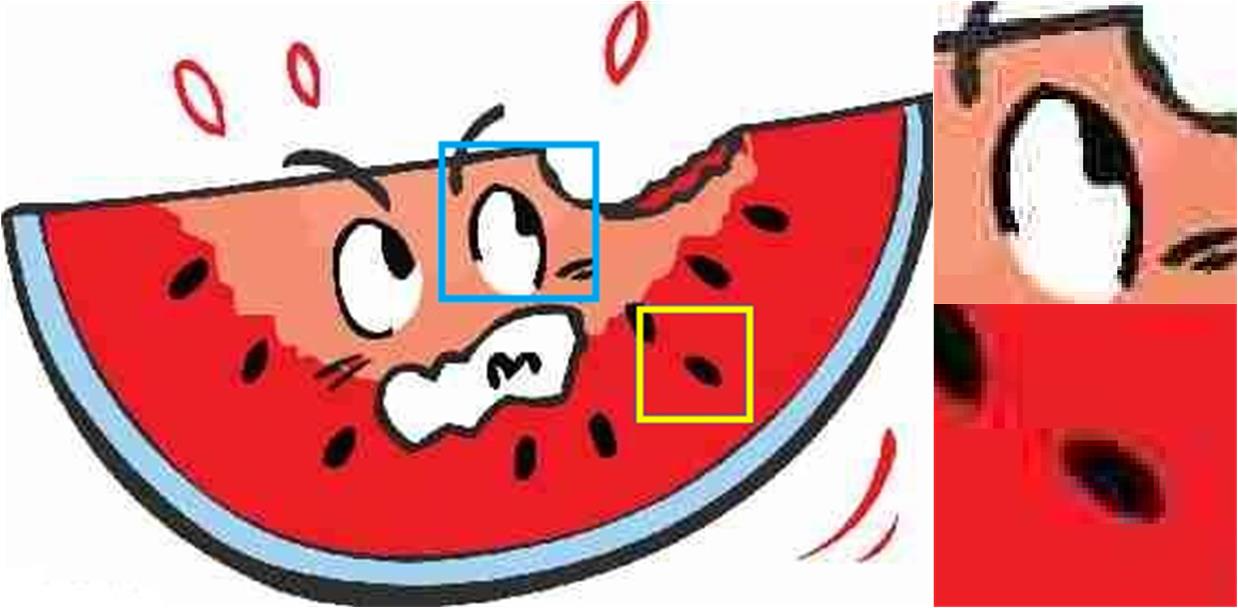}&
  \includegraphics[width=0.33\linewidth]{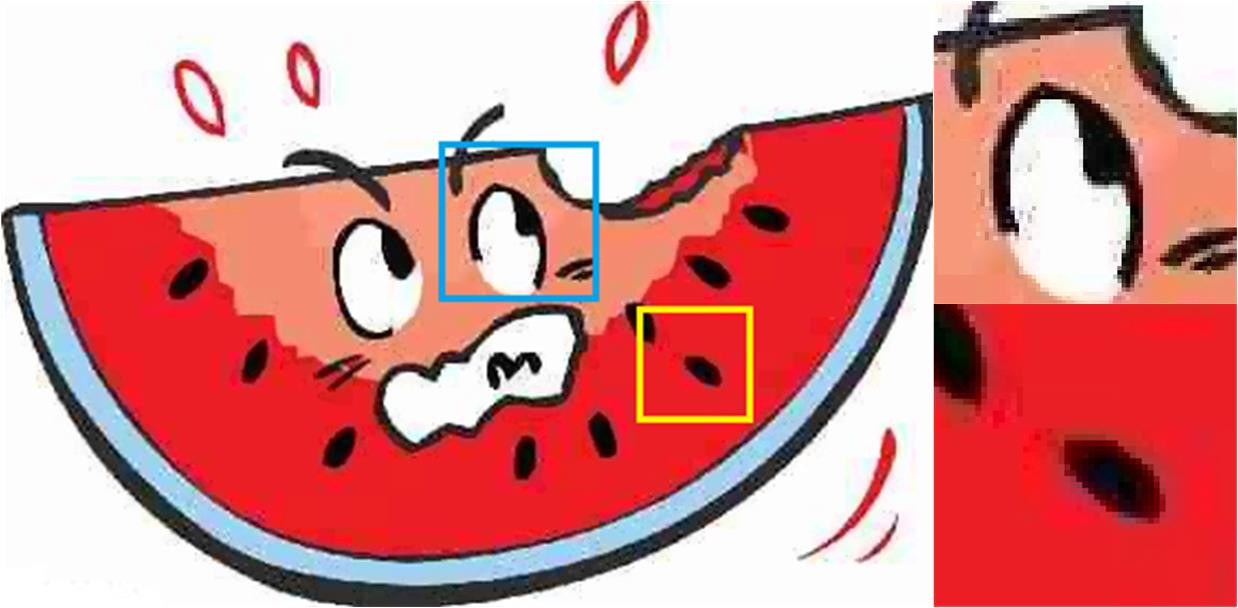}&
  \includegraphics[width=0.33\linewidth]{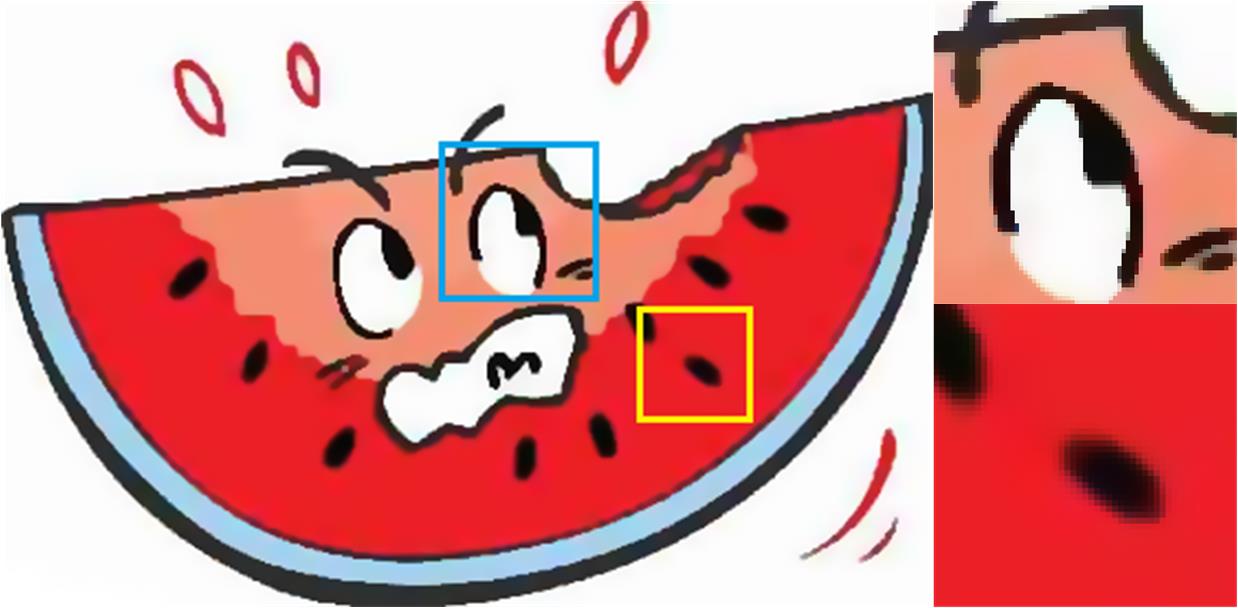}\\
  (a) Input & (b) Wang et~al. & (c) BTF\\

  \includegraphics[width=0.33\linewidth]{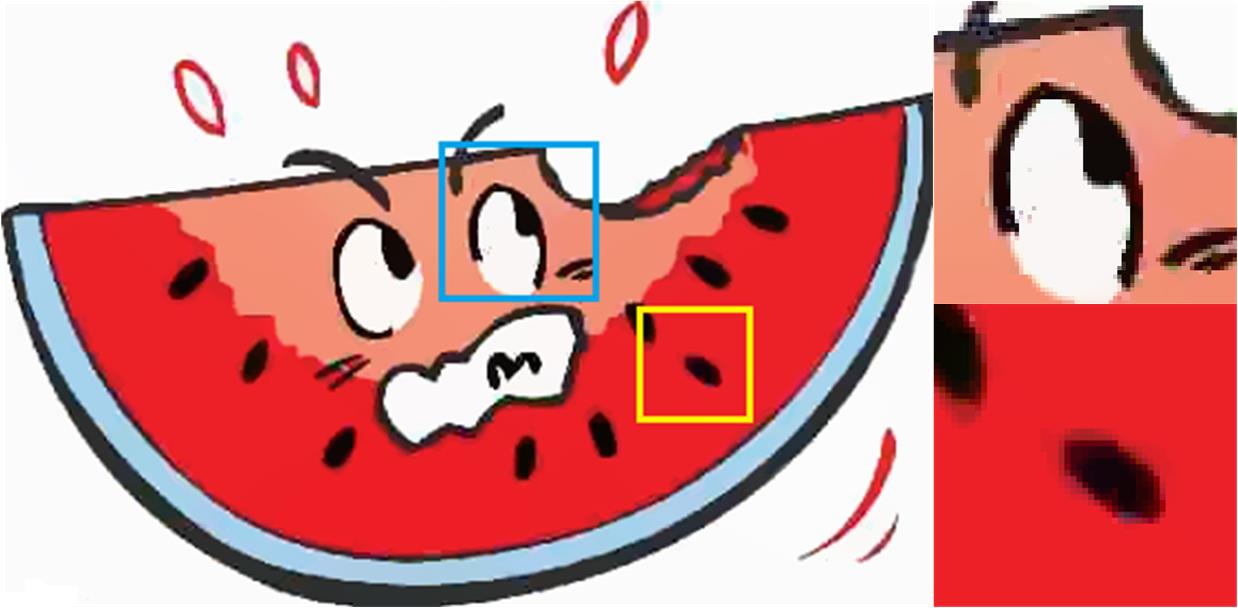}&
  \includegraphics[width=0.33\linewidth]{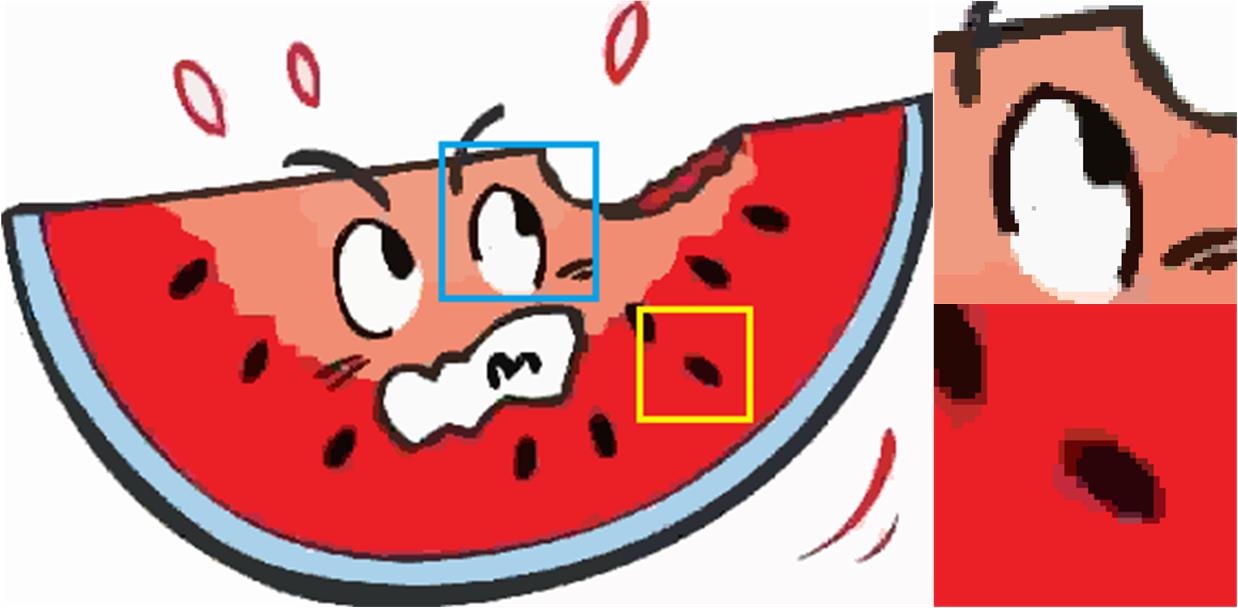}&
  \includegraphics[width=0.33\linewidth]{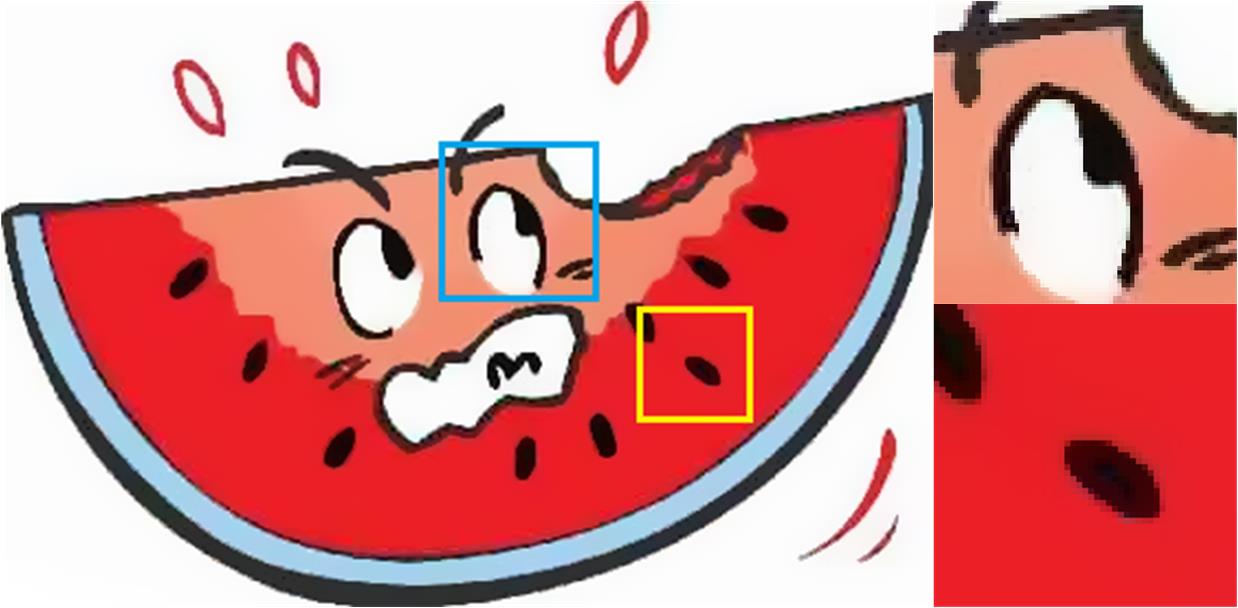}\\
  (d) $L_0$ norm & (e) Region fusion & (f) NC-LS
  \end{tabular}
  \caption{Clip-art compression artifacts removal results of different methods. (a) Input image with compression artifacts. Artifacts removed by (b) the method proposed by Wang et~al. \cite{wang2006deringing}, (c) bilateral texture filtering \cite{cho2014bilateral} ($k=3, n_{itr=3}$), (d) gradient $L_0$ norm smoothing \cite{xu2011image} ($\lambda=0.02$), (e) $L_0$ norm region fusion \cite{nguyen2015fast} ($\lambda=0.04$) and (f) our rolling guidance NC-LS ($\sigma_s=6,\sigma_r=0.02,n=2$), the initial guidance image is obtained by smoothing the input with a Gaussian filter of standard deviation $\sigma=0.75$.}\label{FigClipArt}\vspace{-1em}
\end{figure*}

To validate the effectiveness and efficiency of our method, we apply our method to several tasks including image detail enhancement, HDR tone mapping, flash/no flash image filtering, image texture removal and clip-art compression artifacts removal. For the BLF \cite{tomasi1998bilateral}, we adopt its fast implementation in \cite{paris2006fast}. In all the applications, input images and the gradients, are firstly normalized into range $[0, 1]$ before smoothing and then normalized back to their original range after smoothing.

\textbf{Image detail enhancement} is a fundamental application of image smoothing. By subtracting the smoothed image from the original input image, one can obtain the detail layer which contains the high-frequency details of the original input image. Then the detail layer is magnified and added back to the original input image to get the detail enhanced image. If edges in the smoothed image are sharper than those in the original input image, then these edges will be boosted in a reverse direction which causes the gradient reversals \cite{farbman2008edge, he2013guided}. When the edges are heavily blurred, there will be halos along the blurred edges in the detail enhanced image. This usually occurs along salient edges. The GF \cite{he2013guided} does not cause gradient reversals where a theoretical guarantee exists. However, it can cause halos in the enhanced images as shown in the region highlighted with the red box in Fig.~\ref{FigDetailEnhancement}(b). The gradient $L_0$ norm \cite{xu2011image} can cause clear gradient reversals in the result as shown in the region highlighted with the blue box in Fig.~\ref{FigDetailEnhancement}(c). This is due to its smoothing properties which eliminate low-amplitude structures while sharpen salient edges. The WLS \cite{farbman2008edge} can produce results free of gradient reversals and halos as illustrated in Fig.~\ref{FigDetailEnhancement}(d).

As we have stated in Sec.~\ref{SecBLF}, the BLF can cause both gradient reversals and halos in the results. Small $\sigma_s$ and $\sigma_r$ could cause gradient reversals while large ones may result in halos. Fig.~\ref{FigDetailEnhancement}(e) shows the result of the BLF obtained with $\sigma_s=12,\sigma_r=0.08$. Clear gradient reversals exist in the result as shown in the region highlighted with the blue box. When larger $\sigma_s=12,\sigma_r=0.3$ are adopted, halos occur along salient edges as shown in the region highlighted with the red box in Fig.~\ref{FigDetailEnhancement}(g). We show the results of our BLF-LS in Fig.~\ref{FigDetailEnhancement}(f) and (h) which have similar smoothing strength on the input image as those in  Fig.~\ref{FigDetailEnhancement}(e) and (g). However, our method can properly eliminate the gradient reversals and halos. Similar comparison between the results of the AMF \cite{gastal2012adaptive} and our AMF-LS is shown in Fig.~\ref{FigDetailEnhancement}(i)$\sim$(l).

\textbf{HDR tone mapping} is another popular application that needs edge-preserving smoothing. It can be achieved by decomposing the HDR image into a piecewise smooth base layer and a detail layer \cite{durand2002fast,farbman2008edge,li2005compressing}. The base layer is the smoothed output of input HDR image. The base layer is then nonlinearly mapped to a low dynamic range and re-combined with the detail layer to get the final tone mapped image. We adopt the framework in \cite{durand2002fast} where layer decomposition is applied to the logarithmic HDR images. Similar to image detail enhancement, gradient reversals and halos are also the challenging issues in HDR tone mapping. Fig.~\ref{FigHDRToneMapping} shows results of different methods. Similar to the results in image detail enhancement, except for the WLS, all the other compared methods have either gradient reversals (highlighted with the blue box) or halos (highlighted with the red box) in their results. The AMF, BLF and NC filter have both artifacts in their results. On the contrary, our AMF-LS, BLF-LS and NC-LS can properly eliminate these artifacts in the results which are comparable with that of the WLS.

\textbf{Flash/no-flash filtering} was proposed by Petschnigg et~al. \cite{petschnigg2004digital} to smooth a noisy no-flash image with the guidance of the flash image. In \cite{he2013guided}, He et~al. showed that gradient reversal artifacts also existed in the smoothed results of joint bilateral filter \cite{petschnigg2004digital}. Fig.~\ref{FigFlashNoFlash}(d) shows one example. In our experiments, we find that the result of AMF \cite{gastal2012adaptive} also suffers from this problem as shown in Fig.~\ref{FigFlashNoFlash}(c). Our method can also be used to perform this kind of joint image filtering in a similar way. We show the results of our AMF-LS and BLF-LS in Fig.~\ref{FigFlashNoFlash}(e) and (f) which are comparable to that of GF \cite{he2013guided} in Fig.~\ref{FigFlashNoFlash}(b).

In the above paragraphs, we compare our method with other state-of-the-art local and global methods through several applications. Besides the effectiveness in handling gradient reversals and halos, our method can also run in an efficient manner. We summarize the detailed comparison between our method and the other methods in Table~\ref{TabComp}. Especially, our method has the same properties as the WLS \cite{farbman2008edge} but our method is more than $10\times$ faster. As a global method, the computational cost of our method is even close to that of the GF \cite{he2013guided} which is a local method. The running time of $L_0$ norm smoothing and WLS is based on the implementations provided by the authors. For GF, AMF and NC, we adopt their official released version in OpenCV. Our method is implemented in Python, and we import the ``cv2'' package to make use of the implementation of AMF and NC in OpenCV. We re-implement the BLF with Python based on its fast approximation proposed by Paris et~al. \cite{paris2006fast}\footnote{Their MATLAB implementation can be download here: \url{http://people.csail.mit.edu/jiawen/software/bilateralFilter.m}.}.

Besides the above applications, our rolling guidance NC-LS can also be applied to more complex tasks: image texture removal and clip-art compression artifacts removal. We adopt the implementations, which are provided by the authors, of all the compared methods for comparison.

\textbf{Image texture removal} aims at extracting semantically meaningful structures under the complex texture patterns, which could be regular, near-regular or irregular \cite{xu2012structure}. Fig.~\ref{FigTextureSmooth} shows texture removal results of different methods. The relative total variation (RTV) \cite{xu2012structure} can efficiently smooth the textures while properly preserve edges. However, it can also smooth out some small structures as labeled with the red arrow in Fig.~\ref{FigTextureSmooth}(c). Besides, it is prone to smooth out the shading on the object surface \cite{cho2014bilateral}. The RGF \cite{zhang2014rolling} and bilateral texture filtering (BTF) \cite{cho2014bilateral} perform better in restoring the shading on the object surface but they cannot completely remove the image textures as shown in Fig.~\ref{FigTextureSmooth}(b) and (d). In contrast, our rolling guidance NC-LS can properly smooth out the image textures while preserve small structures. It also performs better than the RTV in restoring the shading on the object surface as shown in Fig.~\ref{FigTextureSmooth}(e). In addition, our method are more efficient than the RTV and BTF. They take 3.2 seconds and 5.3 seconds, respectively, to process the image. In contrast, our method only needs 0.27 seconds. The RGF needs 0.21 seconds which is a little faster than ours, however, our method shows better performance.
\begin{table*}[!ht]
  \centering
  \caption{Property comparison summary of different methods. The running time is measured on processing a $1024\times1024$ RGB color image.}\label{TabComp}
  \resizebox{1\textwidth}{!}
  {
  \begin{tabular}{|c|c|c|c|c|c|c|c|c|c|}
     \hline
      & GF \cite{he2013guided} & $L_0$ norm \cite{xu2011image} & WLS \cite{farbman2008edge} & BLF \cite{tomasi1998bilateral} & AMF \cite{gastal2012adaptive} & NC filter \cite{gastal2011domain} & BLF-LS & AMF-LS & NC-LS \\
      \hline
      runtime efficiency & 0.32s & 4.76s & 6.83s & 0.16s & 0.2s & 0.13s & 0.48s & 0.56s & 0.42s\\
      halos & $\checkmark$ & $\times$ & $\times$ & $\checkmark$ & $\checkmark$ & $\checkmark$ & $\times$ & $\times$ & $\times$ \\
      gradient reversals & $\times$ & $\checkmark$ & $\times$ & $\checkmark$ & $\checkmark$ & $\checkmark$ & $\times$ & $\times$ & $\times$ \\
      local/global & local & global & global & local & local & local & global & global & global\\
     \hline
   \end{tabular}
   }
\end{table*}

\textbf{Clip-art compression artifacts removal} aims to remove compression artifacts in compressed clip-art/carton images. This kind of images are piecewise constant with sharp edges. Clear compression artifacts exist around edges when they are compressed. General denoising approaches do not suit this task as the compression artifacts are strongly correlated with edges. The early work would be the image analogies approach proposed by Wang et~al. \cite{wang2006deringing}, but their method cannot handle images with a high compression rate as shown in Fig.~\ref{FigClipArt}(b). The gradient $L_0$ norm smoothing \cite{xu2011image} cannot completely remove the strong noise near edges as shown in the region highlighted with the blue box in Fig.~\ref{FigClipArt}(d). The $L_0$ norm region fusion \cite{nguyen2015fast} also shares the same problem. In addition, it can also cause clear artifacts along salient edges as shown in the highlighted region in the red box in Fig.~\ref{FigClipArt}(e). The BTF \cite{cho2014bilateral} was also applied to remove the compression artifacts. As shown in Fig.~\ref{FigClipArt}(c), it can properly eliminate the strong noise near edges but it cannot preserve sharp edges. Fig.~\ref{FigClipArt}(f) shows the result of our rolling guidance NC-LS. As can be observed from the figure, strong noise near edges are well smoothed while sharp edges are preserved. In term of processing time, the gradient $L_0$ norm smoothing takes 0.51 seconds, the $L_0$ norm region fusion takes 0.12 seconds, the BTF takes 1.8 seconds and our rolling guidance NC-LS only takes 0.1 seconds.

\section{Conclusion}
\label{SecConclusion}
In this paper, we propose a new global method that embeds the bilateral filter \cite{tomasi1998bilateral} and its variants \cite{gastal2012adaptive} \cite{gastal2011domain} in the least squares framework. The proposed method can have comparable performance with the state-of-the-art global method in producing results free of gradient reversals and halos. In addition, since the proposed method can take advantages of the efficiency of the bilateral filter, its variants and the least squares model, it can run much faster. The computational cost of our method is even comparable with the state-of-the-art local methods. We validate the effectiveness and efficiency of the proposed method in a number of applications with comprehensive experimental results. We also show the flexibility of our method with several extensions in handling complex applications in an efficient manner.

\bibliographystyle{unsrt}
\bibliography{refs}


\begin{IEEEbiography}[{\includegraphics[width=1in,height=1.25in,clip,keepaspectratio]{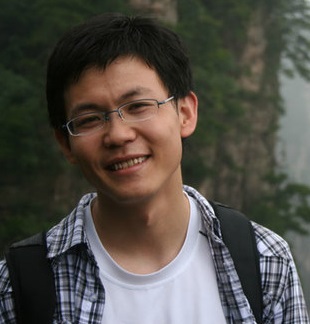}}]{Wei Liu} received the BS degree from the department of Automation of Xi`an Jiaotong University in 2012, Currently he is a Ph.D candidate of Shanghai Jiao Tong University at the Institute of Image Processing and Pattern Recognition under the supervision of professor Jie Yang. His current research interests mainly focus on image filtering and computational graphics.
\end{IEEEbiography}

\begin{IEEEbiography}[{\includegraphics[width=1in,height=1.25in,clip,keepaspectratio]{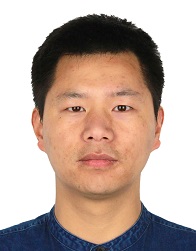}}]{Pingping Zhang} received his B.E. degree in mathematics and applied mathematics, Henan Normal University (HNU), Xinxiang, China, in 2012. He is currently a Ph.D. candidate in the School of Information and Communication Engineering, Dalian University of Technology (DUT), Dalian, China. His research interests are in deep learning, saliency detection, object tracking and semantic segmentation.
\end{IEEEbiography}

\begin{IEEEbiography}[{\includegraphics[width=1in,height=1.25in,clip,keepaspectratio]{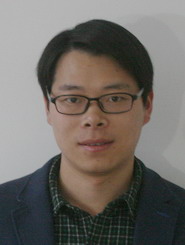}}]{XiaogangChen} (M'11) received the Ph.D. degree from Shanghai Jiao Tong University, Shanghai, China, in 2013. Then he joined the University of Shanghai for Science and Technology, Shanghai, as an Assistant Professor. His current research interests mainly focus on low level computer vision including image/video processing.
\end{IEEEbiography}

\begin{IEEEbiography}[{\includegraphics[width=1in,height=1.25in,clip,keepaspectratio]{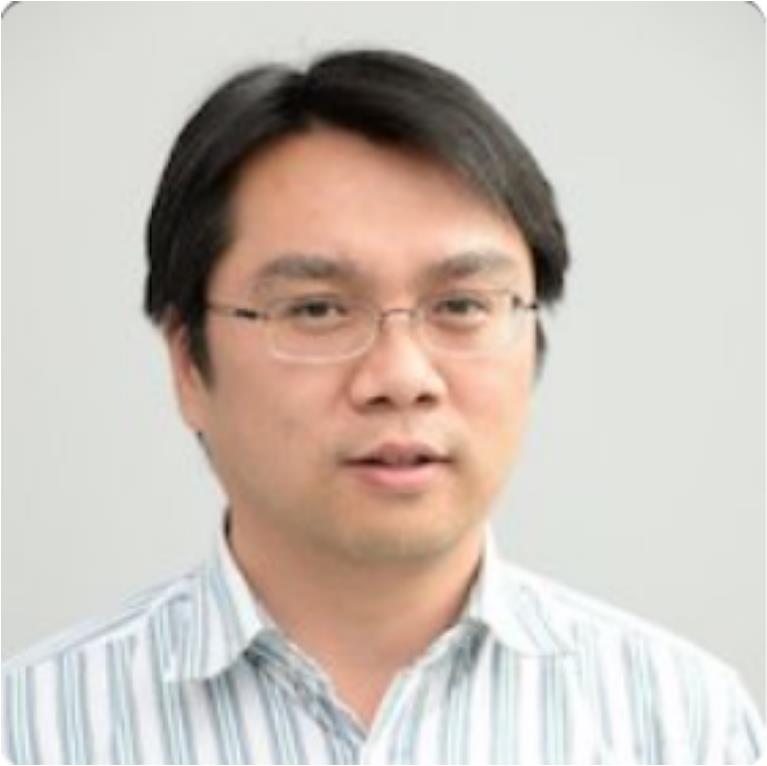}}]{Chunhua Shen} studied at Nanjing University, at Australian National University, and received the PhD degree from University of Adelaide. He is a professor at School of Computer Science, The University of Adelaide. His research interests are in the intersection of computer vision and statistical machine learning. In 2012, he was awarded the Australian Research Council Future Fellowship.
\end{IEEEbiography}

\begin{IEEEbiography}[{\includegraphics[width=1in,height=1.25in,clip,keepaspectratio]{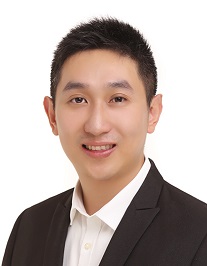}}]{Xiaolin Huang} (S'10-M'12-SM'18)  received the B.S. degree in control science and engineering, and the B.S. degree in applied mathematics  from Xi'an Jiaotong University, Xi'an, China in 2006. In 2012, he received the Ph.D. degree in control science and engineering from Tsinghua University, Beijing, China. From 2012 to 2015, he worked as a postdoctoral researcher in ESAT-STADIUS, KU Leuven, Leuven, Belgium. After that he was selected as an Alexander von Humboldt Fellow and working in Pattern Recognition Lab, the Friedrich-Alexander-Universit\"{a}t Erlangen-N\"{u}rnberg, Erlangen, Germany. From 2016, he has been an Associate Professor at Institute of Image Processing and Pattern Recognition, Shanghai Jiao Tong University, Shanghai, China. In 2017, he was awarded by ``1000-Talent Plan'' (Young Program). His current research areas include machine learning and optimization.

\end{IEEEbiography}

\begin{IEEEbiography}[{\includegraphics[width=1in,height=1.25in,clip,keepaspectratio]{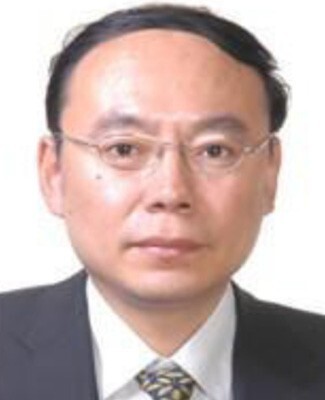}}]{Jie Yang}
received his PhD from the Department of Computer Science, Hamburg University, Germany, in 1994. Currently, he is a professor at the Institute of Image Processing and Pattern Recognition, Shanghai Jiao Tong University, China. He has led many research projects (e.g.,National Science Foundation, 863 National High Tech. Plan), had one book published in Germany, and authored more than 200 journal papers. His major research interests are object detection and recognition, data fusion and data mining, and medical image processing.
\end{IEEEbiography}


\end{document}